\documentclass[useAMS,usenatbib]{mn2e}
\usepackage{epsfig}
\usepackage{float}
\usepackage[T1]{fontenc}
\usepackage{graphicx}
\usepackage{epstopdf}
\usepackage{natbib}
\usepackage{color}
\usepackage{aas_macros}
\usepackage{epstopdf}
\usepackage{pdfpages}

\usepackage{caption}
\usepackage{amsmath}
\usepackage{amsfonts}
\usepackage{amssymb}
\usepackage{threeparttable}
\usepackage{gensymb}
\usepackage{kantlipsum}
\usepackage{etoolbox}
\usepackage{color}
\usepackage{tikz}
\usepackage{tabularx}
\usepackage{booktabs}

\usepackage[position=b, captionskip=-0.4cm, farskip=0.4cm]{subfig}

\usepackage{array}
\newcolumntype{L}[1]{>{\raggedright\let\newline\\
\arraybackslash\hspace{0pt}}m{#1}}
\newcolumntype{C}[1]{>{\centering\let\newline\\
\arraybackslash\hspace{0pt}}m{#1}}
\newcolumntype{R}[1]{>{\raggedleft\let\newline\\
\arraybackslash\hspace{0pt}}m{#1}}

\def\sigs{\mbox{$\sigma_\star$}}

\def\Msun{\mbox{$M_\odot$}}

\def\Mdyn{\mbox{$M_{\rm dyn}$}}

\def\mst{\mbox{$M_{\star}$}}

\def\lsim{\mathrel{\rlap{\lower3.5pt\hbox{\hskip0.5pt$\sim$}}
    \raise0.5pt\hbox{$<$}}}                
\def\gsim{~\rlap{$>$}{\lower 1.0ex\hbox{$\sim$}}}

\def\SN{\mbox{$S/N$}}
\def\Mauto{\mbox{{\tt MAG\_AUTO}}}
\def\Mautor{\mbox{{\tt MAG\_AUTO\_r}}}
\def\MErrautor{\mbox{{\tt MAGERR\_AUTO\_r}}}
\def\zp{\mbox{$z_{\rm phot}$}}
\def\zs{\mbox{$z_{\rm spec}$}}

\def\Fig{\mbox{Figure~}}

\def\Tab{\mbox{Table~}}
\def\Tabs{\mbox{Tables~}}
\def\Sec{\mbox{Section~}}
\def\Secs{\mbox{Sections~}}
\def\App{\mbox{Appendix~}}

\def\magapfour{\mbox{{\tt MAGAP\_4}}}
\def\magapsix{\mbox{{\tt MAGAP\_6}}}
\def\maggaap{\mbox{{\tt MAG\_{GAaP}}}}

\def\UCMG{\mbox{{\tt UCMG}}}
\def\UCMGs{\mbox{{\tt UCMGs}}}

\def\rband{\mbox{$r$-band}}

\def\Ksband{\mbox{$Ks$-band}}

\def\Zsun{\mbox{$Z_{\rm \odot}$}}

\def\Re{\mbox{$R_{\rm e}$}}
\def\Te{\mbox{$\Theta_{\rm e}$}}

\def\Temaj{\mbox{$\Theta_{\rm e, maj}$}}

\def\sqd{\mbox{~sq.~deg.}}
\def\muemed{\mbox{$\langle\mu_{\rm e}\rangle$}}

\def\lephare{\mbox{\textsc{le phare}}}
\def\twodphot{\mbox{\textsc{2dphot}}}
\def\ppgui{\mbox{\textsc{PPGUI}}}
\def\ppxf{\mbox{\textsc{pPXF}}}

\def\iraf{\mbox{\textsc{iraf}}}
\def\python{\mbox{\textsc{python}}}
\def\astropy{\mbox{\textsc{astropy}}}

\def\identify{\mbox{\emph{IDENTIFY}}}
\def\twodspec{\mbox{\emph{TWODSPEC.LONGSLIT}}}
\def\background{\mbox{\emph{BACKGROUND}}}

\def\MzptML{\mbox{{\tt MFREE-zpt-phot}}}
\def\Mzptspec{\mbox{{\tt MFREE-zpt-spec}}}

\def\MML{\mbox{{\tt MFREE-phot}}}
\def\Mspec{\mbox{{\tt MFREE-spec}}}

\def\modMnozpt{\mbox{{\tt MFREE}}}
\def\modMzpt{\mbox{{\tt MFREE-zpt}}}

\def\MLone{\mbox{{\tt ML1}}}
\def\MLtwo{\mbox{{\tt ML2}}}

\def\CF{\mbox{$\mathcal{C}_F$}}
\def\IF{\mbox{$\mathcal{I}_F$}}

\def\UP{\mbox{{\tt UCMG\_PHOT}}}
\def\USS{\mbox{{\tt UCMG\_SPEC\_SPEC}}}
\def\UPS{\mbox{{\tt UCMG\_PHOT\_SPEC}}}
\def\UNEW{\mbox{{\tt UCMG\_NEW}}}
\def\UTNG{\mbox{{\tt UCMG\_TNG}}}
\def\UNTT{\mbox{{\tt UCMG\_NTT}}}

\title[UCMGs in KiDS]{The first sample of spectroscopically confirmed ultra-compact massive galaxies in the Kilo Degree Survey}

\author[Tortora C. et al.]{\noindent
C.~Tortora$^{1}$\thanks{E-mail: ctortora@astro.rug.nl},
N.R.~Napolitano$^{2}$, M.~Spavone$^{2}$, F.~La~Barbera$^{2}$,
G.~D'Ago$^{2}$, C.~Spiniello$^{2}$, \and K.~H.~Kuijken$^{3}$,
N.~Roy$^{2,4}$, M.~A.~Raj$^{2}$, S.~Cavuoti$^{2,4}$,
M.~Brescia$^{2}$, G.~Longo$^{4}$, \and V.~Pota$^{2}$,
C.~E.~Petrillo$^{1}$, M.~Radovich$^{5}$, F.~Getman$^{2}$,
L.V.E.~Koopmans$^{1}$, I.~Trujillo$^{6,7}$, \and G.~Verdoes
Kleijn$^{1}$, M.~Capaccioli$^{4}$, A.~Grado$^{2}$,
G.~Covone$^{4}$, D.~Scognamiglio$^{2}$, \and C.~Blake$^{8}$,
K.~Glazebrook$^{8}$, S.~Joudaki$^{8,9,10}$, C.~Lidman$^{11}$,
C.~Wolf$^{12}$
\\~\\
$^1$ Kapteyn Astronomical Institute, University of Groningen, P.O.
Box 800, 9700 AV Groningen, the Netherlands \\
$^2$ INAF -- Osservatorio Astronomico di
Capodimonte, Salita Moiariello, 16, 80131 - Napoli, Italy\\
$^3$ Leiden Observatory, Leiden University, P.O. Box 9513, 2300 RA
Leiden, the Netherlands\\
$^4$ Dipartimento di Scienze Fisiche,
Universit\`{a} di Napoli Federico II, Compl. Univ. Monte S.
Angelo, 80126 - Napoli,
Italy\\
$^5$ INAF -- Osservatorio Astronomico di Padova, Vicolo
Osservatorio 5, 35122 - Padova, Italy\\
$^6$ Instituto de Astrof\'{\i}sica de Canarias, c/ V\'{\i}a
L\'actea s/n,
E-38205, La Laguna, Tenerife, Spain\\
$^7$ Departamento de Astrof\'{\i}sica, Universidad de La Laguna,
E-38206, La Laguna, Tenerife, Spain\\
$^8$ Centre for Astrophysics and Supercomputing, Swinburne
University of Technology, P.O. Box 218, Hawthorn, VIC 3122,
Australia\\
$^{9}$ ARC Centre of Excellence for All-sky Astrophysics (CAASTRO)\\
$^{10}$ Department of Physics, University of Oxford, Denys
Wilkinson
Building, Keble Road, Oxford OX1 3RH, U.K.\\
$^{11}$ Australian Astronomical Observatory, North Ryde, NSW 2113,
Australia\\
$^{12}$ Research School of Astronomy and Astrophysics, The
Australian National University, Canberra, ACT 2611, Australia\\}

\begin{document}
\date{Accepted  Received }
\pagerange{\pageref{firstpage}--\pageref{lastpage}} \pubyear{xxxx}
\maketitle

\label{firstpage}
\begin{abstract}
We present results from an ongoing investigation using the Kilo
Degree Survey (KiDS) on the VLT Survey Telescope (VST) to provide
a census of ultra-compact massive galaxies (\UCMGs), defined as
galaxies with stellar masses $\mst > 8 \times 10^{10} \rm  \Msun$
and effective radii $\Re < 1.5\,\rm kpc$. \UCMGs, which are
expected to have undergone very few merger events, provide a
unique view on the accretion history of the most massive galaxies
in the Universe. Over an effective sky area of nearly 330 square
degrees, we select \UCMG\ candidates from KiDS multi-colour
images, which provide high quality structural parameters,
photometric redshifts and stellar masses. Our sample of $\sim
1000$ photometrically selected \UCMGs\ at $z < 0.5$ represents the
largest sample of \UCMG\ candidates assembled to date over the
largest sky area. In this paper we present the first effort to
obtain their redshifts using different facilities, starting with
first results for 28 candidates with redshifts $z < 0.5$, obtained
at NTT and TNG telescopes. We confirmed, as bona fide \UCMGs, 19
out of the 28 candidates with new redshifts. A further 46 \UCMG\
candidates are confirmed with literature spectroscopic redshifts
(35 at $z < 0.5$), bringing the final cumulative sample of
spectroscopically-confirmed lower-z \UCMGs\ to 54 galaxies, which
is the largest sample at redshifts below $0.5$. We use these
spectroscopic redshifts to quantify systematic errors in our
photometric selection, and use these to correct our \UCMG\ number
counts. We finally compare the results to independent datasets and
simulations.
\end{abstract}


\section{Introduction}\label{sec:intro}

The ``zoo'' of galaxies we observe in the present-day Universe
reflects a variety of physical processes that have shaped galaxies
across the ages. Galaxies fall into two main, broad classes:
star-forming blue and passive red galaxies
(\citealt{Kauffmann+03}). At redshifts $z>2$, the most massive
star-forming and passive galaxies also have systematically
different structural properties, indicating that they have
undergone different physical processes. Whereas the massive blue
star-forming disks at these redshifts have effective radii of
several kpc (\citealt{Genzel+08}), the passive, quenched spheroids
(the so called ``red nuggets'') have small effective radii, of
about 1 kpc. Galaxies in this massive red population at $z>2$ are
thought to have undergone a sequence of processes: a)
accretion-driven violent disc instability, b) dissipative
contraction resulting in the formation of compact, star-forming
``blue nuggets'', c) quenching of star formation (see
\citealt{Dekel_Burkert14} for further details). At lower
redshifts, corresponding to the last $10$ Gyr of evolution,
massive red galaxies are considerably larger, as revealed in
detailed studies of the local population of early-type galaxies
(ETGs, ellipticals and lenticulars; \citealt{Daddi+05};
\citealt{Trujillo+06, Trujillo+07}; \citealt{vanderWel+08}).

Dry merging has long been advocated as the dominant mechanism with
which to explain the size and stellar mass growth of massive
galaxies (\citealt{Cox+06}; \citealt{Khochfar_Burkert03,
Khochfar_Silk06}; \citealt{Cenarro_Trujillo09}). This process is
believed to be common for very massive systems at high redshifts.
On one side, for the most massive galaxies, different simulations
predict major merger rates (mergers per galaxy per Gyr) in the
range $0.3-1 \, \rm Gyr^{-1}$ at $z \sim 2$ and smaller than $0.2
\, \rm Gyr^{-1}$ at $z  \lsim 0.5$
(\citealt{Hopkins+10_Mergers_LCDM}). On the other side, more
recently various theoretical and observational studies, focussing
on the finer details of the galaxy mass build-up, have started to
exclude major mergers as the leading process in the formation of
massive ETGs, favoring minor mergers instead. Such a scenario can
provide the modest stellar mass accretion with the strong size
evolution that is observed (\citealt{Naab+09};
\citealt{vanDokkum+10}; \citealt{Trujillo+11}; \citealt{Hilz+13};
\citealt{BNE14}; \citealt{Ferreras+14};
\citealt{Tortora+14_DMevol, Tortora+18_KiDS_DMevol}).

Over cosmic time, most of the high-z compact galaxies evolve into
present-day, massive and big galaxies. However, might a fraction
of these objects survive intact till the present epoch, resulting
in compact, old, relic systems in the nearby Universe? An
increasing number of results at low/intermediate redshifts seems
to indicate that this could be the case, with different studies
aiming at increasing the size of \UCMG\ datasamples and at
analyzing in detail the stellar/structural/dynamical properties of
compact galaxies in relation to their environment
(\citealt{Trujillo+09_superdense, Trujillo+12_compacts,
Trujillo+14}; \citealt{Taylor+10_compacts};
\citealt{Valentinuzzi+10_WINGS}; \citealt{Shih_Stockton11};
\citealt{Ferre-Mateu+12, Ferre-Mateu+15};
\citealt{Lasker+13_IMF_compact}; \citealt{Poggianti+13_low_z,
Poggianti+13_evol}; \citealt{Damjanov+13_compacts,
Damjanov+14_compacts, Damjanov+15_compacts,
Damjanov+15_env_compacts}; \citealt{Gargiulo+16_dense,
Gargiulo+17_dense}; \citealt{Hsu+14_compacts};
\citealt{Stockton+14_compacts}; \citealt{Saulder+15_compacts};
\citealt{Stringer+15_compacts}; \citealt{Yildirim+15};
\citealt{Wellons+15_lower_z}; \citealt{Tortora+16_compacts_KiDS};
\citealt{Charbonnier+17_compact_galaxies}; \citealt{Beasley+18};
\citealt{Buitrago+18_compacts}).

On the theoretical side, simulations predict that the fraction of
objects that survive without undergoing any significant
transformation since $z \sim 2$ is about $1-10\%$
(\citealt{Hopkins+09_DELGN_IV}; \citealt{Quilis_Trujillo13}), and
at the lowest redshifts (i.e., $z \lsim 0.2$), they predict
densities of relics  of $10^{-7}-10^{-5}\, \rm Mpc^{-3}$. Thus, in
local wide surveys, as the Sloan Digital Sky Survey (SDSS), we
would expect to find few of these objects.
\cite{Trujillo+09_superdense} have originally found 29 young
ultra-compact ($\Re < 1.5$ kpc), massive ($\mst > 8 \times
10^{10}\, \rm \Msun$) galaxies (\UCMGs, hereafter) in SDSS--DR6 at
$z \lsim 0.2$ and no old systems at all (see also
\citealt{Taylor+10_compacts}; \citealt{Ferre-Mateu+12}). However,
the recent discovery that NGC~1277 in the Perseus cluster may be
an example of a true relic galaxy has re-opened the issue
(\citealt{Trujillo+14}; \citealt{Martin-Navarro+15_IMF_relic}).
Very recently, the same group, relaxing the constraint on the size
(i.e. taking larger values for this quantity) added two further
relic galaxies, Mrk 1216 and PGC 032873, setting the number
density of these compact galaxies within a distance of 106 Mpc at
the value $\sim 6 \times 10^{-7}\, \rm Mpc^{-3}$
(\citealt{Ferre-Mateu+17}). Other candidates have been found by
\cite{Saulder+15_compacts}, although only a few of them are
ultra-compact and massive, and none of them have $z < 0.05$.
\cite{Poggianti+13_low_z} have found, in the local Universe, 4 old
\UCMGs\ within 38\sqd\ in the WINGS survey. In contrast to these
poor statistics, the number of (young and old) compact systems at
lower masses ($< 10^{11}\, \rm \Msun$) is larger, independently of
the compact definition (\citealt{Valentinuzzi+10_WINGS};
\citealt{Poggianti+13_low_z}).

In the intermediate redshift range ($0.2 \lsim z \lsim 0.8$),
compacts have been investigated in detail by
\cite{Damjanov+14_compacts} within the 6373.2\sqd\ of the BOSS
survey. The first systematic and complete analysis was performed
in \cite{Damjanov+15_compacts}, who analyzed F814W HST images for
the COSMOS field, providing robust size measurements for a sample
of 1599 compact systems in the redshift range $0.2 \lsim  z \lsim
0.8$. 45 out of 1599 of their galaxies are \UCMGs\ ($\sim 10$
\UCMGs\ at $z \lsim 0.5$). Recently,
\cite{Charbonnier+17_compact_galaxies} have scanned the $\sim
170\sqd$ of the CFHT equatorial SDSS Stripe 82 (CS82) survey,
finding thousands of compact galaxies, according to different mass
and size selection criteria, and about 1000 photometrically
selected compact galaxies with $\Re < 2 \, \rm kpc$ (and $\sim 20$
galaxies with available SDSS spectra).

The population of such dense passively evolving galaxies in this
intermediate redshift range represents a link between the red
nuggets at high z, and their relics in the nearby Universe. This
is why a large sample of compact galaxies, with high-quality
photometry (to derive reliable structural parameters) and
spectroscopic data, are actually necessary to better trace this
transition.

In \cite{Tortora+16_compacts_KiDS} we have provided an independent
contribution to this field by starting a first census of \UCMGs\
in the Kilo Degree Survey (KiDS; \citealt{deJong+15_KiDS_paperI,
deJong+17_KiDS_DR3}). KiDS is one of the ESO public surveys being
carried out with the VLT Survey Telescope (VST;
\citealt{Capaccioli_Schipani11}), aiming at observing 1500 square
degrees of the sky, in four optical bands ($ugri$), with excellent
seeing (e.g. $0.65''$ median FWHM in \rband). Among other
advantages, the KiDS image quality makes the data very suitable
for measuring structural parameters of galaxies, including compact
ones. The \cite{Tortora+16_compacts_KiDS} study used the first
$\sim 150$\sqd\ of KiDS data (data release DR1/2), and found $\sim
100$ new \UCMG\ candidates at $z \lsim 0.7$.

According to predictions from simulations, we can expect to find
$\sim 0.3-3.5$ relic \UCMGs\ per square degree, at redshift $z
<0.5$ (\citealt{Quilis_Trujillo13}). This prediction does
critically depend on the physical processes shaping size and mass
evolution of galaxies, such as the relative importance of major
and minor galaxy merging. At such low densities, gathering large
samples across wide areas is essential to reduce Poisson errors
and Cosmic Variance. This makes possible to compare with
theoretical predictions for \UCMG\ number counts, and to
investigate the role of the environment in shaping their
structural and stellar population properties. Scanning KiDS images
to pick up photometrically selected \UCMG\ candidates yields a
useful sample size, but it requires a second step consisting of
the spectroscopic validation of (at least a fraction of) our
candidates. This massive effort can be faced only using a
multi-site and multi-facility approach in the North and South
hemisphere: the multi-site will allow to cover the two KiDS
patches, while the multi-facility will allow to optimise the
exposure time according to the target brightness. In this paper we
present the first results of our spectroscopic campaign, with
observations obtained at Telescopio Nazionale Galileo (TNG) and
New Technology Telescope (NTT). We finally update the results in
\cite{Tortora+16_compacts_KiDS}, calculating number counts across
an area of 333\sqd\ of the KiDS survey.

The paper is organized as follows. In \Sec\ref{sec:sample} we
present the KiDS sample of high signal-to-noise ratio galaxies,
and the sub-samples of our spectroscopically and photometrically
selected \UCMGs. Strategy, status of the spectroscopic campaign
and first observations at TNG and NTT are discussed in
\Sec\ref{sec:spectroscopy}. We analyze the spectroscopically
confirmed \UCMG\ sample in \Sec\ref{sec:validation}, investigating
the source of systematics in the selection procedure of \UCMGs\
and the impact on the number counts. Number counts are presented
and discussed in \Sec\ref{sec:number_counts}. A discussion of the
results and future prospects are outlined in
\Sec\ref{sec:conclusions}.

Decimal logarithms are used in the paper. To convert radii in
physical scales and redshifts in distances we adopt a cosmological
model with $(\Omega_{m},\Omega_{\Lambda},h)=(0.3,0.7,0.7)$, where
$h = H_{0}/100 \, \textrm{km} \, \textrm{s}^{-1} \,
\textrm{Mpc}^{-1}$ (\citealt{Komatsu+11_WMAP7}).

\section{Sample selection}\label{sec:sample}

The galaxy samples presented in this work are part of the data
included in the first, second and third data releases of KiDS,
presented in \cite{deJong+15_KiDS_paperI} and
\cite{deJong+17_KiDS_DR3}, consisting of 440 total survey tiles
(corresponding to a total survey area of $\sim 447 \sqd$). We
refer the interested reader to these papers for more details.

We list in the following section the main steps for the galaxy
selection procedure and the determination of galaxy physical
quantities such as structural parameters, photometric redshifts
and stellar masses. The whole procedure was also outlined in
\cite{Tortora+16_compacts_KiDS}.

\subsection{Galaxy data sample}\label{subsec:datasample}

We started from the KiDS multi-band source catalogs, where the
photometry has been obtained with S-Extractor
(\citealt{Bertin_Arnouts96_SEx}) in dual image mode, using as
reference the positions of the sources detected in the r-band
images, which has the best image quality among KiDS filters.
Star/galaxy separation is based on the distribution of the
S-Extractor parameters {\tt CLASS\_STAR} and \SN\ (signal-to-noise
ratio) of a number of sure stars (see
\citealt{LaBarbera_08_2DPHOT};
\citealt{deJong+15_KiDS_paperI,deJong+17_KiDS_DR3}). Image defects
such as saturated pixels, star spikes, reflection halos, satellite
tracks, etc. have been masked using both a dedicated automatic
procedure and visual inspection. We have discarded all sources in
these areas. After masking of bad areas, we collected a catalog
consisting of $\sim 5$ millions of galaxies within an effective
area of 333 \sqd.

Relevant properties for each galaxy have been derived as described
here below:
\begin{itemize}
\item {\it Integrated optical photometry.} For our analysis we have adopted Kron-like total magnitude, \Mauto, aperture
magnitudes \magapfour\ and \magapsix, measured within circular
apertures of 4 and 6 arcsec of diameter, respectively. We also use
Gaussian Aperture and PSF (GAaP) magnitudes, \maggaap\ (see
\citealt{deJong+17_KiDS_DR3} for further details).
\item {\it KiDS structural parameters.} Surface photometry has
been performed using the \twodphot\ environment. \twodphot\
produces a local PSF model from a series of identified {\it sure
stars}, by fitting the two closest stars to that galaxy with a sum
of two two-dimensional Moffat functions. Then galaxy snapshots are
fitted with PSF-convolved S\'ersic models having elliptical
isophotes plus a local background value (see
\citealt{LaBarbera_08_2DPHOT} for further details). The fit
provides the following parameters for the four wavebands: average
surface brightness \muemed, major-axis effective radius, \Temaj,
S\'ersic index, $n$, total magnitude, $m_{S}$, axis ratio, $q$,
and position angle. In the paper we use the circularized effective
radius, \Te, defined as $\Te = \Temaj \sqrt{q}$. Effective radius
are converted to the physical scale value \Re\ using the measured
(photometric or spectroscopic) redshift (see next items). To judge
the quality of the fit, we also computed a reduced $\chi^2$, and a
modified version, $\chi^{\prime 2}$, which accounts for the
central image pixels only, where most of the galaxy light is
concentrated. Large values for $\chi^{2}$ (typically $> 1.5$)
correspond to strong residuals, often associated to spiral arms
(\citealt{Roy+18}).
\item {\it Spectroscopic redshifts.}
We have cross-matched our KiDS catalog with overlapping
spectroscopic surveys to obtain spectroscopic redshift for the
objects in common. In the Northern cap we use redshifts from the
Sloan Digital Sky Survey data release 9 (SDSS-DR9;
\citealt{Ahn+12_SDSS_DR9, Ahn+14_SDSS_DR10}) and Galaxy And Mass
Assembly data release 2 (GAMA-DR2; \citealt{Driver+11_GAMA}). GAMA
also provides information about the quality of the redshift
determination by using the probabilistically defined normalized
redshift quality scale $nQ$. When selecting \UCMGs\ we only
consider the most reliable GAMA redshifts with $nQ
> 2$. We also match with 2dFLenS fields
(\citealt{Blake+16_2dflens}), selecting only those redshifts with
quality flag $\geq 3$. SDSS, GAMA and 2dFLenS fields overlap with
$\sim 64\%$, $\sim 49\%$ and $\sim 36\%$ of our KiDS tiles, with
overlapping regions among SDSS and GAMA, and most of the matched
tiles for 2dFLenS are in the Southern cap (i.e. $\sim 93\%$ of the
total tiles in the South). In total we find $\sim 77000$ galaxies.
\item {\it Photometric redshifts.} Photometric redshifts, \zp,
are determined not with the classical SED fitting approach (e.g.,
\citealt{Ilbert+06}), but with a machine learning (ML) technique,
and in particular with the Multi Layer Perceptron with Quasi
Newton Algorithm (MLPQNA) method (\citealt{Brescia+13,
Brescia+14}; \citealt{Cavuoti+15_PhotoRApToR}) and presented in
\cite{Cavuoti+15_KIDS_I} and \cite{Cavuoti+17_KiDS}, to which we
refer the reader for all details. We use \zp\ from two distinct
networks\footnote{We used two different networks, since galaxy
samples for spectroscopic runs were extracted at two different
epochs, when the latest version of redshift released in
\cite{deJong+17_KiDS_DR3} were not available.}, which we quote as
\MLone\ and \MLtwo. Samples of spectroscopic redshifts, \zs, from
the literature, are cross-matched with KiDS sample to gather the
knowledge base (KB) and train the network.
\begin{itemize}
\item \MLone. This network was trained in the early 2015 using a mixture of the 149 survey
tiles from KiDS--DR1/2, plus few tiles from KiDS--DR3 and the
results are discussed in \cite{Cavuoti+15_KIDS_I}. Both sets of
magnitudes \magapfour\ and \magapsix\ are used. As KB we used a
sample with spectroscopic redshift from the SDSS and GAMA which
together provide redshifts up to $z \lsim 0.8$. The $1 \, \rm
\sigma$ scatter in the quantity $\Delta z \equiv (\zs -
\zp)/(1+\zs)$ is $\sim 0.03$ and the bias, defined as the absolute
value of the mean of $\Delta z$, is $\sim 0.001$.
\item \MLtwo. We gather a sample of photometrically selected \UCMGs\ using the whole KiDS--DR1/2/3 dataset.
For this sample we rely on the MLPQNA redshifts presented in
\cite{deJong+17_KiDS_DR3}. In this case we use \magapfour,
\magapsix\ and \maggaap\ magnitudes. The KB is composed by the
same spectroscopic data used for \MLone\ (i.e., spectroscopic
redshifts from SDSS and GAMA), but based on the whole 440 survey
tiles from the last public KiDS release. The statistical
indicators provide performances similar to the ones reached by
\MLone\ redshifts.
\end{itemize}
\item {\it Stellar masses.} We have used the software \lephare\
(\citealt{Arnouts+99}; \citealt{Ilbert+06}), which performs a
simple $\chi^{2}$ fitting method between the stellar population
synthesis (SPS) theoretical models and data. Single burst models
from \citet[BC03 hereafter]{BC03}, covering all the range of
available metallicities ($0.005 \leq Z/\Zsun \leq 2.5$), with $\rm
age \leq \rm age_{\rm max}$ and a \cite{Chabrier01} IMF, are
used\footnote{We find that constraining the parameter range to the
higher Z (i.e., $> 0.004\, \rm \Zsun$) and ages ($> 3$ Gyr), as
done in \cite{Tortora+18_KiDS_DMevol}, have a negligible impact on
most of the results produced in this paper.}. The maximum age,
$\rm age_{\rm max}$, is set by the age of the Universe at the
redshift of the galaxy, with a maximum value at $z=0$ of $13\, \rm
Gyr$. Age and metallicity are left free to vary in the fitting
procedure. Models are redshifted using the MLPQNA photometric
redshifts or the spectroscopic ones when available from the
literature or our spectroscopic campaign. We adopt the observed
$ugri$ magnitudes \magapsix\ (and related $1\, \sigma$
uncertainties $\delta u$, $\delta g$, $\delta r$ and $\delta i$),
which are corrected for Galactic extinction using the map in
\cite{Schlafly_Finkbeiner11}. Total magnitudes derived from the
S\'ersic fitting, $m_{S}$, are used to correct the \mst\ outcomes
of \lephare\ for missing flux. The single burst assumption is
suitable to describe the old stellar populations in the compact
galaxies we are interested in (\citealt{Thomas+05};
\citealt{Tortora+09}). We also discuss the results when
calibration zero-point errors are added in quadrature to the
uncertainties of the magnitudes derived from SExtractor
(\citealt{Bertin_Arnouts96_SEx}). In \Tab\ref{tab:tab_legend} we
list the different sets of masses used, quoting if: a) calibration
errors in the photometry zero-point $\delta_{\rm zp} \equiv
(\delta u_{zp}, \, \delta g_{zp}, \, \delta r_{zp}, \, \delta
i_{zp}) = (0.075, \, 0.074, \, 0.029, \, 0.055)$ are added in
quadrature to the uncertainties of magnitudes and b) photometric
redshift, \zp, or spectroscopic one, \zs, are used. Optical
photometry cannot efficiently break the age-metallicity
degeneracy, making the estimates of these quantities more
uncertain than stellar mass values. For this reason, and for the
main scope of the paper, we will not discuss age and metallicity
in what follows, postponing this kind of analysis to future works.
\item {\it "Galaxy classification".} Using \lephare, we have
also fitted the observed magnitudes \magapsix\ with a set of $66$
empirical spectral templates used in \cite{Ilbert+06}, in order to
determine a qualitative galaxy classification. The set is based on
the four basic templates (Ell, Sbc, Scd, Irr) described in
\cite{CWW80}, and star burst models from \cite{Kinney+96}. GISSEL
synthetic models \citep{BC03} are used to linearly extrapolate
this set of templates into ultraviolet and near-infrared. The
final set of $66$ templates ($22$ for ellipticals, $17$ for Sbc,
$12$ for Scd, $11$ for Im, and $4$ for starburst) is obtained by
linearly interpolating the original templates, in order to improve
the sampling of the colour space. The best fitted template is
considered.
\item {\it VIKING near-infrared data.}
The optical KiDS \maggaap\ magnitudes are complemented by
five-band near-infrared (NIR) magnitudes (zYJHKs) from the VISTA
Kilo-degree Infrared Galaxy (VIKING) Survey, exploited by the
VISTA telescope (\citealt{Edge+14_VIKING-DR1}). We have extracted
this NIR photometry from the individual exposures that are
pre-reduced by the Cambridge Astronomy Data Unit (CASU). After an
additional background subtraction we run GAaP with the same
matched apertures as for the optical KiDS data. As most objects
are covered by multiple exposures in a given band we have averaged
these multiple measurements. Details of the VIKING data reduction
and photometry will be presented in a forthcoming paper (Wright et
al. in preparation).
\end{itemize}

\begin{table}
\centering \caption{Parameters adopted in the calculation of the
various sets of masses used in this paper. The SPS models and the
range of fitted parameters are the same for all the sets. Then, we
include calibration errors in the photometric zero-points $\rm
\delta_{\rm zp}$, quadratically added to the SExtractor magnitude
errors. Masses are calculated using \zp\ and \zs. See text for
details.}\label{tab:tab_legend}
\begin{tabular}{lccc}
\hline
\rm Set & SPS models & $\rm \delta_{\rm zp}$ & Redshift  \\
\hline
\MML & (age, Z) free & NO & \zp \\
\Mspec & (age, Z) free & NO & \zs \\
\MzptML & (age, Z) free & YES & \zp \\
\Mzptspec & (age, Z) free & YES & \zs \\
\hline
\end{tabular}
\end{table}

We have set a threshold on the \SN\ of r-band images to retain the
highest-quality sources: we have kept only those systems with
$\SN_r \equiv$ 1/\MErrautor $> 50$, where \MErrautor\ is the error
of r-band \Mauto\ (\citealt{LaBarbera_08_2DPHOT, SPIDER-I};
\citealt{Roy+18}). The \SN\ threshold has been set on the basis of
a test performed on simulated galaxies which shows that with $\SN
\gsim 50$ we are able to perform accurate surface photometry and
to determine reliable structural parameters. The sample of
high-\SN\ galaxies is complete down to a magnitude of $\Mautor
\sim 21$ and a stellar mass of $\sim 3 \times 10^{10} \, \Msun$ up
to redshift $z \sim 0.5$. Similar values are found for the samples
of \UCMGs\ introduced in the next sections. More details are
provided in \App\ref{app:completeness}.

\begin{table*}
\centering \caption{Integrated photometry for the first 28 \UCMG\
candidates from our spectroscopic program, 6 in \UTNG\ sample and
22 in \UNTT\ sample (for each subsample the galaxies are ordered
by Right Ascension). From left we show: a) galaxy identifier; b)
galaxy name; c) r-band KiDS \Mauto, corrected for Galactic
extinction; d-g) u-, g-, r- and i-band KiDS magnitudes measured in
an aperture of 6 arcsec of diameter (i.e. \magapsix), corrected
for Galactic extinction, with 1 $\sigma$ errors; h) photometric
redshift, determined using machine learning; i) stellar mass,
determined fitting the aperture photometry using a set of
synthetic models from BC03. To correct for Galactic extinction the
\citet{Schlafly_Finkbeiner11} maps are
used.}\label{tab:phot_parameters}
\resizebox{\textwidth}{!}{\begin{tabular}{ccccccccc} \hline
\rm ID & name &  \Mautor & $u_{\rm 6''}$ & $g_{\rm 6''}$ & $r_{\rm 6''}$ & $i_{\rm 6''}$ & \zp & $\log \mst/\Msun$ \\
%
\hline
1   &   KIDS J091834.71+012246.12   &   19.13   &   23.11   $\pm$   0.25    &   20.69   $\pm$   0.01    &   19.15   $\pm$   0.003   &   18.59   $\pm$   0.008   &   0.29    &   10.97   \\
2   &   KIDS J112821.24-015320.63   &   18.56   &   21.6    $\pm$   0.07    &   19.91   $\pm$   0.001   &   18.6    $\pm$   0.002   &   18.12   $\pm$   0.005   &   0.22    &   11.12   \\
3   &   KIDS J114810.66-014447.79   &   19.87   &   22.64   $\pm$   0.18    &   21.34   $\pm$   0.02    &   19.87   $\pm$   0.007   &   19.36   $\pm$   0.013   &   0.35    &   11.     \\
4   &   KIDS J115446.15-001640.53   &   19.52   &   22.79   $\pm$   0.22    &   20.88   $\pm$   0.02    &   19.49   $\pm$   0.005   &   18.65   $\pm$   0.011   &   0.31    &   11.15   \\
5   &   KIDS J121233.85+013518.69   &   20.78   &   23.09   $\pm$   0.27    &   22.45   $\pm$   0.07    &   20.74   $\pm$   0.018   &   20.09   $\pm$   0.029   &   0.42    &   11.02   \\
6   &   KIDS J142332.83-000013.69   &   20.01   &   23.22   $\pm$   0.35    &   21.54   $\pm$   0.05    &   19.97   $\pm$   0.013   &   19.41   $\pm$   0.02    &   0.36    &   10.95   \\
\hline
7   &   KIDS J021135.09-315540.60   &   19.78   &   23.81   $\pm$   0.49    &   21.3    $\pm$   0.02    &   19.8    $\pm$   0.006   &   19.3    $\pm$   0.012   &   0.32    &   10.94   \\
8   &   KIDS J022421.66-314328.17   &   19.25   &   22.69   $\pm$   0.13    &   20.91   $\pm$   0.01    &   19.24   $\pm$   0.003   &   18.62   $\pm$   0.006   &   0.35    &   11.37   \\
9   &   KIDS J022602.62-315851.65   &   19.25   &   22.17   $\pm$   0.1     &   20.62   $\pm$   0.01    &   19.24   $\pm$   0.003   &   18.74   $\pm$   0.008   &   0.28    &   10.91   \\
10  &   KIDS J024001.94-314142.15   &   19.05   &   22.43   $\pm$   0.13    &   20.61   $\pm$   0.001   &   19.09   $\pm$   0.003   &   18.62   $\pm$   0.009   &   0.29    &   11.01   \\
11  &   KIDS J030324.75-312718.12   &   19.47   &   23.06   $\pm$   0.21    &   21.01   $\pm$   0.02    &   19.45   $\pm$   0.004   &   18.91   $\pm$   0.007   &   0.31    &   11.01   \\
12  &   KIDS J031422.62-321547.76   &   19.57   &   24.5    $\pm$   1.04    &   21.     $\pm$   0.01    &   19.57   $\pm$   0.005   &   19.07   $\pm$   0.008   &   0.27    &   10.95   \\
13  &   KIDS J031645.51-295300.91   &   19.66   &   22.99   $\pm$   0.23    &   21.17   $\pm$   0.02    &   19.64   $\pm$   0.005   &   19.1    $\pm$   0.009   &   0.31    &   10.95   \\
14  &   KIDS J031739.38-295722.23   &   19.1    &   22.5    $\pm$   0.12    &   20.51   $\pm$   0.001   &   19.11   $\pm$   0.003   &   18.64   $\pm$   0.006   &   0.25    &   10.9    \\
15  &   KIDS J032110.91-321319.66   &   19.23   &   22.79   $\pm$   0.18    &   20.69   $\pm$   0.01    &   19.24   $\pm$   0.004   &   18.74   $\pm$   0.007   &   0.27    &   10.97   \\
16  &   KIDS J032603.37-330314.56   &   19.48   &   22.9    $\pm$   0.18    &   20.94   $\pm$   0.01    &   19.47   $\pm$   0.005   &   18.99   $\pm$   0.007   &   0.28    &   10.91   \\
17  &   KIDS J220211.35-310106.17   &   19.43   &   23.01   $\pm$   0.23    &   20.92   $\pm$   0.02    &   19.43   $\pm$   0.004   &   18.93   $\pm$   0.005   &   0.29    &   10.98   \\
18  &   KIDS J220924.49-312052.89   &   19.78   &   23.47   $\pm$   0.44    &   21.31   $\pm$   0.03    &   19.78   $\pm$   0.005   &   19.2    $\pm$   0.02    &   0.34    &   10.98   \\
19  &   KIDS J224431.17-300204.04   &   19.     &   22.48   $\pm$   0.11    &   20.35   $\pm$   0.001   &   19.03   $\pm$   0.003   &   18.51   $\pm$   0.007   &   0.22    &   10.92   \\
20  &   KIDS J225735.20-330652.00   &   19.42   &   23.09   $\pm$   0.25    &   20.78   $\pm$   0.02    &   19.41   $\pm$   0.005   &   18.93   $\pm$   0.011   &   0.25    &   10.91   \\
21  &   KIDS J230520.56-343611.13   &   19.69   &   23.24   $\pm$   0.24    &   21.22   $\pm$   0.02    &   19.67   $\pm$   0.006   &   19.09   $\pm$   0.011   &   0.34    &   11.03   \\
22  &   KIDS J231257.34-343854.93   &   19.32   &   22.94   $\pm$   0.33    &   20.85   $\pm$   0.02    &   19.28   $\pm$   0.005   &   18.75   $\pm$   0.013   &   0.31    &   10.96   \\
23  &   KIDS J232757.84-331202.74   &   19.35   &   23.56   $\pm$   0.38    &   21.     $\pm$   0.02    &   19.35   $\pm$   0.004   &   18.8    $\pm$   0.007   &   0.32    &   11.22   \\
24  &   KIDS J234508.13-321740.12   &   19.65   &   23.     $\pm$   0.2     &   21.19   $\pm$   0.02    &   19.65   $\pm$   0.005   &   19.13   $\pm$   0.01    &   0.33    &   10.96   \\
25  &   KIDS J234547.90-314817.27   &   19.21   &   22.78   $\pm$   0.15    &   20.65   $\pm$   0.01    &   19.26   $\pm$   0.003   &   18.81   $\pm$   0.007   &   0.27    &   11.     \\
26  &   KIDS J235022.88-324037.54   &   18.78   &   22.19   $\pm$   0.09    &   20.13   $\pm$   0.001   &   18.78   $\pm$   0.002   &   18.29   $\pm$   0.005   &   0.23    &   10.92   \\
27  &   KIDS J235630.27-333200.51   &   19.81   &   23.07   $\pm$   0.25    &   21.27   $\pm$   0.02    &   19.79   $\pm$   0.006   &   19.23   $\pm$   0.011   &   0.34    &   10.99   \\
28  &   KIDS J235956.44-332000.90   &   19.59   &   23.47   $\pm$   0.37    &   21.11   $\pm$   0.02    &   19.58   $\pm$   0.005   &   19.04   $\pm$   0.011   &   0.31    &   11.09   \\
\hline
\end{tabular}}
\end{table*}

\begin{table*}
\centering \caption{Structural parameters derived from running
\twodphot\ on g-, r- and i-bands. For each band we show: a)
circularized effective radius \Te, measured in arcsec, b)
circularized effective radius \Re, measured in kpc (calculated
using \zp\ values listed in \Tab\ref{tab:phot_parameters}), c)
S\'ersic index n, d) axis ratio q, e) $\chi^{2}$ of the surface
photometry fit, f) $\chi^{\prime 2}$ of the surface photometry fit
including only central pixels and g) signal-to-noise ratio
\SN.}\label{tab:struc_parameters} \resizebox{\textwidth}{!}{
\begin{tabular}{cccccccccccccccccccccc}
\hline
\rm & \multicolumn{7}{c}{g-band} & \multicolumn{7}{c}{r-band} & \multicolumn{7}{c}{i-band} \\
 \cmidrule(lr){2-8} \cmidrule(lr){9-15} \cmidrule(lr){16-22}
\rm ID & \Te\ & \Re\ & n & q & $\chi^{2}$ & $\chi^{\prime 2}$ & \SN\ & \Te\ & \Re\ & n & q & $\chi^{2}$ & $\chi^{\prime 2}$ & \SN\ & \Te\ & \Re\ & n & q & $\chi^{2}$ & $\chi^{\prime 2}$ & \SN\ \\
\hline
1   &   0.46    &   2.02    &   6.26    &   0.54    &   1.  &   0.9 &   80. &   0.33    &   1.43    &   6.06    &   0.51    &   1.  &   1.1 &   298.    &   0.3 &   1.3 &   5.95    &   0.51    &   1.  &   1.  &   116.    \\
2   &   0.38    &   1.37    &   6.15    &   0.3     &   1.  &   1.  &   163.    &   0.35    &   1.26    &   8.22    &   0.33    &   1.1 &   1.7 &   473.    &   0.3 &   1.07    &   6.69    &   0.31    &   1.  &   1.2 &   175.    \\
3   &   0.14    &   0.71    &   5.4     &   0.05    &   1.  &   1.2 &   46. &   0.22    &   1.08    &   7.45    &   0.18    &   1.1 &   2.  &   148.    &   0.22    &   1.1 &   5.32    &   0.07    &   1.  &   1.2 &   82. \\
4   &   0.22    &   1.      &   4.36    &   0.19    &   1.  &   1.  &   77. &   0.17    &   0.77    &   2.51    &   0.06    &   1.1 &   1.4 &   235.    &   0.26    &   1.2 &   4.61    &   0.29    &   1.  &   0.9 &   103.    \\
5   &   0.21    &   1.18    &   1.7     &   0.47    &   1.  &   0.9 &   22. &   0.14    &   0.77    &   3.25    &   0.38    &   1.  &   1.2 &   87. &   0.04    &   0.23    &   5.56    &   0.02    &   1.1 &   1.  &   48. \\
6   &   0.13    &   0.65    &   1.87    &   0.17    &   1.  &   0.9 &   29. &   0.29    &   1.48    &   3.47    &   0.64    &   1.  &   1.2 &   106.    &   0.26    &   1.32    &   7.75    &   0.6 &   1.  &   1.  &   68. \\
\hline
7   &   0.37    &   1.71    &   5.56    &   0.47    &   1.  &   1.  &   42. &   0.24    &   1.11    &   8.1 &   0.5 &   1.  &   1.1 &   155.    &   0.11    &   0.54    &   8.15    &   0.48    &   1.  &   0.9 &   78. \\
8   &   0.36    &   1.78    &   4.3     &   0.38    &   1.  &   1.  &   72. &   0.25    &   1.23    &   6.5 &   0.39    &   1.  &   1.1 &   354.    &   0.29    &   1.45    &   6.06    &   0.42    &   1.  &   1.  &   161.    \\
9   &   0.38    &   1.61    &   3.65    &   0.6     &   1.  &   1.  &   90. &   0.34    &   1.42    &   3.65    &   0.59    &   1.  &   1.4 &   336.    &   0.35    &   1.47    &   4.04    &   0.6 &   1.  &   1.  &   136.    \\
10  &   0.28    &   1.22    &   5.      &   0.27    &   1.  &   1.1 &   97. &   0.19    &   0.81    &   8.2 &   0.29    &   1.  &   1.3 &   336.    &   0.15    &   0.65    &   8.1 &   0.25    &   1.  &   1.  &   102.    \\
11  &   0.2     &   0.89    &   2.73    &   0.14    &   1.  &   1.  &   74. &   0.29    &   1.29    &   3.  &   0.3 &   1.1 &   1.3 &   291.    &   0.22    &   1.01    &   3.68    &   0.24    &   1.  &   1.  &   170.    \\
12  &   0.27    &   1.12    &   1.35    &   0.39    &   1.  &   1.2 &   82. &   0.15    &   0.61    &   6.36    &   0.38    &   1.  &   1.2 &   222.    &   0.15    &   0.62    &   5.54    &   0.41    &   1.  &   1.1 &   129.    \\
13  &   0.07    &   0.31    &   5.12    &   0.2     &   1.  &   1.1 &   67. &   0.2 &   0.92    &   2.54    &   0.31    &   1.  &   1.1 &   239.    &   0.21    &   0.95    &   3.52    &   0.33    &   1.  &   1.  &   123.    \\
14  &   0.31    &   1.21    &   3.33    &   0.18    &   1.  &   1.  &   102.    &   0.26    &   1.02    &   5.01    &   0.21    &   1.  &   1.2 &   319.    &   0.23    &   0.91    &   6.15    &   0.23    &   1.  &   1.  &   158.    \\
15  &   0.39    &   1.61    &   4.59    &   0.38    &   1.  &   1.1 &   75. &   0.28    &   1.17    &   5.72    &   0.4 &   1.  &   1.1 &   264.    &   0.31    &   1.29    &   4.93    &   0.39    &   1.  &   0.9 &   145.    \\
16  &   0.36    &   1.55    &   3.24    &   0.38    &   1.  &   1.  &   74. &   0.32    &   1.36    &   3.66    &   0.35    &   1.  &   1.  &   216.    &   0.31    &   1.3 &   3.77    &   0.35    &   1.  &   1.  &   144.    \\
17  &   0.39    &   1.71    &   5.67    &   0.45    &   1.  &   1.  &   66. &   0.31    &   1.36    &   4.24    &   0.38    &   1.  &   1.2 &   267.    &   0.28    &   1.23    &   4.15    &   0.39    &   1.  &   0.9 &   196.    \\
18  &   0.21    &   1.04    &   3.44    &   0.18    &   1.  &   1.  &   41. &   0.27    &   1.33    &   2.98    &   0.23    &   1.  &   1.  &   192.    &   0.16    &   0.77    &   5.25    &   0.25    &   1.  &   1.  &   51. \\
19  &   0.41    &   1.45    &   4.16    &   0.68    &   1.  &   0.9 &   103.    &   0.28    &   0.99    &   8.81    &   0.63    &   1.1 &   1.2 &   317.    &   0.31    &   1.11    &   4.75    &   0.69    &   1.  &   0.9 &   124.    \\
20  &   0.35    &   1.37    &   4.31    &   0.38    &   1.  &   1.  &   62. &   0.16    &   0.65    &   5.19    &   0.41    &   1.1 &   1.2 &   230.    &   0.29    &   1.15    &   3.01    &   0.41    &   1.  &   0.9 &   93. \\
21  &   0.42    &   2.      &   3.41    &   0.5     &   1.  &   0.9 &   54. &   0.29    &   1.41    &   4.78    &   0.4 &   1.1 &   1.2 &   186.    &   0.31    &   1.47    &   3.89    &   0.39    &   1.  &   0.8 &   99. \\
22  &   0.84    &   3.81    &   0.9     &   0.74    &   1.  &   1.1 &   68. &   0.24    &   1.1 &   2.25    &   0.43    &   1.  &   1.2 &   226.    &   0.2 &   0.89    &   3.36    &   0.4 &   1.  &   0.9 &   90. \\
23  &   0.38    &   1.78    &   4.46    &   0.61    &   1.  &   1.1 &   63. &   0.28    &   1.29    &   6.63    &   0.69    &   1.  &   1.1 &   253.    &   0.25    &   1.18    &   5.94    &   0.67    &   1.  &   0.9 &   137.    \\
24  &   0.16    &   0.74    &   4.16    &   0.18    &   1.  &   1.  &   54. &   0.3 &   1.46    &   2.96    &   0.36    &   1.  &   1.1 &   208.    &   0.26    &   1.27    &   3.22    &   0.39    &   1.  &   1.  &   105.    \\
25  &   0.61    &   2.54    &   6.73    &   0.41    &   1.  &   0.9 &   80. &   0.28    &   1.16    &   7.35    &   0.44    &   1.  &   1.3 &   262.    &   0.36    &   1.49    &   6.95    &   0.38    &   1.  &   1.  &   134.    \\
26  &   0.37    &   1.34    &   2.65    &   0.25    &   1.  &   1.  &   151.    &   0.3 &   1.1 &   2.9 &   0.26    &   1.1 &   1.3 &   438.    &   0.21    &   0.76    &   3.72    &   0.19    &   1.  &   0.9 &   206.    \\
27  &   0.29    &   1.41    &   4.28    &   0.4     &   1.  &   1.  &   55. &   0.22    &   1.05    &   4.18    &   0.33    &   1.  &   1.  &   183.    &   0.15    &   0.75    &   4.41    &   0.34    &   1.  &   1.1 &   98. \\
28  &   0.43    &   1.94    &   4.38    &   0.42    &   1.  & 0.9 & 63. &   0.24    &   1.08    &   7.22    &   0.38    &   1. & 1. & 199.    &   0.2 &   0.92    &   4.49    &   0.39    & 1.  &   1.1 &   94. \\
\hline
\end{tabular}
}
\end{table*}

\subsection{\UCMG\ selection criteria}\label{subsec:UCMG_criteria}

From our large sample of high \SN\ galaxies, we select the
candidate \UCMGs, using the following criteria:
\begin{enumerate}
\item {\it Massiveness}. The most massive galaxies
with $\mst
> 8 \times 10^{10}\, \rm \Msun$ are taken, as done in the
literature (\citealt{Trujillo+09_superdense};
\citealt{Tortora+16_compacts_KiDS}).
\item {\it Compactness}. We select the densest galaxies by following recent
literature (\citealt{Trujillo+09_superdense};
\citealt{Tortora+16_compacts_KiDS}). To take into account the
impact of colour gradients and derive more robust quantities, we
first calculate a median circularized radius, \Re, as median
between circularized radii in g-, r- and i-bands, and then we
select galaxies with $\Re < 1.5 \, \rm kpc$. Note that in a few
cases the $\Re$ values are derived from images with $\SN$ somewhat
lower than 50 (mainly in g band). However, since in general r band
structural parameters fall between those from g and i band (e.g.,
\citealt{Vulcani+11}), for most of the cases our median \Re\ is
equivalent to the r-band \Re\ which, by selection, is
characterized by $\SN
> 50$, indicating that our selection is robust.
\item {\it Best-fit structural parameters}. The best-fit structural parameters
are considered, taking those systems with a reduced $\chi^{2}$
from \twodphot\ smaller than $1.5$ in g, r and i filters
(\citealt{SPIDER-I}). To avoid any accidental wrong fit, we have
also removed galaxies with unreasonable r-band best-fitted
parameters\footnote{We notice that the criteria applied to r-band
structural parameters are valid for the other two bands for most
of the selected candidates.}, applying a minimum value for the
size ($\Te = 0.05$ arcsec), the axial ratio ($q = 0.1$) and the
S\'ersic index ($n>0.5$). Although the effective radius is only a
parameter of a fitting function, and thus potentially can assume
any value, we remove very small values, which would correspond to
unrealistically small and quite uncertain radii. The limit on the
axis ratio is used to avoid wrong fits or remove any edge-on-like
disks. The minimum value in the S\'ersic index is meant to
possibly remove misclassified stars, which are expected to be
fitted by a box-like profile\footnote{Also if PSF is taken into
account in our procedure, due to the limited spatial resolution of
the observations, the star light profile resembles a step
function.} (mimicked by a S\'ersic profile with $n \to 0$). But
there is also a physical reason to assume this lower limit, since
a S\'ersic profile with $n < 0.5$ present a central depression in
the luminosity density, which is clearly unphysical
(\citealt{Trujillo+01}). Measuring the sizes of \UCMGs\ is a
challenging task, since their \Re\ are of the order of the pixel
scale and smaller than the PSF FWHM. To demonstrate the validity
and robustness of our estimates and to access the uncertainties on
the effective radii, we performed two tests, which are discussed
in details in \App\ref{app:syst_uncertainties}. First, we have
determined the reliability of the effective radii adopted for the
\UCMG\ selection, simulating mock galaxies. We find that, on
average, the input \Re\ are recovered quite well (with an average
difference between input and output \Re\ of $\sim - 6\%$), with
shifts of $\sim - 30\%$ in the smallest galaxies with $\Re \sim
0.05''$. Second, we have quantified the uncertainties on the \Re\
and in particular on the median \Re\ adopted for the
classification. The errors stay around $20\%$ for most of the
cases, reaching values of $\sim 80\%$ only for the smallest
galaxies. We take into account these effects in the number
densities.

\item  We have adopted a morphological criterion to perform the star-galaxy classification
(\citealt{Bertin_Arnouts96_SEx}; \citealt{LaBarbera_08_2DPHOT}).
However, based on optical data only, a star can be still
misclassified as a galaxy on the basis of its morphology, and this
issue can be dramatic for very compact objects (generally with
size comparable or smaller than the seeing). In absence of
spectroscopic information, optical+NIR colour-colour diagrams can
provide a strong constraint on the nature of the candidates. We
use g, J and \Ksband\ \maggaap\ magnitudes for this purpose,
plotting datapoints on the $g-J$ vs. $J-Ks$ plane. Stars and
galaxies are located in different regions of this plane
(\citealt{Maddox+08}; \citealt{Muzzin+13}). We discuss further
this selection on our data in the next section.
\end{enumerate}

\subsection{Selected samples}\label{subsec:selected_samples}

We define different samples of \UCMGs, all satisfying the criteria
described in the previous section, but split in different groups,
according to the type of redshift determination used to derive the
masses and sizes in physical units (photometric or spectroscopic,
from the literature or from our dedicated spectroscopic follow-up)
and to select them.

This grouping is necessary a) to define a sample of
photometrically selected \UCMG\ candidates to derive total \UCMG\
number counts, and b) to gather subsamples with available
spectroscopic redshifts to evaluate systematics affecting the
selection.

In what follows, we will present samples of galaxies with
redshifts up to $z=1$, but, we limit the analysis of number counts
to the redshifts range $z < 0.5$, where our KiDS high-\SN\ sample
is complete (see \Sec\ref{subsec:datasample}). This allows us to avoid selection effects which could
bias our research to blue (non passive) systems at $z
> 0.5$ (e.g. \citealt{Cebrian_Trujillo14}).

\begin{table}
\centering \caption{Number of selected \UCMGs\ in the samples
presented and discussed in \Secs\ref{sec:sample} and
\ref{sec:validation}.}\label{tab:samples}
%
\begin{tabular}{lcccc} \hline
\rm Sample & \multicolumn{2}{c}{\modMnozpt} & \multicolumn{2}{c}{\modMzpt}\\
 \cmidrule(lr){2-3} \cmidrule(lr){4-5}
\rm  & \zp\ & \zs\ & \zp\ & \zs\ \\
\hline
\UP\ ($\zp < 1$) & 1527 & - & 1378 & - \\
\UP\ ($\zp < 1$, NIR) & 1382 & - & 1240 & - \\
\UP\ ($\zp < 0.5$) & 1000 & - & 896 & - \\
\UP\ ($\zp < 0.5$, NIR) & 995 & - & 891 & - \\
\USS\ ($\zs < 1$) & - & 46 & - & 27\\
\USS\ ($\zs < 0.5$) & - & 35 & - & 18 \\
\UPS\ ($\zs < 1$) & 45 & 26 & 24 & 12 \\
\UPS\ ($\zs < 0.5$) & 29  & 16  & 15  & 8 \\
\UNEW\ & 28 & 19  & 13 & 9 \\
\hline
\end{tabular}
%
\end{table}

\begin{figure}
\includegraphics[width=0.45\textwidth]{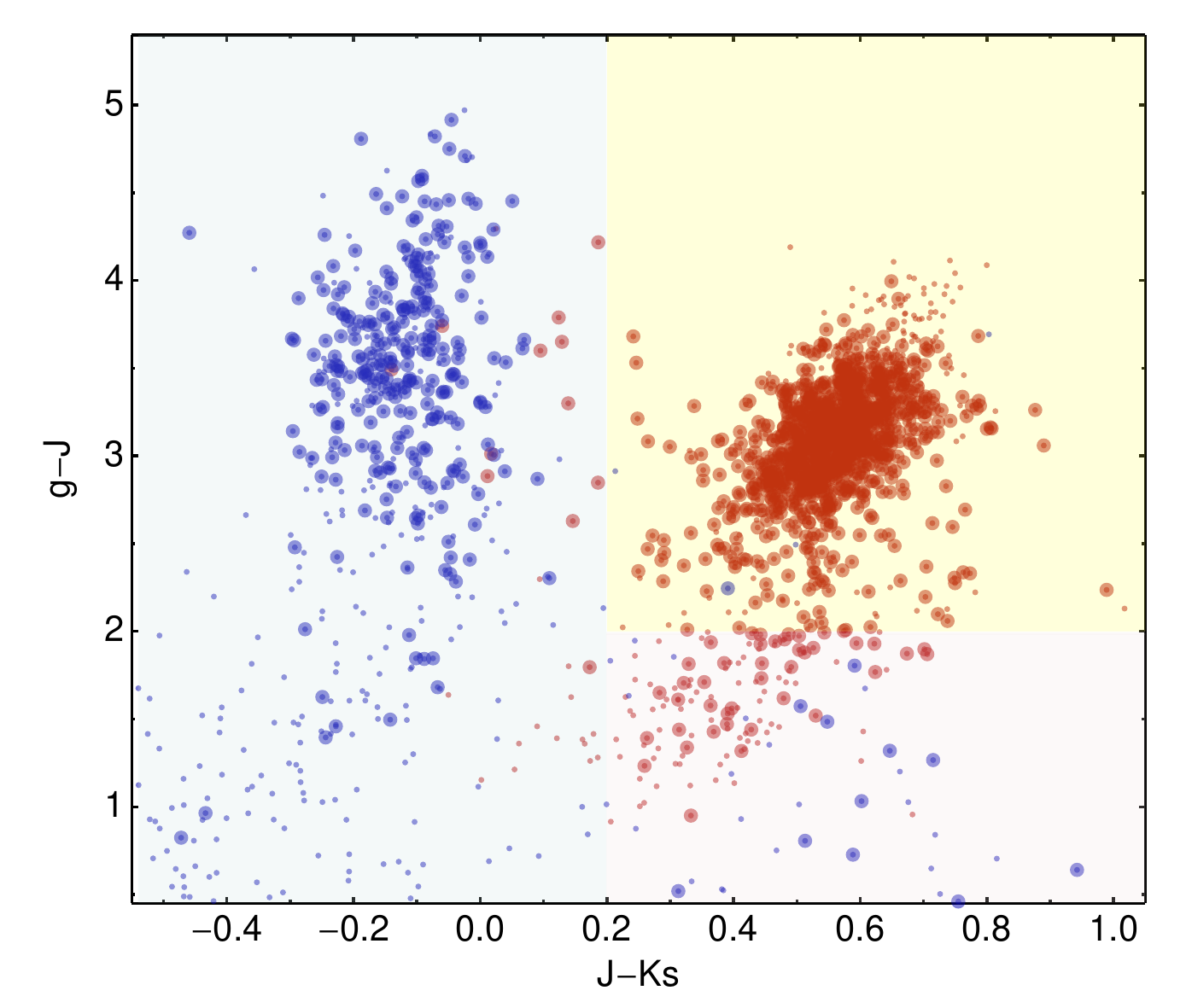}
\caption{$J-Ks$ vs. $g-J$ diagram for the \UP\ sample selected
using \modMnozpt\ masses. \maggaap\ magnitudes are adopted. Blue
symbols are for high-confidence stars, while red points are for
the photometrically selected \UCMGs. Larger symbols are for
stars/galaxies with the best photometry, i.e. with errors $\delta
g, \, \delta J, \, \delta Ks < 0.05$. We highlight the regions
which are populated by stars (blue), red galaxies (yellow) and
QSO-like objects, or blue ($z \gsim 0.5$) galaxies (purple). We
have considered as sure \UCMG\ candidates those objects with
colours $J-Ks> 0.2$ and $g-J
> 2$ (yellow shaded region).}\label{fig: J-K_vs_g-J}
\end{figure}

We start defining the sample of \UCMGs\ we use to plot number
counts in terms of redshift in \Sec\ref{sec:number_counts}.
\begin{itemize}
\item \UP. This sample contains all the photometrically selected \UCMGs\ from 440 DR1+DR2+DR3 survey tiles,
corresponding to an effective area of 333\sqd, after masking. We
use \zp\ obtained with the trained network \MLtwo\ discussed in
\Sec\ref{sec:sample}. Assuming the set of masses \modMnozpt\ (see
\Tab\ref{tab:tab_legend}), the sample contains 1527 \UCMGs\ at
$\zp < 1$ (1000 at $\zp < 0.5$). Instead, using the \modMzpt\
values, the number reduces to 1378 (896 at $\zp < 0.5$). This
difference in numbers is due to the fact that including the
calibration errors gives higher metallicites and smaller ages,
which result in lower masses, causing the reduced number of
\UCMGs. Using the "classification" scheme discussed in
\Sec\ref{sec:sample}, the fraction of galaxies well fitted by
spectral models of ellipticals are $80-85\%$ of the total.
Instead, at $z < 0.5$ $\sim 98\%$ of the candidates are classified
as ellipticals, potentially most of them are passive systems.
However, a more accurate stellar population analysis and spectral
classification is needed, using high-resolution spectra and/or
inclusion of NIR photometry.

As discussed in the previous section, for a subsample of
candidates we can also rely on VIKING NIR data, thus we combine
optical+NIR photometry to reduce the fraction of contaminants,
i.e. misclassified stars, quasars and higher-z/blue galaxies
(\citealt{Maddox+08}; \citealt{Muzzin+13}). Stars and galaxies
with the best photometry (i.e., with $\delta g, \, \delta J, \,
\delta Ks < 0.05$) are also considered. For the \UCMG\ sample
selected using \modMnozpt\ masses we find VIKING photometry for
1337 \UCMG\ candidates at $\zp < 1$ (874 at $\zp < 0.5$), instead
if we use \modMzpt\ masses these numbers are 1196 at $\zp < 1$
(774 at $\zp < 0.5$). The $J-Ks$ vs. $g-J$ diagram for these
galaxies is shown in \Fig\ref{fig: J-K_vs_g-J} for the \modMnozpt\
case. Stars (which are represented as blue dots in the figure)
have blue $J-Ks$ colours (i.e., $J-Ks \lsim 0.2$, see light blue
shaded region in \Fig\ref{fig: J-K_vs_g-J}). However, also some of
our candidates (red points) have $J-Ks \lsim 0.2$. These indeed
are stars that have been erroneously classified as galaxies. We
take as compact ($z \lsim 0.5$) candidates those systems with
$J-Ks> 0.2$ and $g-J
> 2$ (see light-yellow shaded region in \Fig\ref{fig:
J-K_vs_g-J}).

After this selection we are left with 975 \UCMGs\ at $\zp < 1$
(869 at $\zp < 0.5$) when \modMnozpt\ masses are used, and 845
\UCMGs\ at $\zp < 1$ (769 at $\zp < 0.5$) when \modMzpt\ masses
are used. If the whole sample with $\zp < 1$ and VIKING photometry
is considered, then the contamination would amount to about 10\%,
due to mainly $z \gsim 0.5$ \UCMG\ candidates with $g-J < 2$.
Fortunately, in the redshift range we are mostly interested in,
i.e. at $\zp < 0.5$, the contamination is less than $1\%$, which
confirms the goodness of KiDS S/G separation and our selection
procedure. The results are independent of the mass definition
adopted. We will remove contaminants in the discussions that
follow, summing up the cleaned sample just discussed to the rest
of the galaxies without NIR photometry. We are finally left with
1382 \UCMGs\ at $\zp < 1$ (995 at $\zp < 0.5$) when \modMnozpt\
masses are used, and 1240 \UCMGs\ at $\zp < 1$ (891 at $\zp <
0.5$) if \modMzpt\ masses are used. We will use these samples in
\Sec\ref{sec:number_counts}, where we will study number counts at
$z < 0.5$.
\end{itemize}

One of the main systematics in our selection of \UCMGs\ is induced
by wrong redshift determination, which can affect both the
(linear) effective radii and stellar masses, moving the compact
out of our selection criteria. If $\zs > \zp$ ($\zs < \zp$), then
if we re-calculate \Re\ and \mst\ using \zs, \Re\ gets larger
(smaller) and in most of the cases also \mst\ get systematically
larger (smaller). Although the photometric redshifts approximate
quite well the spectroscopic ones (\Sec\ref{sec:validation}; see
more details in \citealt{Cavuoti+15_KIDS_I}), also small changes
in \zp\ can induce changes in \Re\ and \mst\ large enough to find
$\Re
> 1.5 \, \rm kpc$ and /or $\mst < 8 \times 10^{10} \, \rm \Msun$.
Thus, because of "wrong" \zp\ values, two effects should be taken
into account when estimating \UCMG\ number counts: 1) we are
including some "contaminants", i.e., galaxies which are selected
as \UCMGs\ according to their photometric redshift, but would not
result ultra-compact and massive on the basis of the more accurate
spectroscopic value (see \citealt{Tortora+16_compacts_KiDS}); 2)
we are "missing" some objects, i.e., those galaxies which are not
selected as \UCMGs\ according to their photometric redshift, but
would be selected using the spectroscopic value\footnote{The
present analysis improves the one performed in
\cite{Tortora+16_compacts_KiDS}, where we have taken into account
only the former effect and not the latter.} (i.e., they are real
\UCMGs). Following a more conventional terminology in statistics,
"contaminants" and "missing objects" are also referred to as
"false positives" and "false negatives". We therefore define the
{\it contamination factor}, \CF, to account for the number of
"contaminants" and the {\it incompleteness factor}, \IF, to
estimate the incompleteness of the sample, quantifying the number
of "missing" objects. To quantify these effects we need to collect
1) photometrically selected samples of \UCMG\ candidates with
known spectroscopic redshifts from the literature and new
observations, and 2) spectroscopically selected samples of \UCMGs\
from the literature.

Therefore, we now define two further samples, with measured
spectroscopic redshifts from the literature, which are used to
quantify "missing" objects and "contaminants".
\begin{itemize}
\item \UPS. This is a subsample of \UP\
(i.e., selected using the measured \zp) with measured
spectroscopic redshifts from SDSS,
GAMA or 2dFLenS
(\citealt{Blake+16_2dflens}), which overlap the KiDS fields in the
Northern and Southern caps. We are left with a sample of 45 \UCMG\
candidates using \modMnozpt\ masses and 22 using \modMzpt\ masses.
This sample is useful to quantify the number of \UCMGs\ which we
have missed in the photometric selection.
\item \USS. Within the 440 DR1+DR2+DR3 fields
we have also selected a sample of galaxies with spectroscopic
redshifts from the literature (from SDSS, GAMA or 2dFLenS), and we
have used this time directly \zs\ to select \UCMGs\, instead of
\zp\ as done for \UPS. The sample comprises 46 confirmed \UCMGs\
using \modMnozpt\ masses and 27 using \modMzpt\ masses.
\end{itemize}

Extrapolating the numbers of confirmed \UCMGs\ in \USS\ to the
full survey area (i.e. $1500 \, \sqd$), we would already expect to
find $\sim 170$ ($\sim 100$) \UCMGs\ with known spectroscopic
redshift from SDSS, GAMA and 2dFLenS using \modMnozpt\ (\modMzpt)
masses. However, to avoid any residual selection effect in the
galaxy targeting made in the above mentioned surveys and aiming at
further increasing the sample size of spectroscopically confirmed
\UCMGs, we have started a program to obtain spectra on hundreds of
candidates, as we will discuss in the next section. We started
observing with the Telescopio Nazionale Galileo (TNG) for the
\UCMG\ candidates in the North and the New Technology Telescope
(NTT) for those in the South hemisphere. These two samples will be
used with the \UPS\ sample to quantify the number of "contaminants".
Accordingly, we selected two subsamples.
\begin{itemize}
\item \UTNG. The first subsample was extracted from
an updated version of the dataset of candidates selected in
\cite{Tortora+16_compacts_KiDS} from the first 156\sqd\ of KiDS
(with observations from KiDS--DR1/2/3), where the first \UCMG\
candidates from KiDS were discussed. We have selected galaxies in
the equatorial strip ($-3 < DEC < 3$ degrees) observed by KiDS. In
\cite{Tortora+16_compacts_KiDS} and in the current paper we use
the photometric redshift catalog based on the trained network
\MLone, presented in \cite{Cavuoti+15_KIDS_I} and structural
parameters (\Re, Sérsic index, etc.) in \cite{Roy+18}. The
follow-up of these galaxies were performed at Canarias Islands
with TNG.
\item \UNTT. The second subsample of galaxies was
collected from 120\sqd\ southern fields in KiDS--DR3. Redshifts
were determined using the same network \MLone\ trained and
discussed in \cite{Cavuoti+15_KIDS_I}, and applied to the new
observed fields in KiDS--DR3. These redshifts are quite consistent
with the newest and public machine learning redshifts presented in
the KiDS--DR3 paper (\Sec\ref{sec:validation} and
\citealt{deJong+17_KiDS_DR3}). This sample has been observed in
Chile, at NTT.
\end{itemize}
We will name this cumulative sample of new \UCMG\ candidates as
\UNEW. Note that only 17 (11) \UCMG\ candidates in \UTNG\ and
\UNTT\ are present in the sample \UP, if \modMnozpt\ (\modMzpt)
masses are used. This is due to the different sets of photometric
redshifts adopted for the two selections (\MLone\ and \MLtwo). In
fact, small changes in \zp\ could push the compact out of our
selection criteria.

\begin{figure*}
\includegraphics[width=0.99\textwidth]{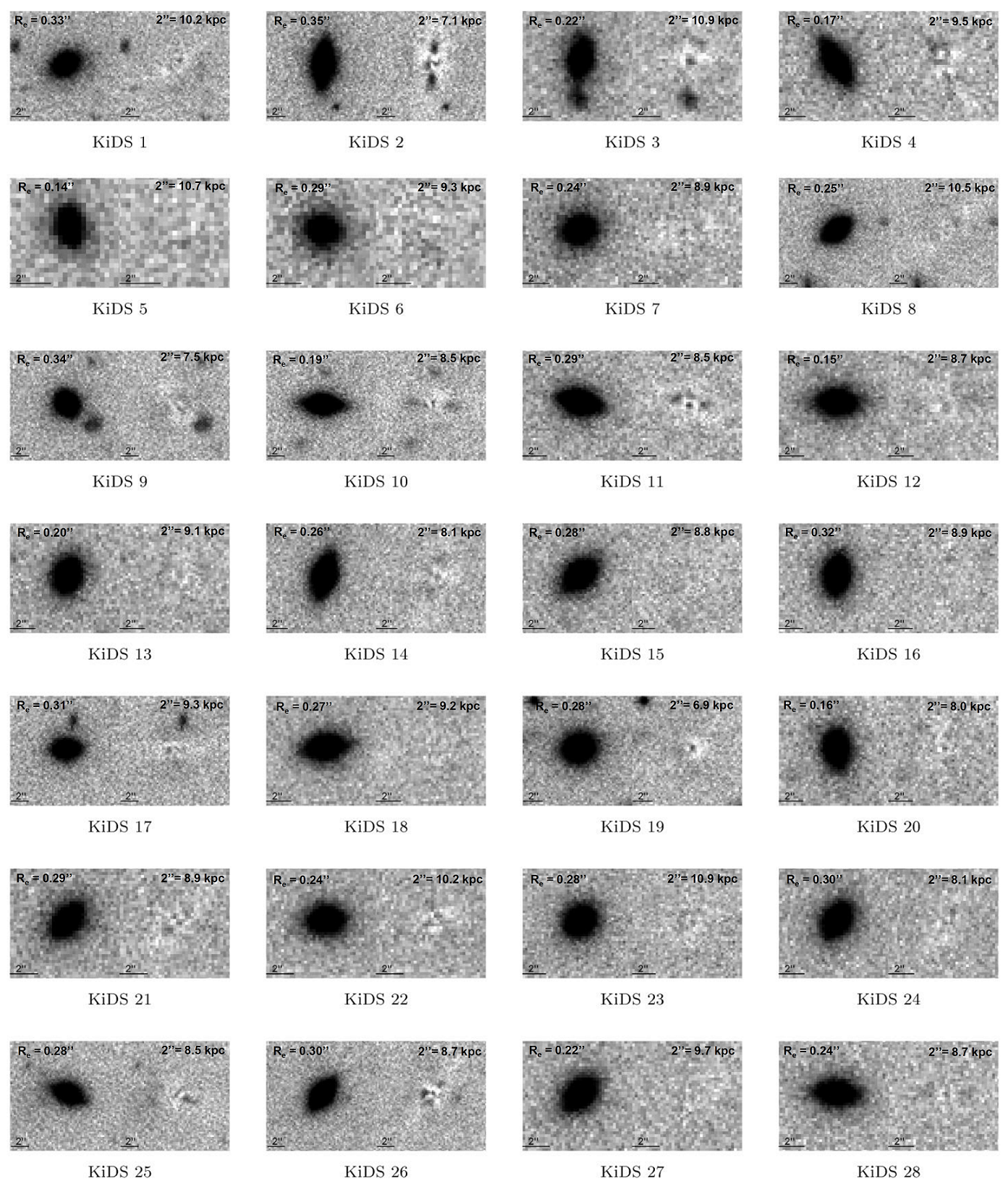}
\caption{2D fit output from \twodphot\ procedure for the new 28
\UCMG\ candidates in \UNEW. For each \UCMG\ candidate, we show
galaxy image (left) and residual after the fit (right). The scale
of 2 arcsec is indicated on the images and for each galaxy the
conversion to the physical scale is also added (we use the
spectroscopic redshifts presented in \Sec\ref{sec:spectroscopy}).
The effective radius value in arcsec is also
shown.}\label{fig:2DPHOT_sample}
\end{figure*}

\begin{figure*}
\includegraphics[width=1.01\textwidth]{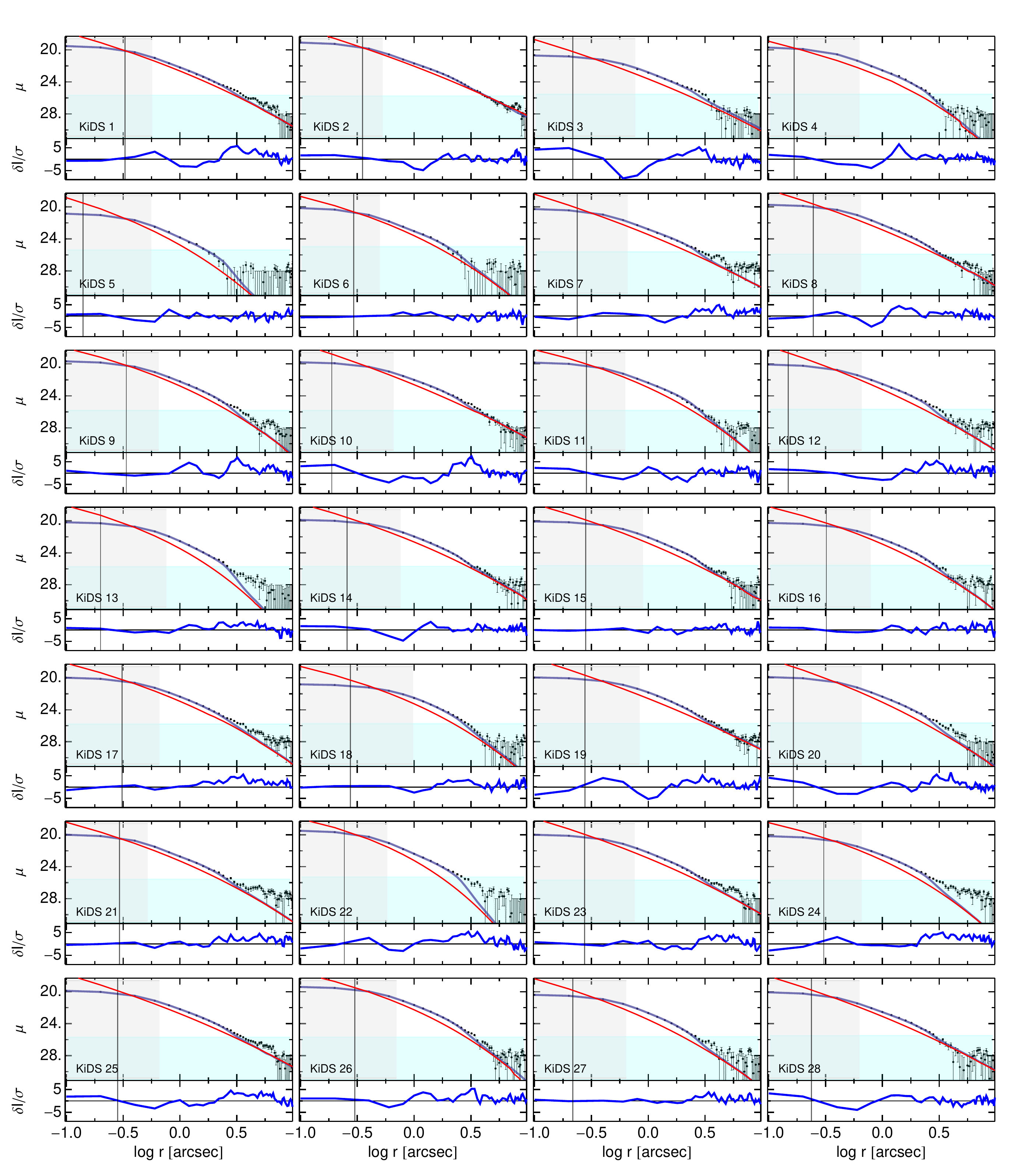}
\caption{The r-band surface brightness profiles and residuals of
the 28 candidates from the \UNEW\ sample. The 1D surface
brightness profiles of the observed image and model from
\twodphot\ are calculated as averages in circles of radius
$r_{i}$. Black points with error bars are for the observed surface
brightness profile. The error calculation is described in the main
text. The blue line show the best-fitted S\'ersic model (convolved
with the PSF). Sky level background has been subtracted from these
two profiles. The red line shows the de-convolved best-fitted
S\'ersic profile. The upper limit of the cyan region corresponds
to the $1 \sigma$ from the sky level. For each galaxy, we also
show in the bottom panel the residuals of the best fit, defined as
the difference of the observed and model profile, normalized by
the noise. The vertical gray line sets the effective radius, and
the grey region sets the FWHM of the average seeing of the KiDS
tile where each galaxy is found. } \label{fig:light_profiles}
\end{figure*}

\subsection{Surface photometry and \muemed--\Re\ plane}

In \Fig\ref{fig:2DPHOT_sample} we present the KiDS r-band images
and residuals after the best fitted S\'ersic profile is subtracted
for the 28 candidates in \UNEW. To better quantify the results
seen in \Fig\ref{fig:2DPHOT_sample}, their surface brightness
profiles are shown in \Fig\ref{fig:light_profiles}. For each
galaxy, we show the 1D brightness profile derived from the
observed KiDS r-band image (black symbols) and the best fitted
convolved S\'ersic model (blue line). These 1D profiles are also
compared with the best-fitted deconvolved S\'ersic profile,
calculated inserting the best fitted parameters in the 2D S\'ersic
analytical formula (red line). To calculate these profiles, first
we numerically interpolate image, 2D convolved and 2D unconvolved
models, removing those pixels corresponding to nearby galaxies,
masked during the fitting procedure performed by \twodphot. Then,
we obtain both the profiles in concentric circles of radius
$r_{i}$, by calculating the interpolating function at different
angles and deriving the average of these values\footnote{Mainly in
the outer regions, some pixels have a negative flux. To avoid
unphysical fluxes at some $r_{i}$, we have fixed the latter values
to a very faint magnitude value ($r=32$), and considered in these
cases an upper/lower limit of 28/40 (see some examples in
\Fig\ref{fig:light_profiles}). Note that these magnitude values
are arbitrary, and will not affect our considerations.}. These
circles are centered on the mean position of the 4 brightest
pixels, corresponding to the maximum of the surface brightness
profile. The errors on each pixel for the observed profiles are
calculated adding in quadrature a) the background noise and b) the
photonic noise in the pixel, calculated as square root of the
ratio of the intensity in that pixel and the effective gain of the
related KiDS tile. The error on the intensity calculated in each
circle is re-scaled by the square root of the number of pixels
within a circumference of radius $r_{i}$ (also in this case
removing the masked pixels). This error is properly propagated to
obtain the uncertainties for the surface brightness in the top
panels. We have subtracted the sky level (estimated from
\twodphot) to 1D profiles from the observed KiDS r-band image and
the best fitted convolved S\'ersic model, in order to have a
homogeneous comparison with the unconvolved S\'erisc profile.

For each galaxy, we also show the difference between the
intensities of observed image and best fitted model. We calculate
the related interpolated function, which is evaluated at radii
$r_{i}$, as done for the surface brightness profiles. This
quantity is normalized to the noise, which is defined above.

The figure shows that we are recovering the brightness profile of
the \UCMG\ candidates from the very center to the outskirts quite
well. Thus, although the size is of the order of the pixel scale
and below the PSF FWHM, these levels of accuracy, allowed by the
high quality of the KiDS images and a proper PSF de-convolution,
assure us about the goodness of our derived effective radii. This
result matches the consideration we make in
\App\ref{app:syst_uncertainties}, where we show, using mock
galaxies, that we are able to recover the input effective radii
with a good level of accuracy. For some galaxies in
\Fig\ref{fig:light_profiles} the observed surface brightness at
radii larger than 3 arcsec (i.e., $\gsim 8 \Re$) is not regular,
generating some residuals. Such residuals are probably related to
very faint sources surrounding the central galaxy, which are not
properly masked, and pop up in our 1D profiles, but do not affect
the \twodphot\ best fits. However, at these radii, the flux of our
galaxies is negligible (as seen from \Fig\ref{fig:2DPHOT_sample}),
and the measured surface brightness reach values which are within
the rms of the sky level. This makes residuals at radii larger
than 3 arcsec not statistically significant.

To further analyze such residuals and quantify the presence of
underlying disks, we have stacked the 28 profiles and compared
observed and fitted model. The stacking procedure increases the
signal-to-noise ratio and can possibly enhance structures in the
external regions. We take these average profiles and calculate the
growth curve (i.e. the projected luminosity) in terms of the
radius. We find that at radii where the sky background starts to
dominate (i.e. $r \sim 3$ arcsec), no difference is found. If we
calculate the growth curves up to $10$ arcsec, then the
discrepancy between the stacked profiles is $\sim 0.05$ dex, which
is comparable to statistical and zero point errors on the
magnitudes. This excludes the presence of diks, and assures us
that our size estimates are not affected by disk contribution not
properly accounted in the fitting procedure.

\begin{figure}
\centering
\includegraphics[width=0.38\textwidth]{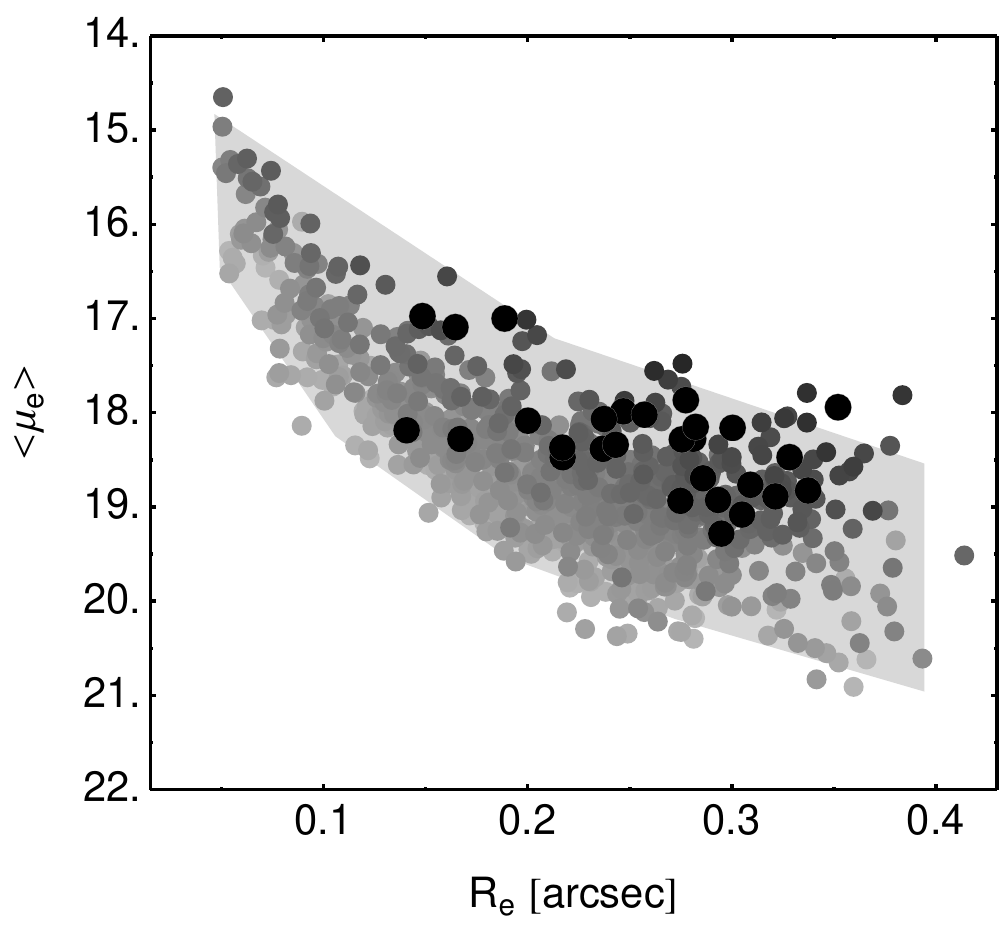}
\includegraphics[trim= 0mm -9mm 5mm 0mm, width=0.06\textwidth]{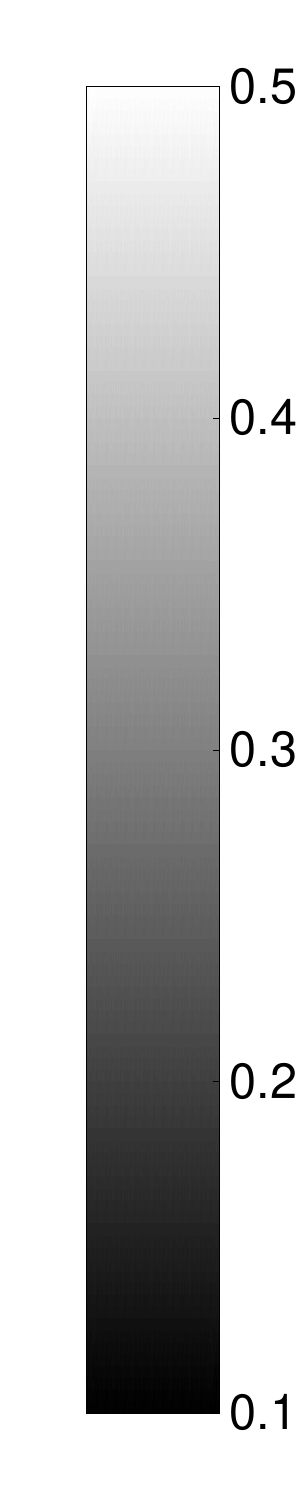}
\caption{\muemed--\Re\ plane. Small dots are for the \UP\ sample
using \modMnozpt\ masses and with $\zp \leq 0.5$. The points are
colour coded accordingly to their photometric redshift (a color
bar is added to the plot). Larger black dots with gray edges are
for all the \UCMG\ candidates in \UNEW. Shaded gray region
encloses the \twodphot\  \muemed\ and \Re\ output values for the
mock galaxies simulated in \App\ref{app:syst_uncertainties}. The
lower bound of the shaded region corresponds to galaxies with
$m_{\rm S} \sim 21$, which is approximatively our magnitude
limit.} \label{fig:mue_vs_Re}
\end{figure}

We show in \Fig\ref{fig:mue_vs_Re} the \muemed--\Re\ plane for our
\UCMG\ candidates. In particular, we show the results for the
\UCMG\ candidates in \UP\ with $\zp < 0.5$. They are colored in
terms of the redshifts. Higher-z galaxies have systematically
fainter surface brightness and magnitude, at fixed \Re\
(\citealt{Kormendy77_II}). We also show the results for the
candidates with new spectroscopic redshifts in the \UNEW\ sample,
plotted as bigger dots with a gray edge. Such galaxies, by
selection, are located at systematically brighter \muemed\ and
lower redshifts, and overlap in the \muemed--\Re\ plane with
galaxies in \UP\ with redshifts in the the same range (i.e., $\sim
(0.25,0.35)$). An interesting comparison of these results is done
with the outputs of simulations which will be discussed in more
details in \App\ref{app:syst_uncertainties}. Those mock galaxies,
simulated as S\'ersic profiles with ranges of parameters similar
to those of the KiDS \UCMG\ candidates, with realistic observing
conditions, are fitted with \twodphot\ to recover the structural
parameters. The outputs of this analysis are enclosed in the
shaded gray area plotted in \Fig\ref{fig:mue_vs_Re}. Since these
are created using parameters in the same range of the
observations, and are recovered by \twodphot\ at a good accuracy,
there is a fine overlap of observations and mock galaxies. The
lower bound of the region corresponds to a S\'ersic magnitude of
$\sim 21$, which is approximatively our completeness magnitude
(see \Sec\ref{sec:sample} and \App\ref{app:completeness}).

\section{New spectroscopy}\label{sec:spectroscopy}

As mentioned, to increase the number of spectroscopically
confirmed \UCMGs\ we have started a multi-site and multi-facility
spectroscopic campaign in the North and South hemisphere, to cover
the whole KiDS area during the entire solar year. The multi-site
approach allows us to cover the two KiDS patches (KiDS-North from
La Palma and KiDS-South from Chile), while the multi-facility
allows to optimize the exposure time according to the target
brightness (ranging from $\Mautor \sim 18.5$ to $\sim 20.5$). We
have planned to observe our \UCMG\ candidates at 3--4m and 8--10m
class telescopes (for brighter and fainter targets, respectively).

In this paper, we first present the results for a sample of
\UCMGs\ with spectroscopic redshifts gathered from the literature
and then we discuss the first results of our spectroscopic
campaign, presenting the new spectroscopic redshifts obtained with
TNG and NTT telescopes during the first two runs performed in 2016
(see \Sec\ref{subsec:UCMG_criteria}).

In \Secs\ref{subsec:TNG_spec} and \ref{subsec:NTT_spec} we provide
some details about the instruments used for spectroscopy,
observational set-up, strategy and quality of the extracted
spectra. The calculation of spectroscopic redshifts is outlined in
\Sec\ref{subsec:zspec_calculation}.

\begin{figure*}
\includegraphics[width=1\textwidth]{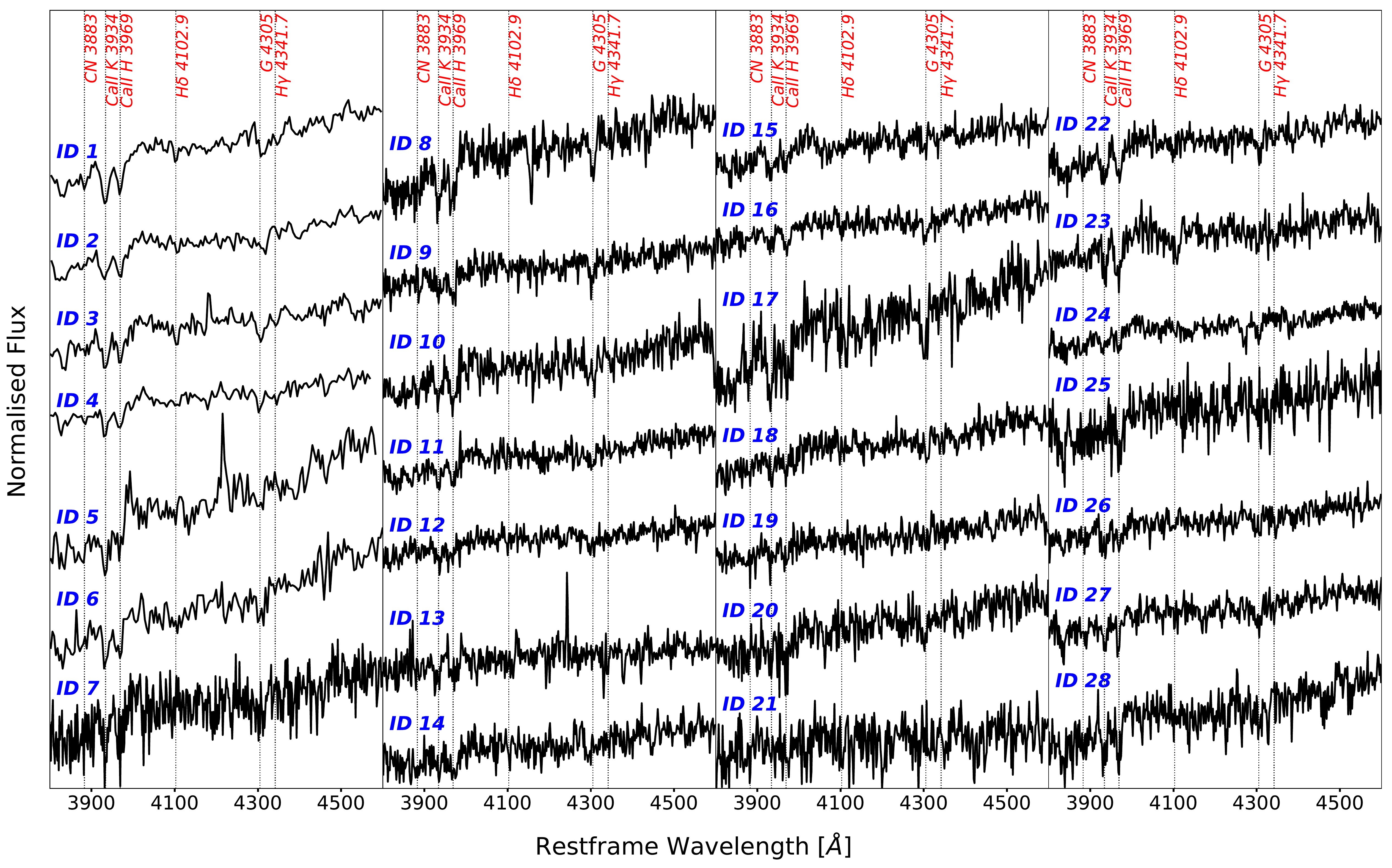}
\caption{First spectra of \UCMG\ candidates observed in our
spectroscopic campaign. Following the ordering in
\Tabs\ref{tab:phot_parameters} and \ref{tab:struc_parameters} we
have plotted the spectra for the 6 candidates observed with TNG
and 22 with NTT. The flux is arbitrarily normalized and plotted
vs. wavelength, restframed using the measured spectroscopic
redshift. We only plot a narrow wavelength region, including CN
3883 band, Ca H and K lines, $H_{\delta}$, G-band and
$H_{\gamma}$. The main spectral features are highlighted in red
and the galaxy ID is reported above each spectrum.}\label{fig:
TNG_NTT_spectra}
\end{figure*}

\subsection{TNG spectroscopy}\label{subsec:TNG_spec}

The first spectra discussed in the present paper are relative to
\UCMG\ candidates selected in \UTNG\ and are obtained with the
Device Optimized for the LOw RESolution (DOLORES) at TNG
telescope, in visitor mode, during the observing run A32TAC\_45 on
March 2016 (proposal title: Spectroscopic follow-up of new massive
compact galaxies selected in the KIDS public survey, PI: C.
Tortora). The detector used for the observations consisted of a
$2048 \times 2048$ E2V 4240 thinned back-illuminated,
deep-depleted, Astro-BB coated CCD with a pixel size of 0.252
arcsec/pixel and a field of view of $8.6 \times 8.6$ arcmin. We
have used the grism LR-B with a dispersion of 2.52 \AA/pixel and
resolution $R = 585$ (calculated within a slit of 1 arcsec width)
in the 3000--8430 \AA\ wavelength range. The average seeing was of
FWHM $\sim$ 1.0 arcsec. The data, consisting of a set of 1 up to 3
single exposures for each source, were acquired with a slit 1.5
arcsec wide.

Spectra were reduced and processed using a suite of
\iraf\footnote{\iraf\ (Image Reduction and Analysis Facility) is
distributed by the National Optical Astronomy Observatories, which
is operated by the Associated Universities for Research in
Astronomy, Inc. under cooperative agreement with the National
Science Foundation.} tools and \python/\astropy. For each night,
the flat-field and the bias images were averaged together,
creating a master flat and a master bias. Scientific spectra were
then divided by the master flat image, while the master bias was
subtracted from them. Wavelength calibration was performed using
the \identify\ task on a Ar, Ne+Hg, and Kr lamps which were
acquired before starting the scientific exposure. Pixels were
mapped to wavelengths using a 5-th order polynomial function.
These spectra were finally resampled to the resolution and scale
of DOLORES.

We have observed 16 candidates: 5 with long-slit and 11 with
multi-object spectroscopy (MOS), the latter configuration is used
to obtain spectroscopic redshifts for compact candidate and
neighbors. The magnitudes of the \UCMG\ candidates within the slit
are of $\lsim 20$ and photometric redshifts are $\zp < 0.5$. The
total exposure time for each candidate is in the range 900-4500s.
Unfortunately, due to weather downtime, we obtained reliable
spectra with a reasonable \SN\ of $\gsim$ 10 for Angstrom only for
6 candidates.

We focus here on the results for the compact galaxies, and we
discuss the role of the environment in a future paper.

\subsection{NTT spectroscopy}\label{subsec:NTT_spec}

The largest part of new spectra analyzed in this work were
obtained with EFOSC2 (ESO Faint Object Spectrograph and Camera
v.2) at ESO-NTT telescope, in visitor mode, during the observing
run 098.B-0563 on October 2016 (title: Spectroscopic follow-up
with NTT and VLT of massive ultra-compact galaxies selected in the
KIDS public survey, PI: C. Tortora). The detector used for the
observations consisted of Loral/Lesser, thinned, AR coated, UV
flooded, MPP chip controlled by ESO-FIERA, with a scale of 0.12
arcsec/pixel and a field of view of $4.1 \times 4.1$ arcmin. We
have used the GR\#4 grism with a dispersion of 1.68 \AA/pixel and
resolution of 12.6 \AA\ (within a slit of 1 arcsec width),
corresponding to $R \sim$300--600 in the 4085--7520 \AA\
wavelength range. The average seeing was FWHM $\sim$ 0.9 arcsec.
The data, consisting of a set of at least 3 spectra for each
source, were acquired with a slit 1.2 arcsec wide.

Individual frames were pre-reduced using the standard \iraf\ image
processing packages. The main strategy adopted included dark
subtraction, flat-fielding correction and sky subtraction.
Wavelength calibration was achieved by means of comparison spectra
of He-Ar lamps acquired for each observing night, using the \iraf\
\twodspec\ package. The sky spectrum was extracted at the outer
edges of the slit, and subtracted from each row of the two
dimensional spectra by using the \iraf\ task \background\ in the
\twodspec\ package. The sky-subtracted frames were co-added to
final averaged 2D spectra, which were used to derive the
spectroscopic redshifts.

We have observed 23 compact candidates, with r-band magnitudes
within the slits $\lsim 20$ and redshifts $\zp \lsim 0.35$. Total
integration times per system ranges between 1200s and 3600s and we
obtained cumulative \SN\ per Angstrom mostly in the range 4-8. 1
out of the 23 candidates was classified as a star from the
spectrum, and thus has been excluded from the discussion in the
next sections, leaving us with a sample of 22 \UCMG\ candidates.
In future spectroscopic follow-ups we will rely on new samples
pre-selected using optical+NIR colour-colour diagrams (as
discussed for \UP\ in \Sec\ref{subsec:selected_samples}), further
reducing the chance to include misclassified stars.

\subsection{Redshift calculation}\label{subsec:zspec_calculation}

Redshifts have been calculated by making use of a graphical user
interface (\ppgui, written by G. D'Ago, to be distributed) based
on the Penalized Pixel-Fitting code (\ppxf,
\citealt{Cappellari17}). In our case, \ppxf\ uses, as templates,
combinations of MILES Simple Stellar Population libraries
\citep{Vazdekis10}, plus an additive polynomial, to fit the
observed spectrum. The resolution of the templates is degraded via
a convolution process to the instrumental resolution of the
spectrograph. \ppgui\ allows the user to visualize and inspect the
observed spectrum, and easily set the \ppxf\ fitting parameters
before running the code. It also allows one to clean up the
spectrum by trimming it and masking wavelengths affected by
typical gas emission, cosmic rays or bad reduction. The spectra
for the 28 observed \UCMG\ candidates (non calibrated in flux) are
shown in \Fig\ref{fig: TNG_NTT_spectra}, where we zoom in the
wavelength region 3800--4500 \AA, highlighting some of the main
absorption features in the plotted wavelength range\footnote{We do
not show the best-fitted models and we only plot a limited range
of wavelength since we are mainly interested to show that
redshifts are finely recovered.}. H and K lines of Calcium doublet
are the most clear features visible in all the spectra, which have
helped us (together with the estimated \zp) to set an initial
guess for the redshift search. The G-band is also prominent in
most of the spectra, as it is typical for passive galaxies
(\citealt{Wang+17}). The other features (i.e. CN 3883 band,
$H_{\delta}$ and $H_{\gamma}$), which are also intrinsically
weaker in high-\SN\ and high-resolution spectra in the literature,
are visible only in a few spectra. For most of the galaxies Mg
5177 lines and/or most of Fe lines (but not shown in \Fig\ref{fig:
TNG_NTT_spectra}) are also strong in our spectra, further
confirming the passive nature of the candidates.

\section{The validated sample and the analysis of systematics}\label{sec:validation}

\begin{figure*}
\includegraphics[width=0.85\textwidth]{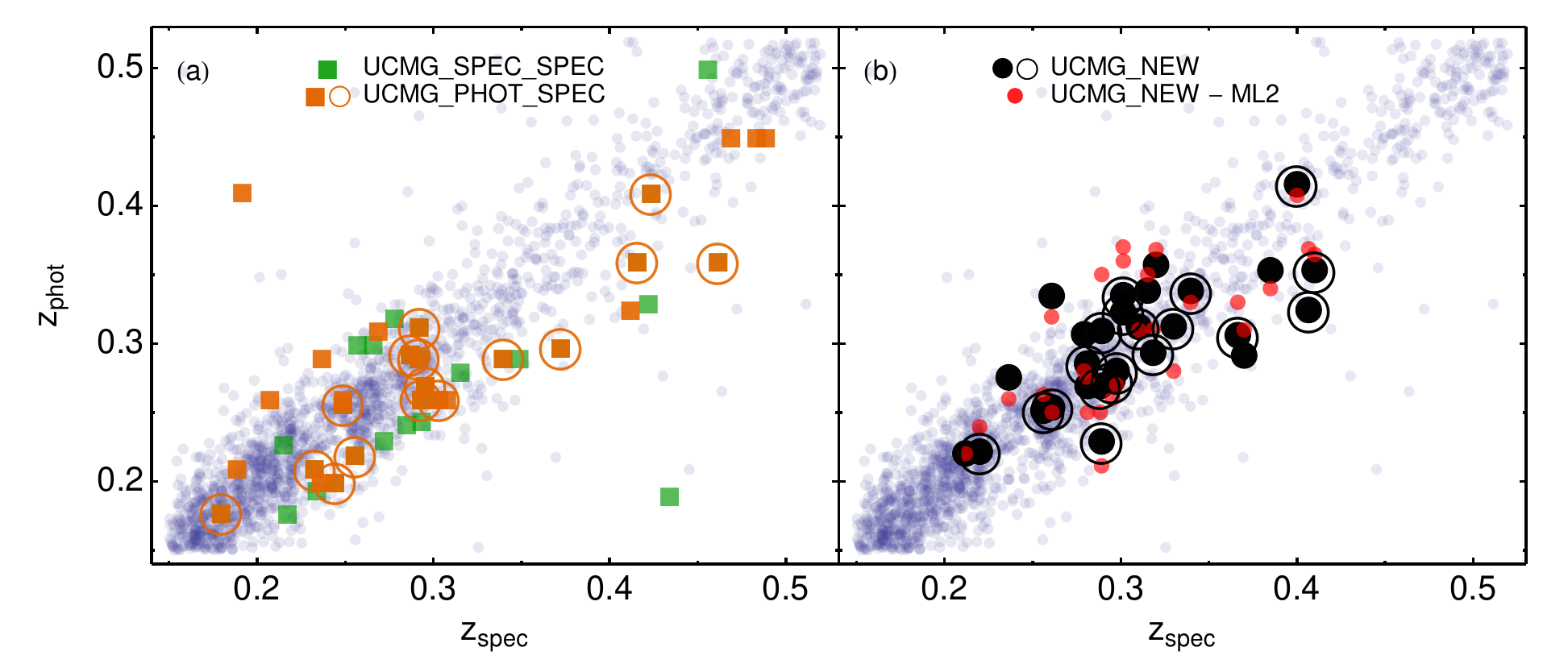}
\caption{Spectroscopic vs. photometric redshifts. Blue points are
relative to the blind test set in \citet{Cavuoti+15_KIDS_I}, who
used SDSS and GAMA spectroscopic redshifts. Redshifts from
different selections are plotted in the two panels. {\it Panel a.}
Green squares are for the sample of \UCMGs\ selected using the
spectroscopic redshifts from the literature (\USS). Orange squares
are relative to the set of \UCMG\ candidates selected using ML
photometric redshifts, but with available measured \zs\ from the
literature (\UPS). Confirmed \UCMGs\ from \UPS, after \zs\ is
used, are drawn as orange circles. {\it Panel b.} Black and red
points are for the 28 new \UCMG\ KiDS candidates with redshifts
measured with observations at TNG and NTT (\UTNG\ and \UNTT,
respectively). In particular, black points are for ML photometric
redshifts used for the selection, while the ML photometric
redshits included in KiDS--DR3 are plotted in red. Black circles
are for confirmed \UCMGs, after \zs\ is used. For all the sets of
redshifts plotted in the two panels, we find a good agreement with
the 1-to-1 relation, with a systematic slight underestimation of
\zp\ at $\zs \gsim 0.35$.}\label{fig: zs_vs_zp_UCMGs}
\end{figure*}

In \Sec\ref{subsec: validated_zs_lit} we start discussing the
results for the sample of \UCMGs\ with \zs\ from the literature
(\USS), studying the success rate of our selection and
systematics, through the comparison with the photometrically
selected sample \UPS. The new results from the observations with
TNG and NTT about the samples \UTNG\ and \UNTT\ are analyzed in
\Sec\ref{subsec: validated_new}. In \Fig\ref{fig: zs_vs_zp_UCMGs}
we plot derived spectroscopic vs. photometric redshifts for the
samples analyzed, and sizes and stellar masses are shown in
\Fig\ref{fig: Re_vs_Mstar_UCMGs}. For most of the samples
discussed we plot $g-i$ colour in terms of redshift in
\Fig\ref{fig: g-i_vs_z_UCMGs}. In \Tab\ref{tab:samples} we present
the numbers of galaxies in the different samples, which we discuss
in this section.

\begin{table*}
\centering \caption{Photometric and spectroscopic parameters for
the validation of the 28 \UCMG\ candidates observed with TNG and
NTT, for the two sets of masses \modMnozpt\ and \modMzpt . We
list: (1) candidate ID, (2) redshifts (\zp\ and \zs), (3)
effective radii calculated in kpc, (4) stellar masses without
errors on the zero-points, (5) relative validation response, (6)
stellar masses including errors on the zero-points and (7)
relative validation response. For all the quantities in columns
(2)--(7), we show the value calculated using \zp\ and \zs .
Finally, for the validation response, we use "YES" or "NO" to
state if a galaxy is a candidate for \MML\ or \MzptML\ or a
confirmed \UCMG\ for \Mspec\ or
\Mzptspec.}\label{tab:Re_Mstar_validation_REF}
\begin{tabular}{ccccccccccccc} \hline
\rm & & & & & \multicolumn{4}{c}{\modMnozpt} &
\multicolumn{4}{c}{\modMzpt} \\
\rm  ID (1) & \multicolumn{2}{c}{z (2)} & \multicolumn{2}{c}{\Re\ (3)} & \multicolumn{2}{c}{$\log \mst/\Msun$ (4)} & \multicolumn{2}{c}{Validation (5)} & \multicolumn{2}{c}{$\log \mst/\Msun$ (6)} & \multicolumn{2}{c}{Validation (7)} \\
 \cmidrule(lr){2-3} \cmidrule(lr){4-5} \cmidrule(lr){6-7}
 \cmidrule(lr){8-9} \cmidrule(lr){10-11} \cmidrule(lr){12-13}
\rm & {\tt ML} & {\tt spec} & {\tt ML} & {\tt spec} & {\tt ML} & {\tt spec}  & {\tt ML} & {\tt spec} & {\tt ML} & {\tt spec}  & {\tt ML} & {\tt spec} \\
\hline
1   &   0.29    &   0.37    &   1.43    &   1.68    &   10.97   &   11.35   &   YES &   NO  &   10.91   &   11.4    &   YES &   NO  \\
2   &   0.22    &   0.22    &   1.28    &   1.27    &   11.12   &   11.11   &   YES &   YES &   11.15   &   11.14   &   YES &   YES \\
3   &   0.35    &   0.41    &   1.09    &   1.19    &   11.     &   11.1    &   YES &   YES &   10.64   &   10.38   &   NO  &   NO  \\
4   &   0.31    &   0.33    &   1.06    &   1.1     &   11.15   &   11.22   &   YES &   YES &   11.16   &   11.21   &   YES &   YES \\
5   &   0.42    &   0.4     &   0.67   &    0.66   &   11.02   &   10.98   &   YES &   YES &   10.81   &   10.77   &   NO  &   NO  \\
6   &   0.36    &   0.32    &   1.46    &   1.36    &   10.95   &   10.87   &   YES &   NO  &   10.95   &   10.81   &   YES &   NO  \\
\hline
7   &   0.32    &   0.3     &   1.11    &   1.06    &   10.94   &   10.94   &   YES &   YES &   10.63   &   10.56   &   NO  &   NO  \\
8   &   0.35    &   0.38    &   1.45    &   1.54    &   11.37   &   11.43   &   YES &   NO  &   11.29   &   11.41   &   YES &   NO  \\
9   &   0.28    &   0.24    &   1.47    &   1.32    &   10.91   &   10.84   &   YES &   NO  &   10.85   &   10.78   &   NO  &   NO  \\
10  &   0.29    &   0.28    &   0.81   &    0.80   &   11.01   &   10.99   &   YES &   YES &   11.05   &   11.03   &   YES &   YES \\
11  &   0.31    &   0.28    &   1.01    &   0.95   &   11.01   &   10.77   &   YES &   NO  &   10.96   &   10.98   &   YES &   YES \\
12  &   0.27    &   0.29    &   0.62   &    0.65   &   10.95   &   11.     &   YES &   YES &   10.72   &   10.71   &   NO  &   NO  \\
13  &   0.31    &   0.31    &   0.92   &    0.91   &   10.95   &   10.94   &   YES &   YES &   10.71   &   10.71   &   NO  &   NO  \\
14  &   0.25    &   0.26    &   1.02    &   1.04    &   10.9    &   10.94   &   YES &   YES &   10.66   &   10.71   &   NO  &   NO  \\
15  &   0.27    &   0.3     &   1.29    &   1.36    &   10.97   &   11.09   &   YES &   YES &   10.75   &   11.09   &   NO  &   YES \\
16  &   0.28    &   0.3     &   1.36    &   1.42    &   10.91   &   10.97   &   YES &   YES &   10.87   &   10.93   &   NO  &   YES \\
17  &   0.29    &   0.32    &   1.36    &   1.43    &   10.98   &   11.04   &   YES &   YES &   10.76   &   11.04   &   NO  &   YES \\
18  &   0.34    &   0.32    &   1.04    &   0.99   &   10.98   &   10.89   &   YES &   NO  &   10.98   &   10.86   &   YES &   NO  \\
19  &   0.22    &   0.21    &   1.11    &   1.08    &   10.92   &   10.7    &   YES &   NO  &   10.75   &   10.77   &   NO  &   NO  \\
20  &   0.25    &   0.26    &   1.15    &   1.16    &   10.91   &   10.93   &   YES &   YES &   10.65   &   10.67   &   NO  &   NO  \\
21  &   0.34    &   0.3     &   1.47    &   1.37    &   11.03   &   10.93   &   YES &   YES &   11.04   &   10.88   &   YES &   NO  \\
22  &   0.31    &   0.37    &   1.1     &   1.24    &   10.96   &   11.13   &   YES &   YES &   10.96   &   11.13   &   YES &   YES \\
23  &   0.32    &   0.41    &   1.29    &   1.5     &   11.22   &   11.2    &   YES &   YES &   10.99   &   11.     &   YES &   YES \\
24  &   0.33    &   0.26    &   1.27    &   1.07    &   10.96   &   10.81   &   YES &   NO  &   10.94   &   10.77   &   YES &   NO  \\
25  &   0.27    &   0.28    &   1.49    &   1.54    &   11.     &   11.04   &   YES &   NO  &   10.66   &   10.7    &   NO  &   NO  \\
26  &   0.23    &   0.29    &   1.1     &   1.3     &   10.92   &   11.08   &   YES &   YES &   10.73   &   10.86   &   NO  &   NO  \\
27  &   0.34    &   0.34    &   1.05    &   1.05    &   10.99   &   10.99   &   YES &   YES &   10.9    &   10.9    &   NO  &   NO  \\
28  &   0.31    &   0.29    &   1.08    &   1.03    &   11.09   &   11.03   &   YES &   YES &   10.93   &   10.74   &   YES &   NO  \\
\hline
\end{tabular}
\end{table*}

\begin{figure*}
\includegraphics[width=0.99\textwidth]{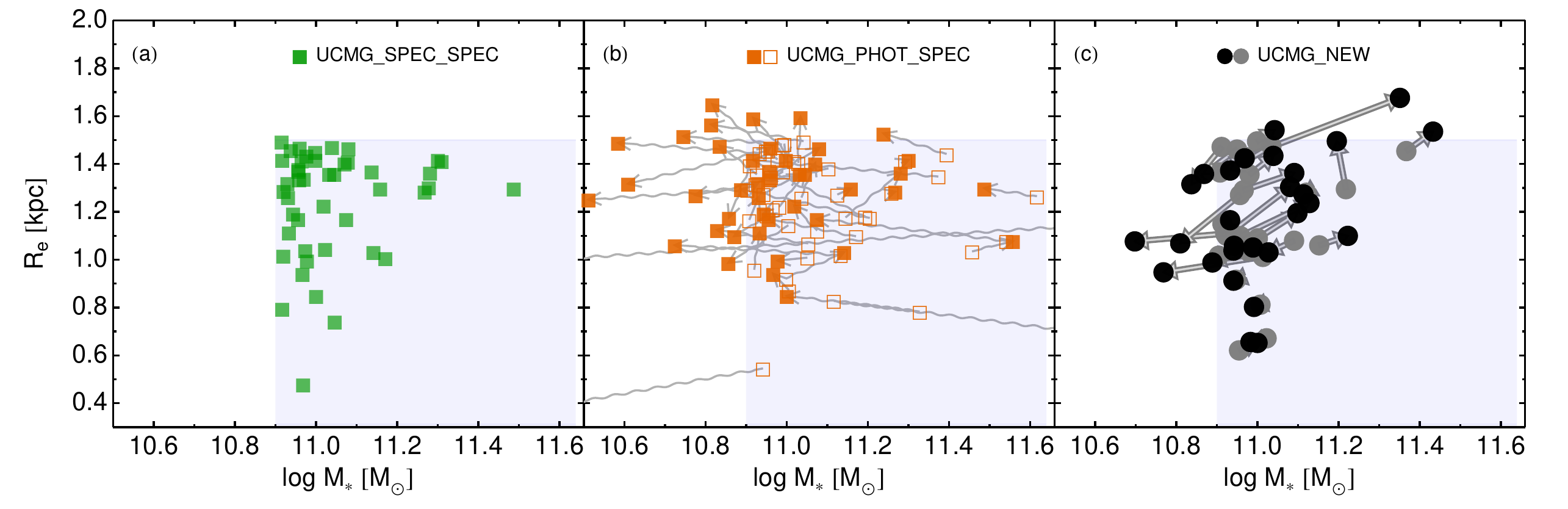}
\caption{Size vs. mass for \UCMGs. Gray shaded region highlights
the area where candidates are selected (i.e., $\Re < 1.5 \, \rm
kpc$ and $\log \mst/\Msun > 10.9$). Symbols are connected by
arrows to highlight the effect of changing the redshift on the
results. {\it Panel a.} Green squares are for the sample of
\UCMGs\ selected using the spectroscopic redshifts from the
literature (\USS). {\it Panel b.} Orange squares are relative to
the set of \UCMGs\ selected using ML photometric redshifts, but
with available measured \zs\ from the literature (\UPS). For open
and filled symbols $(\mst, \, \Re)$ are calculated assuming \zp\
and \zs, respectively. {\it Panel c.} Dots are for the 28 new
\UCMG\ KiDS candidates with \zs\ measured with observations at TNG
and NTT. For gray and black dots $(\mst, \, \Re)$ are calculated
assuming \zp\ and \zs, respectively.}\label{fig:
Re_vs_Mstar_UCMGs}
\end{figure*}

\subsection{The samples with \zs\ from the literature}\label{subsec:
validated_zs_lit}

We show the basic photometric and structural parameters for the 46
\UCMG\ candidates in the spectroscopically selected sample \USS\
in \Tabs\ref{tab:phot_parameters_sample_lit} and
\ref{tab:struc_parameters_sample_lit} in
\App\ref{app:zspec_sample}. In particular, r-band Kron magnitude,
aperture magnitudes used in the SED fitting, spectroscopic
redshifts and stellar masses are shown in
\Tab\ref{tab:phot_parameters_sample_lit}. S\'ersic structural
parameters from the \twodphot\ fit of g-, r- and i-band KiDS
surface photometry, as such as $\chi^{2}$s and \SN s, are
presented in \Tab\ref{tab:struc_parameters_sample_lit}.

We plot in \Fig\ref{fig: zs_vs_zp_UCMGs} the spectroscopic
redshifts, \zs, vs. the photometric values, \zp, for the sample
\USS\ of \UCMGs\ selected using the spectroscopic redshifts from
the literature (green squares) and the set \UPS\ of \UCMGs\
selected using ML photometric redshifts, but with available
measured \zs\ from the literature (orange squares). In the plot we
focus on the redshifts $z < 0.5$, since this is the range where
our photometrically selected sample, \UP, is complete in mass, but
for completeness we also discuss in the rest of this section some
results for galaxies at larger redshifts. The total sample
selected using \zs, taking all the galaxies with $\zs < 1$ has a
redshift bias of 0.029 and standard deviation of 0.042, and these
numbers are 0.027 and 0.038 if we reduce to the smaller redshift
range $0.15 < \zs < 0.45$. If we consider the sample selected
using \zp\ (i.e., \UPS), with $\zs < 1$, the redshifts follow the
1-to-1 relation with a bias of 0.0024, while the standard
deviation is 0.11. If we limit to the redshift range $0.15 < \zs <
0.45$, then the bias is 0.003, while the scatter is 0.049. These
values for the various statistical indicators are worse, but still
acceptable, if compared with those found for the galaxies in the
test sample of the trained network in \cite{Cavuoti+15_KIDS_I},
plotted as blue dots in \Fig\ref{fig: zs_vs_zp_UCMGs}. In fact,
including all the test set with $\zs < 1$, then the bias is 0.001
and standard deviation is 0.031, in the redshift range $0.15 < \zs
< 0.45$, the bias is 0.0025 and standard deviation is 0.029.

These objects in \USS\ are plotted in the \Re--\mst\ plane in the
left panel of \Fig\ref{fig: Re_vs_Mstar_UCMGs}. 19 out of the 46
galaxies in \USS\ are still \UCMGs\ if we include the zero-point
offset errors in the SED fitting (i.e., using \modMzpt). If we
select the set of \UCMGs\ using the zero-point offsets, then we
find a total number of 27 \UCMGs\ with $\zs < 1$, 19 in common
with the sample gathered without including the zero-point offset.

Selecting the \UCMGs\ using their ML photometric redshifts (\UPS)
yields 45 \UCMG\ candidates. 39 of these (i.e. 87 per cent) are
still compact with $\Re < 1.5 \, \rm kpc$, after \Re\ are
re-calculated using \zs\ values. But the impact on stellar masses
is more important, since 26 out of the 45 candidates (i.e. 58 per
cent or equivalently one of every $\CF = 1.73$ galaxies of the
total) are {\em bona fide} \UCMGs, after both \Re\ and \mst\ are
calculated using \zs\ values (middle panel of \Fig\ref{fig:
Re_vs_Mstar_UCMGs}). 21 out of 45 are still \UCMG\ candidates if
\modMzpt\ masses are used, instead of \modMnozpt\ values, and 13
out of 26 galaxies are still confirmed \UCMGs. The success rate
for these new numbers is of $(13/21) \sim 62$ per cent ($\CF
=1.62$). If the selection of \UCMGs\ is directly performed using
\modMzpt\ masses, then we select 24 candidates in total. 21 out of
24 candidates (88 per cent) are still compact if \zs\ is used to
calculate \Re. Instead, 12 out of 24 candidates (50 per cent,
$\CF=2$) are validated \UCMGs, after \zs\ is used for masses and
sizes. If we limit to the redshift range $\zs < 0.5$, where most
of our new observations are located, and where our samples are
complete, we find $\CF = 1.81 \, (1.88)$ if \modMnozpt\ (\modMzpt)
are used. Thus, about 87-88 per cent of candidates is still
compact if sizes are calculated using \zs, while this fraction
decreases to 50-60 per cent if we search for \UCMGs\ when both
sizes and stellar masses are recomputed with \zs. We refine these
statistics and the contamination factor \CF\ using the new
spectroscopic sample discussed in \Sec\ref{subsec: validated_new}.

We can quantify what fraction of \UCMGs\ are missed by our photo-z
based selection by cross-matching the \USS\  and \UPS\ samples. We
find that, in total, using \modMnozpt\ masses, only 26 out of 45
\UCMGs\ (57 per cent) are selected as candidates in the
photometrically selected sample \UPS, too. This means that the
number counts should be corrected by a factor \IF$=1.77$.
Similarly, if we use \modMzpt\ masses, then 17 out of 27 \UCMGs\
(63 per cent) are selected as candidates in \UPS, corresponding to
\IF$=1.59$.

Taking into account both these contrasting systematic effects,
\CF\ and \IF, we calculate the overall correction factor for the
number counts as $\IF / \CF$, finding that the true number counts
for \UCMGs\ at $z<0.5$ would be $\sim 20$ ($\sim 20$) per cent, or
equivalently $\sim 0.1$ ($\sim 0.1$) dex, higher (lower) than the
values found in a photometrically selected sample if \modMnozpt\
(\modMzpt) masses are used. These systematic errors are of the
same order of magnitude of statistical errors arising from Poisson
noise and Cosmic Variance, which we will discuss in the next
section. Though small, we take into account these systematics in
our number count calculation, presented in
\Sec\ref{sec:number_counts}. For simplicity, we will neglect the
uncertainty on these factors.

\subsection{A new confirmed \UCMG\ sample}\label{subsec:
validated_new}

From our spectroscopic campaign we have obtained redshifts for 6
and 22 candidates from the \UTNG\ and \UNTT\ samples,
respectively. The basic photometric properties of these two latter
samples, as r-band Kron magnitude, aperture magnitudes used in the
SED fitting, photometric redshifts from machine learning and
stellar masses, are shown in \Tab\ref{tab:phot_parameters}.
S\'ersic parameters from the \twodphot\ fit of g-, r- and i-band
KiDS surface photometry, as such as $\chi^{2}$s and \SN s, are
presented in \Tab\ref{tab:struc_parameters}. The image outputs of
the S\'ersic fit in the r-band are shown in
\Fig\ref{fig:2DPHOT_sample}: \UCMG\ image and residual image. A
summary of sizes and masses calculated with \zp\ or \zs\ and
results of validation process are provided in
\Tab\ref{tab:Re_Mstar_validation_REF}.

In right panel of \Fig\ref{fig: zs_vs_zp_UCMGs} we compare the new
derived spectroscopic redshifts for the 28 candidates with the
photometric values (black points). We also plot the same galaxies
considering the new machine learning photometric redshifts
(\MLtwo) stored in the last KiDS release (KiDS--DR3,
\citealt{deJong+17_KiDS_DR3}). For our sample of galaxies we find
a bias of 0.0045 and a standard deviation of 0.028. If we use the
new redshifts from KiDS--DR3, then bias and standard deviation are
0.0029 and 0.030, respectively. The new \MLtwo\ photometric
redshifts seem to work better. However, for both the redshift
assumptions (\MLone\ and \MLtwo) at $\zs \lsim 0.35$ our galaxies
exactly follow the average 1-to-1 relation, while at larger \zs, 5
out of 6 candidates have underestimated \zp\ values. Therefore,
the distribution of our redshifts seem quite consistent with what
found using the full sample of galaxies included in the blind test
in \cite{Cavuoti+15_KIDS_I}, reproducing quite well the
spectroscopic redshifts.

\begin{figure}
\includegraphics[width=0.45\textwidth]{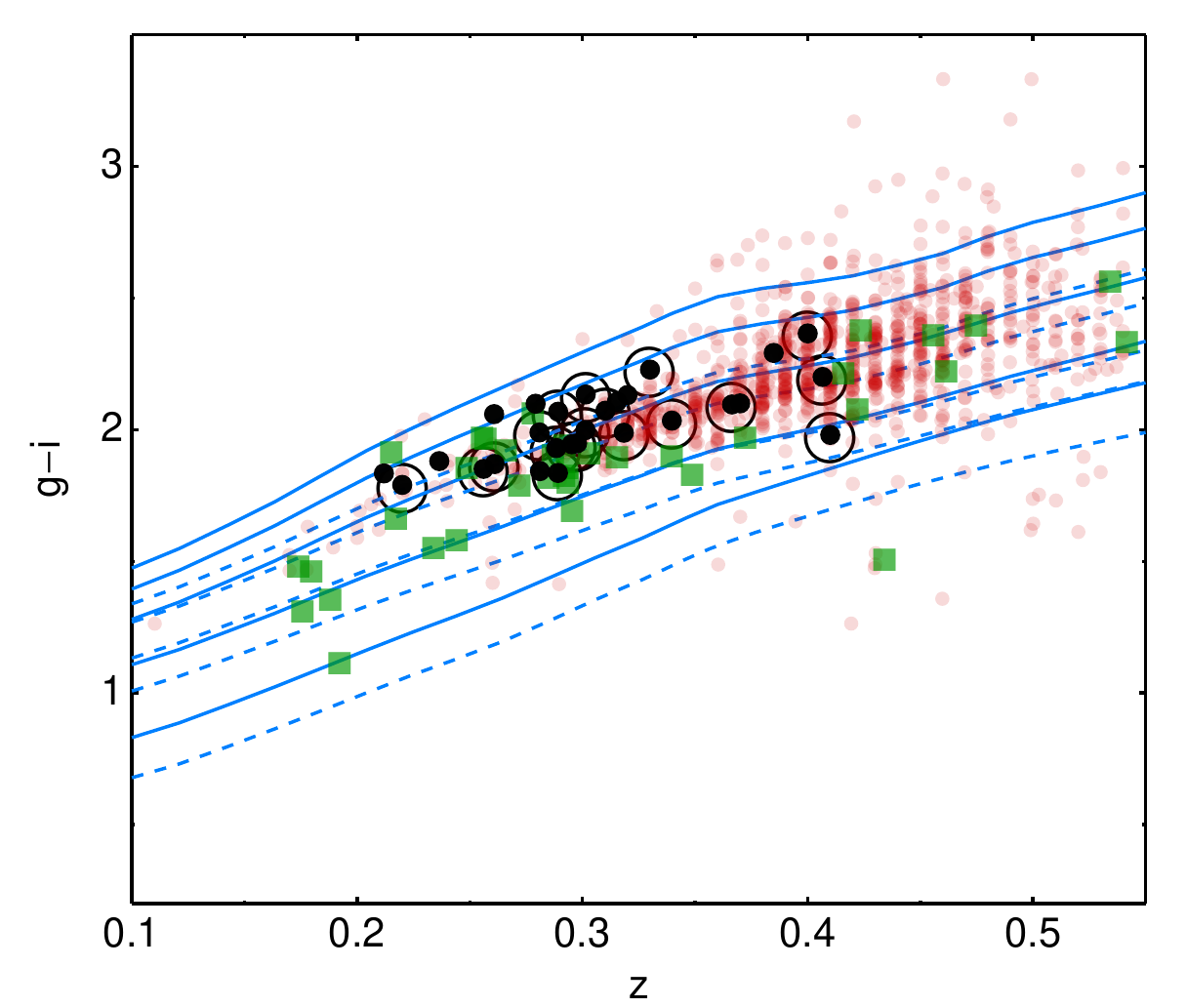}
\caption{$g-i$ vs. redshift for \UP\ sample (red dots), \USS\
(green squares), \UNEW\ (black dots) and validated \UCMGs\ in
\UNEW\ (black circles). The $g-i$ colours are calculated within an
aperture of 6'' of diameter, using \magapsix\ magnitudes.
Photometric redshifts are used for \UP, while for \USS\ and \UNEW\
we use spectroscopic values. Galaxies selected using \modMnozpt\
are plotted. Blue lines represent BC03 single-burst models. Dashed
and solid lines are models with  $Z=0.4\Zsun$ and $\Zsun$,
respectively. For each metallicity, we show models for four
different ages, from $10^{9}$ to $12 \times 10^{9} \, \rm
Gyr$.}\label{fig: g-i_vs_z_UCMGs}
\end{figure}

After recalculating sizes and masses with the new measured
spectroscopic redshifts from TNG and NTT, we find that 19 out of
the 28 \UCMG\ candidates survive as confirmed \UCMGs, which
translates to a success rate of $ 66$ per cent, or $\CF =1.52$.
Adopting \modMzpt\ masses instead, 13 out of 28 are \UCMG\
candidates, and 9 out of 13 (69 per cent, i.e. $\CF \sim 1.44$)
are still confirmed \UCMGs. As seen in \Sec\ref{subsec:
validated_zs_lit}, it is interesting to note that our selection in
size is very robust, since with spectroscopic redshifts 26 out of
28 (90 per cent) are still compact with $\Re \leq 1.5$ kpc.
Moreover, the three galaxies which have fallen out of the
compactness region are consistent within the errors with the
threshold at $1.5 \, \rm kpc$. Also in this case, it is the
stellar mass determination which is strongly affecting the
selection, as already discussed in \Sec\ref{subsec:
validated_zs_lit}. Most of our selected galaxies have masses which
are close to the mass threshold of $8 \times 10^{10}\, \rm \Msun$,
for this reason, even a very small change in the redshift could
induce changes in stellar mass which can move the galaxies out of
the range of \mst\ values for a confirmed \UCMG. However, as for
the size criterion, the uncertainties (of about 0.1-0.2 dex) make
most of these galaxies consistent within the errors with being
classified as \UCMG.

The colours of the \UCMGs\ in the spectroscopic samples (i.e.
\USS\ and \UNEW) and the photometric sample \UP\ are shown in
\Fig\ref{fig: g-i_vs_z_UCMGs}, where the colour $g-i$ is plotted
vs. redshift, and it is compared with single-burst (metal rich)
BC03 synthetic models.

\section{\UCMG\ number counts}\label{sec:number_counts}

The number counts of compact massive galaxies as a function of
redshift provide an important constraint on models of galaxy
assembly. In recent years there have been different efforts to
produce a census of such systems in different redshift bins
(\citealt{Trujillo+09_superdense, Trujillo+12_compacts,
Trujillo+14}; \citealt{Taylor+10_compacts};
\citealt{Valentinuzzi+10_WINGS}; \citealt{Poggianti+13_low_z,
Poggianti+13_evol}; \citealt{Damjanov+13_compacts,
Damjanov+14_compacts, Damjanov+15_compacts,
Damjanov+15_env_compacts}; \citealt{Gargiulo+16_dense,
Gargiulo+17_dense}; \citealt{Saulder+15_compacts};
\citealt{Tortora+16_compacts_KiDS};
\citealt{Charbonnier+17_compact_galaxies}). In the following
Section, we will compute the number counts of the sample of
\UCMGs, up to $z=0.5$, comparing our results with the ones in the
literatures.

\subsection{KiDS number counts}

We have introduced in the previous sections a set of samples of
compact galaxies, which allow us, first to quantify the \UCMG\
number counts observed in KiDS, and secondly, to correct these
numbers for systematics. We take into account the two systematics
effects discussed in \Sec\ref{sec:validation}, which would affect
the number of selected \UCMGs, considering that a) only a fraction
$1/\CF$ of photometrically selected \UCMG\ are validated after
\zs\ is measured, but b) we miss some galaxies which are not
\UCMGs\ adopting photometric redshifts, thus real numbers would be
\IF\ times larger. We correct our number counts for the factor
$\IF/\CF$. We calculate \CF\ and \IF, using the results shown in
\Sec\ref{sec:validation} (including the samples with new measured
redshifts and those from the literature), in different redshift
bins, to correct the observed number counts in terms of redshift.

\begin{figure*}
\includegraphics[width=0.95\textwidth]{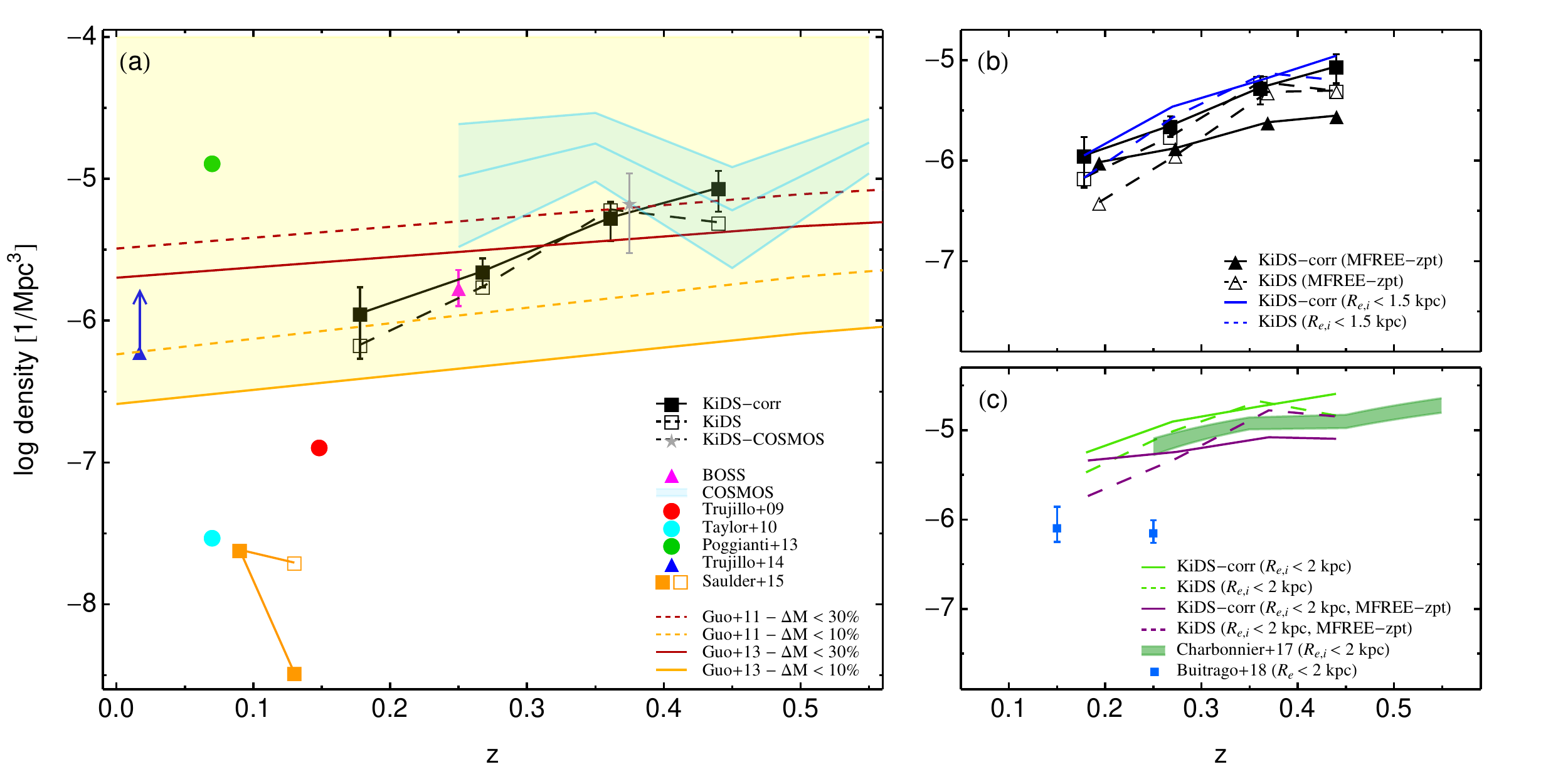}
\caption{Number density of \UCMGs\ vs. redshift. {\it Panel a.}
Open (filled) black squares, with dashed (solid) line, quoted as
KiDS and KiDS-corr in the legend, plot the number density before
(after) correction for systematics, for the selected sample
assuming \modMnozpt\ masses. Error bars denote $1\sigma$
uncertainties, taking into account Poisson noise, Cosmic Variance
and errors on \mst\ and \Re\ (see the text for more details). The
gray star is for the 4 \UCMG\ candidates at $z<0.5$ found in the
tile KIDS\_150.1\_2.2 centered on COSMOS field. The magenta
triangle with error bar shows the number counts of galaxies at
$z\sim 0.25$, with $\Re < 1.5\, \rm kpc$ and $\Mdyn
> 8 \times 10^{10}\, \rm \Msun$, from
\citet{Damjanov+14_compacts}. The cyan line with lighter cyan
region plot number counts for compacts in the COSMOS area
\citep{Damjanov+15_compacts}, selected with the same criteria as
in the present work. Red, cyan and green points are the results
for compact galaxies from \citet{Trujillo+09_superdense},
\citet{Taylor+10_compacts} and \citet{Poggianti+13_low_z},
respectively. Orange boxes show the number counts for compacts in
SDSS area from \citet{Saulder+15_compacts}, adopting our same
criteria on mass and size. Filled boxes plot the results using
S\'ersic profiles, while open boxes are for the de Vaucouleurs
profile (note that the results for the two profiles in the lowest
redshift bin are superimposed). The blue triangle and arrow are
for the lower limit at $z \sim 0$ provided by \citet{Trujillo+14}.
Dashed and solid lines are extracted from \citet{Guo+11_sims} and
\citet{Guo+13_sims} SAMs, respectively
(\citealt{Quilis_Trujillo13}). The shaded yellow region highlight
the regions allowed by the predictions from simulations. {\it
Panel b.} Number counts calculated assuming \modMnozpt\ masses
(open and filled squares) are compared with number counts when
\modMzpt\ masses are used (open and filled triangles for results
before and after the correction for systematics). Blue symbols
(open and filled before and after correction for systematics),
plot the selection of galaxies using i-band \Re, instead of the
median of g-, r- and i-band \Re\ (see text for details). {\it
Panel c.} Compacts are selected using a set of criteria similar to
the ones used in Figure~16 in
\citet{Charbonnier+17_compact_galaxies}, i.e. $\mst > 8 \times
10^{10}\, \rm \Msun$ and i-band $\Re < 2$ kpc. As in the other
panels, dashed and solid symbols are before and after correction
for systematics. Light green and violet symbols are for samples
done using \modMnozpt\ and \modMzpt\ masses. Dark green region is
for the results in Figure~16 in
\citet{Charbonnier+17_compact_galaxies}. The number densities for
the 22 objects with $\mst
> 8 \times 10^{10}\, \rm \Msun$ and $\Re \lsim 2$ kpc from Table~3
of \citet{Buitrago+18_compacts} are shown as blue boxes
with error bars. In most of the results we have omitted error bars
to not clutter the plots. In the redshift bin $(0.15-0.2)$, no
\UCMG\ candidates from \UP\ and \UNEW\ sample are found using
\modMzpt\ masses, thus we set $\CF=1$.}\label{fig: abundances}
\end{figure*}

In \Fig\ref{fig: abundances} we first plot the number counts of
the sample of photometrically selected \UCMG\ candidates
(collected in \UP) using our reference \modMnozpt\ masses. The
results for the uncorrected and corrected number counts are
plotted as open and filled symbols in the left panel of
\Fig\ref{fig: abundances}. To determine the number counts we have
binned galaxies with respect to redshift and normalized to the
comoving volume corresponding to the observed KiDS effective sky
area\footnote{Following \cite{Tortora+16_compacts_KiDS} we
multiply the number of candidates by $f_{\rm area} = A_{\rm
sky}/A_{\rm survey}$, where $A_{\rm sky}$ ($=41253$~sq.~deg.) is
the full sky area and $A_{\rm survey}$ is the effective KiDS area
(333 \sqd\ for the area analyzed in this paper). Then, the density
is derived by dividing for the comoving volume corresponding to
each redshift bin.}. The redshift bins have width of $0.1$, except
for the lowest-z bin corresponding to the redshift interval
$(0.15-0.2)$. The errors on number counts take into account
fluctuations due to Poisson noise, as well as those due to
large-scale structure (i.e. the Cosmic Variance). Following
\cite{Tortora+16_compacts_KiDS}, they are calculated with the
online {\textsc
CosmicVarianceCalculator}\footnote{http://casa.colorado.edu/$\sim$trenti/CosmicVariance.html}
tool (\citealt{Trenti_Stiavelli08}). For doing this calculation we
only use the number of spectroscopically validated \UCMGs\ from
\USS\ and \UNEW\ in each redshift bin, to take into account, in a
proper statistical way, only the confirmed \UCMGs. We have also
included in the error budget uncertainties in stellar mass and
\Re\ measurements. We build a set of 1000 Monte Carlo realizations
of the \UCMGs\ from \UP, varying both stellar mass and size of our
selected galaxies, assuming Gaussian errors of a) $\delta \log
\mst/\Msun$ equal to the uncertainty of stellar masses (on
average, $\sim 0.15\, \rm dex$) and b) $\delta \Re/kpc$ changing
with \Re\ (from $80\%$ at $\Re = 0.05''$ to $20\%$ at $\Re \gsim
0.3''$), following the results discussed in
\App\ref{app:syst_uncertainties}. We calculate the standard
deviation of the resulting number count distributions in each
redshift bin, and sum it in quadrature to the relative value from
Poisson noise and Cosmic Variance. The errors from the different
sources are of the same order of magnitude. The total relative
error on number densities is in the range $25-45$ per cent in the
bins at $z
> 0.2$. In the lowest redshift bin (0.15-0.2), due to the low
statistics, the error is $\sim 70$ per cent. These error estimates
are quite conservative, and will be reduced when larger samples of
spectroscopically validated \UCMGs\ will be collected. We find
number counts which are decreasing with cosmic time, from $\sim 9
\times 10^{-6} \, \rm Mpc^{-3}$ at $z \sim 0.5$, to $\sim 10^{-6}
\, \rm Mpc^{-3}$ at $z \sim 0.15$, which corresponds to a decrease
of $\sim 9$ times in about 3 Gyr. If we remove the lowest redshift
bin, since it is the most uncertain due to the low statistics, the
densities are 4 times less from $z\sim 0.5$ and $z \sim 0.25$
(i.e. in $\sim 2$ Gyr). In \UP\ we find just 8 photometrically
selected \UCMGs\ at $z \lsim 0.2$, and 7 of them are concentrated
in the range $0.15-0.2$ and the last one in the bin $0.1-0.15$.
Fewer (only 5 with $z \sim 0.17-0.20$) confirmed \UCMGs\ are found
in \USS\ with none among the new spectroscopically confirmed
galaxies in \UTNG\ and \UNTT.

We find larger number densities of those in
\cite{Tortora+16_compacts_KiDS}, particularly for higher-z bins,
and an inverted trend with redshift. The new results supersede the
previous one, due to some improvements implemented in the present
analysis. These improvements consist in a larger area covered (3
times more) and the larger number of candidates found (10 times
more), which provide more stable results in terms of Poisson
uncertainties and Cosmic Variance. Improvements have been also
obtained by updated NIR data and finally by the spectroscopic
sample, which has given a first constraint on incompleteness and
contaminants. In addition, a source of difference with respect to
our previous compilation is also residing in the different stellar
mass calculations, which rely, in the present analysis, on updated
KiDS filter throughput. We further test homogeneity of number
densities across the KiDS area, in connection with Poisson noise
and Cosmic Variance, in \App\ref{app:syst}.

\subsection{Comparison with literature}

At redshifts $z \lsim 0.15$, we see a lack of candidates. This is
only apparently contrasting the results of
\citet{Trujillo+09_superdense} who found, within the 6750\sqd\ of
SDSS--DR6, 29 secure \UCMGs\ at $z<0.2$ fulfilling our same
criteria, almost all of them having young ages $\lsim 4\, \rm Gyr$
(see also \citealt{Ferre-Mateu+12}). In fact, since our survey's
effective area is about 20 times smaller, these numbers suggest we
would find $\sim 1\pm 1$ candidates in our surveyed area, which is
indeed in good statistical agreement with our findings. One should
also notice that out of the 29 \UCMGs\ of
\citet{Trujillo+09_superdense}, only one is at redshift $<0.1$,
still pointing to the extreme paucity of such systems in the
nearby Universe, and consistent with our result. Similarly,
\cite{Taylor+10_compacts} found one possible old \UCMG\ at low
redshift, using a more relaxed criterion for the size, than the
one we adopt here.

Restricting to high velocity dispersions ($\sigs > 323.2\, \rm km
s^{-1}$) and sizes $\Re < 2.18$ kpc (and without any explicit cut
on stellar mass), \cite{Saulder+15_compacts} have found a sample
of 76 compact galaxies over an area of $6373\sqd$ in SDSS at $0.05
< z < 0.2$. These galaxies resemble quiescent galaxies at high-z,
i.e. systems with small effective radii and large velocity
dispersions. In this sample, 1 galaxy at $z<0.1$ and 6 at $z>0.1$
satisfy our \UCMG\ cuts (using $\Re$ from a de Vaucouleurs profile
fit; the latter number drops to only 1 if a S\'ersic profile is
fitted instead). These numbers correspond to number counts of
$2.4\times 10^{-8}\, \rm Mpc^{-3}$ in the redshift range $0.05 < z
< 0.1$, and $2\times 10^{-8}\, \rm Mpc^{-3}$ and $3.3\times
10^{-9}\, \rm Mpc^{-3}$ at $0.1 < z < 0.2$, if de Vaucouleurs or
S\'ersic profile are fitted, respectively. As mentioned in
\Sec\ref{sec:intro}, these findings seem to trouble the current
hierarchical paradigm of galaxy formation, where some relic
systems at $z \sim 0$ are actually expected to be found. In
contrast, over an area of $38\sqd$, \cite{Poggianti+13_low_z} have
found 4 galaxies fulfilling our same criteria, and all of these
galaxies are old, with mass-weighted ages older than $8\, \rm
Gyr$. These numbers translate into a very large number count of
$\sim 10^{-5}\, \rm Mpc^{-3}$ (and larger number counts should be
found including younger systems). Recently, based on theoretical
calculations, \cite{Trujillo+14} find that there should be $\sim
60$ \UCMGs\ at $z<0.1$ in the $8032\sqd$ covered by the
spectroscopic SDSS Legacy DR7, which would translate to a number
of $\sim 3 \pm 2$ in our KiDS area, still consistent with our
non-detection at $z<0.1$. However, these authors added a new
element to the story, finding one relic compact in the nearby
Perseus cluster (the only one within a distance of $73$ Mpc), i.e.
NGC 1277, reconciling the observations for relic \UCMGs\ at $z
\lsim 0.2$ with predictions from simulations. Relaxing the
constraint on the size, allowing for less compact galaxies,
\cite{Ferre-Mateu+17} confirmed two further relic galaxies, i.e.
Mrk 1216 and PGC 032873, with $\Re = 2.3$ and $1.8$ kpc,
respectively. The inclusion of these new galaxies sets the number
count of local compact galaxies at the value $\sim 6 \times
10^{-7}\, \rm Mpc^{-3}$.

The reason for the absence of relics in most of the recent studies
(which rely on very large areas) is not clear. It could be related
to spectroscopic incompleteness in some areas of the sky. Some
results point to an overabundance of \UCMGs\ in dense cluster
regions (\citealt{Valentinuzzi+10_WINGS};
\citealt{Poggianti+13_low_z}; \citealt{Damjanov+15_env_compacts};
\citealt{Stringer+15_compacts}; \citealt{PeraltadeArriba+16}). In
these dense environments the spatial proximity of the galaxies
could have prevented proper spectroscopic coverage of the targets
in SDSS and could be actually under-represented over the area
currently mapped by KiDS.

At $z > 0.2$ we find a good agreement with results from
\cite{Damjanov+14_compacts}, who select stellar-like objects
having spectroscopic redshifts from BOSS-DR10, and use a criterion
on dynamical instead of stellar mass, which is not exactly similar
to the one we apply (the purple triangle in the left panel of
\Fig\ref{fig: abundances} plots the number density of galaxies
with $\Re<1.5$ kpc and $\Mdyn
> 8 \times 10^{10}\, \rm \Msun$). The cyan region in the left panel of \Fig\ref{fig: abundances}
plots number densities for galaxies in the COSMOS survey
\citep{Damjanov+15_compacts}\footnote{These data are kindly
computed for us by I. Damjanov (private communication) by applying
the same size and mass selection criteria as in the present
work.}. Remarkably, no evolution with redshift is found in COSMOS
(on average $\sim 10^{-5}\, \rm Mpc^{-3}$). Moreover, we are
consistent with COSMOS number counts in the highest redshift bin,
but our number counts are systematically lower at lower-z, with
differences of about 1 order of magnitude in the lowest-z
bin\footnote{\cite{Damjanov+15_compacts} uses F814W effective
radii in their selection, the change of waveband would provide
smaller sizes, and thus increase the number of compact galaxies of
$\sim 0.1$ dex, as we will discuss later.}. Since
\cite{Damjanov+15_compacts} claim to find consistent density
estimates between COSMOS and BOSS (the latter having an area 4000
times larger than COSMOS), Cosmic Variance seems not to be
responsible for the above discrepancy. However, we cannot exclude
some role from the environment, which could also be the origin of
the scatter at $z \lsim 0.2$ (\citealt{Trujillo+09_superdense,
Trujillo+14}; \citealt{Taylor+10_compacts};
\citealt{Poggianti+13_low_z}). We probe the effect of Cosmic
Variance considering the tile KIDS\_150.1\_2.2, which is
overlapping with the COSMOS area. We find 4 \UCMG\ candidates
across this area (using \modMnozpt\ masses) and plot the average
number density as a gray star in the left panel of \Fig\ref{fig:
abundances} (only Poisson noise and Cosmic Variance are included
in the error budget). The results are perfectly consistent with
KiDS densities calculated across the whole DR1/2/3 area, and
within the error with \cite{Damjanov+15_compacts} results. In
\App\ref{app:syst} we further investigate the impact of Poisson
noise and Cosmic Variance, selecting samples of \UCMGs\ in random
regions. We are collecting data to study the environment in some
of our galaxies, we will investigate this issue in future papers.
Our results are also in a qualitative agreement with
\cite{Carollo+13_quenched} and \cite{Cassata+13}, which find that
the evolution of ETGs is strongly size-dependent, with a faster
decrease of the number counts for the most compact galaxies, with
respect to bigger ones. A direct comparison is not possible since
mass and size criteria from the aforementioned works are different
from ours.

Finally, we compare \UCMG\ number densities with predictions from
semi-analytical models\footnote{We caution the reader that stellar
masses and sizes are measured in a different way between
simulations and observations, hampering a straightforward
comparison of the two.} (SAMs). \cite{Quilis_Trujillo13} have
determined the evolution of the number counts of compact galaxies
from SAMs based on Millennium N-body simulations
(\citealt{Guo+11_sims, Guo+13_sims}), where relic compacts are
defined as galaxies which have barely increased their stellar mass
between $z\sim 2$ and $z\sim 0$. Operationally, they selected from
the merger tree those objects that have increased their mass since
$z=2$ by less than 10 and 30 per cent, respectively. However,
theoretical predictions should be actually considered as upper
limits, as \cite{Quilis_Trujillo13} did not apply any precise
selection in size, since the resolution in the simulations does
not allow reliable estimates of galaxy effective radii to be
obtained. On the other hand, considering that some of the \UCMGs\
in our sample may have a formation redshift $z_{\rm f} < 2$, then,
our number counts are an upper limit for number counts of relic
\UCMGs. For this reason, when compared with our data in
\Fig\ref{fig: abundances}, simulations from
\cite{Quilis_Trujillo13} have to be considered as a lower limit.

Our number counts present an evolution with redshift steeper than
predictions from simulations, being consistent with the most
(less) efficient (in terms of merging occurrence) model
predictions from \cite{Guo+11_sims} and \cite{Guo+13_sims} at low
(high) redshifts.

In {\it Panel b} of \Fig\ref{fig: abundances} we first investigate
the impact of zero-point calibration errors in the determination
of stellar masses, finding that \modMzpt\ masses decrease our
numbers, in particular for the corrected number counts. Moreover,
we study the impact of using the i-band \Re\ (using the reference
\modMnozpt\ masses), instead of our median \Re, which usually is
associated to the r-band value (the median of the KiDS g-, r- and
i-bands). At the wavelength of KiDS i-band the galaxies are known
to have smaller sizes (e.g., \Tabs\ref{tab:struc_parameters} and
\ref{tab:struc_parameters_sample_lit}; \citealt{Vulcani+14}). For
this reason, more galaxies enter in our \UCMG\ selection. Our
number counts are shifted upward of 1.3 times (i.e. $\sim 0.1$
dex).

In the right-bottom panel ({\it panel c}), we investigate the
impact on our densities of the compactness criterion, selecting
those galaxies with i-band $\Re < 2 \, \rm kpc$, assuming
\modMnozpt\ masses and using the same corrections adopted for the
sample of \UCMGs\ with r-band $\Re < 1.5 \, \rm kpc$. We find
$\sim 3.5-4$ times more galaxies ($\sim 0.55-0.6$ dex) than those
found using our size criterion. Our number counts using
\modMnozpt\ and \modMzpt\ are quite consistent with the results
from \cite{Charbonnier+17_compact_galaxies}, bracketing their
findings. The two sets of results, obtained on two different
surveys (CFHT equatorial SDSS Stripe 82, CS82, vs. KiDS) and on
different areas (their effective area of $83 \sqd$ vs our 333\sqd,
$\sim 4$ times more) are quite consistent, for what concern both
the normalization and the trend with redshift, indicating smaller
number counts at lower z, and a milder change with redshift if
compared with the results obtained when $\Re < 1.5 \, \rm kpc$. We
also show the recent results from \cite{Buitrago+18_compacts}, who
found, in 180 \sqd\ of the GAMA survey, a sample of 22 objects
with $\mst > 8 \times 10^{10}\, \rm \Msun$ and $\Re < 2 \, \rm
kpc$ in at least two bands at $z < 0.3$. They use a) KiDS g-, r-
and i-photometry and VIKING Z-band to derive the sizes and b) GAMA
stellar masses, corrected to total flux using their best fitted
S\'ersic profiles. We compare their number densities as blue boxes
with error bars with our results. We find that their densities are
systematically lower than ours of $\gsim 0.5 \, \rm dex$ at $z =
0.25$. Instead, due to the large uncertainties in our lowest
redshift bin and the very few spectroscopically validated
candidates used to correct for systematics in that bin, we cannot
exclude a marginal agreement with their number density at $z =
0.15$. However, if we take their galaxies with $\Re < 1.5$ kpc,
extrapolate the number densities, and compare the new results with
our \UCMGs\ adopting the same cut in size, then the disagreement
should be reduced.

\section{Conclusions and future prospects}\label{sec:conclusions}

Thanks to the large area covered, high image quality, excellent
spatial resolution and seeing, the Kilo Degree Survey (KiDS)
provides a unique opportunity to study the properties of
ultra-compact massive galaxies (\UCMGs). In particular, the oldest
\UCMGs\ play a key role in our understanding of galaxy formation
and evolution, sitting in the transition region between the two
different phases of the so-called "two-phase" formation scenario.
They are believed to have missed the channels of galaxy size
growth and are therefore unique systems to shed lights on the
mechanism that regulates the mass accretion history of the most
massive galaxies in our Universe.

We have started a systematic census of \UCMGs\ in
\cite{Tortora+16_compacts_KiDS} and followed up the work in this
paper, by starting a spectroscopic campaign to validate a large
subsample of candidates to have the purest sample of \UCMGs. The
present analysis improves, in terms of numbers, covered area and
analysis the one performed in \cite{Tortora+16_compacts_KiDS}.

\begin{itemize}
\item Our spectroscopic campaign has started with the observations
made with TNG and NTT telescopes of 28 candidates (19 of these 28
candidates are confirmed). Including a sample of 46 galaxies with
spectroscopic redshifts from the literature, we collect a total of
65 confirmed \UCMGs\ at $z<1$, mostly concentrated at $0.15 < z <
0.5$. We have discussed the details of our campaign, the
spectroscopic set-up and the new redshifts for the 28 candidates.
\item We have also provided a first detailed investigation of all the
sources of systematics in the search of \UCMGs\ in a photometric
survey as KiDS, which, also providing very precise photometric
redshifts with a scatter of $\sim 0.03$, is unavoidably prone to
systematics induced by small differences between the true
spectroscopic redshift and the more uncertain photometric value.
These effects have been analyzed using subsamples of \UCMGs\ with
spectroscopic redshifts from literature and the new measured
redshifts with TNG and NTT, comparing mass and $\Re$ cuts derived
with spectroscopic and photometric redshifts. These subsamples
provide a unique chance to quantify the systematics. A "wrong"
redshift induces a change in both the size and stellar mass, and
we have seen that stellar mass is more dramatically affected,
representing the more uncertain quantity in our \UCMG\ selection.
We have quantified the effects of contamination and incompleteness
due to the redshift errors via the {\it contamination factor},
\CF, and the {\it incompleteness factor}, \IF, and used them to
correct the final number counts of \UCMGs.
\item We have finally shown \UCMG\ number counts across the
last 5 Gyr, collecting a sample of $\sim 1000$ candidates at
$z<0.5$ (\UP). We find a steep decrease with cosmic time of almost
one order of magnitude, from $\sim 9 \times 10^{-6} \, \rm
Mpc^{-3}$ at $z \sim 0.5$, to $\sim 10^{-6} \, \rm Mpc^{-3}$ at $z
\sim 0.15$. We find a paucity of \UCMGs\ at $z<0.2$ which is
statistically consistent with what found in local surveys.
Although not finding consistent results with
\cite{Damjanov+15_compacts}, we find a good agreement with and an
evolution with redshift similar to the recent results from
\cite{Charbonnier+17_compact_galaxies}, when we adopt exactly
their same compactness criterion (i.e., i-band $R_{\rm e} < 1.5 \,
\rm kpc$). This result, if verified using larger datasamples and
the whole KiDS area, should suggest a size-dependent evolution of
the number count of ETGs, with the smallest and most massive
galaxies progressively reducing their number (e.g.
\citealt{Cassata+13}; \citealt{Carollo+13_quenched}).
\end{itemize}

{\it To our knowledge, our \UP\ sample, with about 1000 galaxies
spread over nearly 330 square degrees of sky, represents the
largest sample of \UCMG\ candidates assembled to date. Moreover,
using archival data as well as first results from our new
spectroscopic campaign, we have gathered the largest sample of
validated \UCMGs\ at redshift below 0.5 (and the first ones in the
Southern hemisphere).}

In a future paper we will analyze the data from new spectroscopic
observations, increasing the sample of spectroscopically validated
\UCMGs\ at redshifts $z < 0.5$. The new datasets will further
improve our knowledge of systematics in derived number counts,
allowing to reduce their uncertainties. We will also rely on
near-infrared photometry from the VIKING@VISTA survey, which we
have used in this paper to study the contamination by stars, but
in future we plan to use the 9-bands from KiDS and VIKING to
improve stellar mass measurement.

Moreover, higher resolution/deeper spectroscopy and photometry
will allow us to further investigate the properties of some
interesting candidates. First, with better spectra, we aim at
measuring absorption features and stellar velocity dispersion if
not available, constraining in this way stellar population
properties and Initial Mass Function
(\citealt{LaBarbera+13_SPIDERVIII_IMF};
\citealt{TRN13_SPIDER_IMF}). With reliable estimates for galaxy
ages, an accurate selection among relic \UCMGs\ and young \UCMGs\
will be also performed. On the other hand, the structural
properties of the \UCMGs\ need to be better understood, by using
a) adaptive optic observations which, relying on a very high
resolution, will allow to measure the small sizes of our \UCMGs\
with an exceptional precision and b) deeper photometry, to scan
their outskirts, to understand if some residual disk structures
can be found. Finally, we have already started to collect data
from multi-object spectroscopy to determine redshifts of nearby
galaxies and study the role of environment on the formation and
evolution of our \UCMGs, which can provide important clues about
the evolution of the most massive galaxies in our neighborhoods.


\section*{Acknowledgments}
We thank the referee for his/her comments, which helped to improve
the manuscript. CT, CEP and LVEK are supported through an NWO-VICI
grant (project number 639.043.308). CS has received funding from
the European Union's Horizon 2020 research and innovation
programme under the Marie Sklodowska-Curie actions (n.664931). KK
acknowledges support by the Alexander von Humboldt Foundation. SC
and MB acknowledges financial contribution from the agreement
ASI/INAF I/023/12/1. MB acknowledges the PRIN--INAF 2014 {\it
Glittering kaleidoscopes in the sky: the multifaceted nature and
role of Galaxy Clusters}. NRN, FLB and IT acknowledges financial
support from the European Union's Horizon 2020 research and
innovation programme under Marie Sk$\l$odowska-Curie grant
agreement No 721463 to the SUNDIAL ITN network and from grant
AYA2016-77237-C3-1-P from the Spanish Ministry of Economy and
Competitiveness (MINECO). GVK acknowledges financial support from
the Netherlands Research School for Astronomy (NOVA) and Target.
Target is supported by Samenwerkingsverband Noord Nederland,
European fund for regional development, Dutch Ministry of economic
affairs, Pieken in de Delta, Provinces of Groningen and Drenthe.
CB acknowledges the support of the Australian Research Council
through the award of a Future Fellowship. Based on observations
made with the Italian Telescopio Nazionale Galileo (TNG) operated
on the island of La Palma by the Fundación Galileo Galilei of the
INAF (Istituto Nazionale di Astrofisica) at the Spanish
Observatorio del Roque de los Muchachos of the Instituto de
Astrofisica de Canarias. Based on data products from observations
made with ESO Telescopes at the La Silla Paranal Observatory under
programme IDs 177.A-3016, 177.A-3017 and 177.A-3018, and on data
products produced by Target/OmegaCEN, INAF-OACN, INAF-OAPD and the
KiDS production team, on behalf of the KiDS consortium. OmegaCEN
and the KiDS production team acknowledge support by NOVA and NWO-M
grants. Members of INAF-OAPD and INAF-OACN also acknowledge the
support from the Department of Physics \& Astronomy of the
University of Padova, and of the Department of Physics of Univ.
Federico II (Naples). 2dFLenS is based on data acquired through
the Australian Astronomical Observatory, under program
A/2014B/008. It would not have been possible without the dedicated
work of the staff of the AAO in the development and support of the
2dF-AAOmega system, and the running of the AAT.


\bibliographystyle{mn2e}   



\begin{thebibliography}{98}
\expandafter\ifx\csname
natexlab\endcsname\relax\def\natexlab#1{#1}\fi

\bibitem[{{Ahn} {et~al}\mbox{.}(2014){Ahn}, {Alexandroff}, {Allende Prieto},
  {Anders}, {Anderson}, {Anderton}, {Andrews}, {Aubourg}, {Bailey}, {Bastien},
  \& et~al.}]{Ahn+14_SDSS_DR10}
{Ahn} C.~P. {et~al.}, 2014, \apjs, 211, 17

\bibitem[{{Ahn} {et~al}\mbox{.}(2012){Ahn}, {Alexandroff}, {Allende Prieto},
  {Anderson}, {Anderton}, {Andrews}, {Aubourg}, {Bailey}, {Balbinot}, {Barnes},
  \& et~al.}]{Ahn+12_SDSS_DR9}
{Ahn} C.~P. {et~al.}, 2012, \apjs, 203, 21

\bibitem[{{Arnouts} {et~al}\mbox{.}(1999){Arnouts}, {Cristiani}, {Moscardini},
  {Matarrese}, {Lucchin}, {Fontana}, \& {Giallongo}}]{Arnouts+99}
{Arnouts} S., {Cristiani} S., {Moscardini} L., {Matarrese} S.,
{Lucchin} F.,
  {Fontana} A., {Giallongo} E., 1999, \mnras, 310, 540

\bibitem[{{Beasley} {et~al}\mbox{.}(2018){Beasley}, {Trujillo}, {Leaman}, \&
  {Montes}}]{Beasley+18}
{Beasley} M.~A., {Trujillo} I., {Leaman} R., {Montes} M., 2018,
\nat, 555, 483

\bibitem[{{Belli} {et~al}\mbox{.}(2014){Belli}, {Newman}, \& {Ellis}}]{BNE14}
{Belli} S., {Newman} A.~B., {Ellis} R.~S., 2014, \apj, 783, 117

\bibitem[{{Bertin} \& {Arnouts}(1996)}]{Bertin_Arnouts96_SEx}
{Bertin} E., {Arnouts} S., 1996, \aaps, 117, 393

\bibitem[{{Blake} {et~al}\mbox{.}(2016){Blake}, {Amon}, {Childress}, {Erben},
  {Glazebrook}, {Harnois-Deraps}, {Heymans}, {Hildebrandt}, {Hinton},
  {Janssens}, {Johnson}, {Joudaki}, {Klaes}, {Kuijken}, {Lidman}, {Marin},
  {Parkinson}, {Poole}, \& {Wolf}}]{Blake+16_2dflens}
{Blake} C. {et~al.}, 2016, \mnras, 462, 4240

\bibitem[{{Brescia} {et~al}\mbox{.}(2013){Brescia}, {Cavuoti}, {D'Abrusco},
  {Longo}, \& {Mercurio}}]{Brescia+13}
{Brescia} M., {Cavuoti} S., {D'Abrusco} R., {Longo} G., {Mercurio}
A., 2013,
  \apj, 772, 140

\bibitem[{{Brescia} {et~al}\mbox{.}(2014){Brescia}, {Cavuoti}, {Longo}, \& {De
  Stefano}}]{Brescia+14}
{Brescia} M., {Cavuoti} S., {Longo} G., {De Stefano} V., 2014,
\aap, 568, A126

\bibitem[{{Bruzual} \& {Charlot}(2003)}]{BC03}
{Bruzual} G., {Charlot} S., 2003, \mnras, 344, 1000

\bibitem[{{Buitrago} {et~al}\mbox{.}(2018){Buitrago}, {Ferreras}, {Kelvin},
  {Baldry}, {Davies}, {Angthopo}, {Khochfar}, {Hopkins}, {Driver}, {Brough},
  {Sabater}, {Conselice}, {Liske}, {Holwerda}, {Bremer}, {Phillipps},
  {Lopez-Sanchez}, {Graham}, \& {Norberg}}]{Buitrago+18_compacts}
{Buitrago} F. {et~al.}, 2018, ArXiv e-prints

\bibitem[{{Capaccioli} \& {Schipani}(2011)}]{Capaccioli_Schipani11}
{Capaccioli} M., {Schipani} P., 2011, The Messenger, 146, 2

\bibitem[{{Cappellari}(2017)}]{Cappellari17}
{Cappellari} M., 2017, \mnras, 466, 798

\bibitem[{{Carollo} {et~al}\mbox{.}(2013){Carollo}, {Bschorr}, {Renzini},
  {Lilly}, {Capak}, {Cibinel}, {Ilbert}, {Onodera}, {Scoville}, {Cameron},
  {Mobasher}, {Sanders}, \& {Taniguchi}}]{Carollo+13_quenched}
{Carollo} C.~M. {et~al.}, 2013, \apj, 773, 112

\bibitem[{{Cassata} {et~al}\mbox{.}(2013){Cassata}, {Giavalisco}, {Williams},
  {Guo}, {Lee}, {Renzini}, {Ferguson}, {Faber}, {Barro}, {McIntosh}, {Lu},
  {Bell}, {Koo}, {Papovich}, {Ryan}, {Conselice}, {Grogin}, {Koekemoer}, \&
  {Hathi}}]{Cassata+13}
{Cassata} P. {et~al.}, 2013, \apj, 775, 106

\bibitem[{{Cavuoti} {et~al}\mbox{.}(2015{\natexlab{a}}){Cavuoti}, {Brescia},
  {De Stefano}, \& {Longo}}]{Cavuoti+15_PhotoRApToR}
{Cavuoti} S., {Brescia} M., {De Stefano} V., {Longo} G.,
2015{\natexlab{a}},
  Experimental Astronomy, 39, 45

\bibitem[{{Cavuoti} {et~al}\mbox{.}(2015{\natexlab{b}}){Cavuoti}, {Brescia},
  {Tortora}, {Longo}, {Napolitano}, {Radovich}, {Barbera}, {Capaccioli}, {de
  Jong}, {Getman}, {Grado}, \& {Paolillo}}]{Cavuoti+15_KIDS_I}
{Cavuoti} S. {et~al.}, 2015{\natexlab{b}}, \mnras, 452, 3100

\bibitem[{{Cavuoti} {et~al}\mbox{.}(2017){Cavuoti}, {Tortora}, {Brescia},
  {Longo}, {Radovich}, {Napolitano}, {Amaro}, {Vellucci}, {La Barbera},
  {Getman}, \& {Grado}}]{Cavuoti+17_KiDS}
{Cavuoti} S. {et~al.}, 2017, \mnras, 466, 2039

\bibitem[{{Cebri{\'a}n} \& {Trujillo}(2014)}]{Cebrian_Trujillo14}
{Cebri{\'a}n} M., {Trujillo} I., 2014, \mnras, 444, 682

\bibitem[{{Cenarro} \& {Trujillo}(2009)}]{Cenarro_Trujillo09}
{Cenarro} A.~J., {Trujillo} I., 2009, \apjl, 696, L43

\bibitem[{{Chabrier}(2001)}]{Chabrier01}
{Chabrier} G., 2001, \apj, 554, 1274

\bibitem[{{Charbonnier} {et~al}\mbox{.}(2017){Charbonnier}, {Huertas-Company},
  {Gon{\c c}alves}, {Men{\'e}ndez-Delmestre}, {Bundy}, {Galliano}, {Moraes},
  {Makler}, {Pereira}, {Erben}, {Hildebrandt}, {Shan}, {Caminha}, {Grossi}, \&
  {Riguccini}}]{Charbonnier+17_compact_galaxies}
{Charbonnier} A. {et~al.}, 2017, \mnras, 469, 4523

\bibitem[{{Coleman} {et~al}\mbox{.}(1980){Coleman}, {Wu}, \& {Weedman}}]{CWW80}
{Coleman} G.~D., {Wu} C.-C., {Weedman} D.~W., 1980, \apjs, 43, 393

\bibitem[{{Cox} {et~al}\mbox{.}(2006){Cox}, {Dutta}, {Di Matteo}, {Hernquist},
  {Hopkins}, {Robertson}, \& {Springel}}]{Cox+06}
{Cox} T.~J., {Dutta} S.~N., {Di Matteo} T., {Hernquist} L.,
{Hopkins} P.~F.,
  {Robertson} B., {Springel} V., 2006, \apj, 650, 791

\bibitem[{{Daddi} {et~al}\mbox{.}(2005){Daddi}, {Renzini}, {Pirzkal},
  {Cimatti}, {Malhotra}, {Stiavelli}, {Xu}, {Pasquali}, {Rhoads}, {Brusa}, {di
  Serego Alighieri}, {Ferguson}, {Koekemoer}, {Moustakas}, {Panagia}, \&
  {Windhorst}}]{Daddi+05}
{Daddi} E. {et~al.}, 2005, \apj, 626, 680

\bibitem[{{Damjanov} {et~al}\mbox{.}(2013){Damjanov}, {Chilingarian}, {Hwang},
  \& {Geller}}]{Damjanov+13_compacts}
{Damjanov} I., {Chilingarian} I., {Hwang} H.~S., {Geller} M.~J.,
2013, \apjl,
  775, L48

\bibitem[{{Damjanov} {et~al}\mbox{.}(2015{\natexlab{a}}){Damjanov}, {Geller},
  {Zahid}, \& {Hwang}}]{Damjanov+15_compacts}
{Damjanov} I., {Geller} M.~J., {Zahid} H.~J., {Hwang} H.~S.,
  2015{\natexlab{a}}, \apj, 806, 158

\bibitem[{{Damjanov} {et~al}\mbox{.}(2014){Damjanov}, {Hwang}, {Geller}, \&
  {Chilingarian}}]{Damjanov+14_compacts}
{Damjanov} I., {Hwang} H.~S., {Geller} M.~J., {Chilingarian} I.,
2014, \apj,
  793, 39

\bibitem[{{Damjanov} {et~al}\mbox{.}(2015{\natexlab{b}}){Damjanov}, {Zahid},
  {Geller}, \& {Hwang}}]{Damjanov+15_env_compacts}
{Damjanov} I., {Zahid} H.~J., {Geller} M.~J., {Hwang} H.~S.,
  2015{\natexlab{b}}, ArXiv e-prints

\bibitem[{{Dawson} {et~al}\mbox{.}(2013){Dawson}, {Schlegel}, {Ahn},
  {Anderson}, {Aubourg}, {Bailey}, {Barkhouser}, {Bautista}, {Beifiori},
  {Berlind}, {Bhardwaj}, {Bizyaev}, {Blake}, {Blanton}, {Blomqvist}, {Bolton},
  {Borde}, {Bovy}, {Brandt}, {Brewington}, {Brinkmann}, {Brown}, {Brownstein},
  {Bundy}, {Busca}, {Carithers}, {Carnero}, {Carr}, {Chen}, {Comparat},
  {Connolly}, {Cope}, {Croft}, {Cuesta}, {da Costa}, {Davenport}, {Delubac},
  {de Putter}, {Dhital}, {Ealet}, {Ebelke}, {Eisenstein}, {Escoffier}, {Fan},
  {Filiz Ak}, {Finley}, {Font-Ribera}, {G{\'e}nova-Santos}, {Gunn}, {Guo},
  {Haggard}, {Hall}, {Hamilton}, {Harris}, {Harris}, {Ho}, {Hogg}, {Holder},
  {Honscheid}, {Huehnerhoff}, {Jordan}, {Jordan}, {Kauffmann}, {Kazin},
  {Kirkby}, {Klaene}, {Kneib}, {Le Goff}, {Lee}, {Long}, {Loomis}, {Lundgren},
  {Lupton}, {Maia}, {Makler}, {Malanushenko}, {Malanushenko}, {Mandelbaum},
  {Manera}, {Maraston}, {Margala}, {Masters}, {McBride}, {McDonald}, {McGreer},
  {McMahon}, {Mena}, {Miralda-Escud{\'e}}, {Montero-Dorta}, {Montesano},
  {Muna}, {Myers}, {Naugle}, {Nichol}, {Noterdaeme}, {Nuza}, {Olmstead},
  {Oravetz}, {Oravetz}, {Owen}, {Padmanabhan}, {Palanque-Delabrouille}, {Pan},
  {Parejko}, {P{\^a}ris}, {Percival}, {P{\'e}rez-Fournon},
  {P{\'e}rez-R{\`a}fols}, {Petitjean}, {Pfaffenberger}, {Pforr}, {Pieri},
  {Prada}, {Price-Whelan}, {Raddick}, {Rebolo}, {Rich}, {Richards}, {Rockosi},
  {Roe}, {Ross}, {Ross}, {Rossi}, {Rubi{\~n}o-Martin}, {Samushia},
  {S{\'a}nchez}, {Sayres}, {Schmidt}, {Schneider}, {Sc{\'o}ccola}, {Seo},
  {Shelden}, {Sheldon}, {Shen}, {Shu}, {Slosar}, {Smee}, {Snedden}, {Stauffer},
  {Steele}, {Strauss}, {Streblyanska}, {Suzuki}, {Swanson}, {Tal}, {Tanaka},
  {Thomas}, {Tinker}, {Tojeiro}, {Tremonti}, {Vargas Maga{\~n}a}, {Verde},
  {Viel}, {Wake}, {Watson}, {Weaver}, {Weinberg}, {Weiner}, {West}, {White},
  {Wood-Vasey}, {Yeche}, {Zehavi}, {Zhao}, \& {Zheng}}]{Dawson+13_GAMA}
{Dawson} K.~S. {et~al.}, 2013, \aj, 145, 10

\bibitem[{{de Jong} {et~al}\mbox{.}(2017){de Jong}, {Kleijn}, {Erben},
  {Hildebrandt}, {Kuijken}, {Sikkema}, {Brescia}, {Bilicki}, {Napolitano},
  {Amaro}, {Begeman}, {Boxhoorn}, {Buddelmeijer}, {Cavuoti}, {Getman}, {Grado},
  {Helmich}, {Huang}, {Irisarri}, {La Barbera}, {Longo}, {McFarland},
  {Nakajima}, {Paolillo}, {Puddu}, {Radovich}, {Rifatto}, {Tortora},
  {Valentijn}, {Vellucci}, {Vriend}, {Amon}, {Blake}, {Choi}, {Conti}, {Gwyn},
  {Herbonnet}, {Heymans}, {Hoekstra}, {Klaes}, {Merten}, {Miller}, {Schneider},
  \& {Viola}}]{deJong+17_KiDS_DR3}
{de Jong} J.~T.~A. {et~al.}, 2017, \aap, 604, A134

\bibitem[{{de Jong} {et~al}\mbox{.}(2015){de Jong}, {Verdoes Kleijn},
  {Boxhoorn}, {Buddelmeijer}, {Capaccioli}, {Getman}, {Grado}, {Helmich},
  {Huang}, {Irisarri}, {Kuijken}, {La Barbera}, {McFarland}, {Napolitano},
  {Radovich}, {Sikkema}, {Valentijn}, {Begeman}, {Brescia}, {Cavuoti}, {Choi},
  {Cordes}, {Covone}, {Dall'Ora}, {Hildebrandt}, {Longo}, {Nakajima},
  {Paolillo}, {Puddu}, {Rifatto}, {Tortora}, {van Uitert}, {Buddendiek},
  {Harnois-D{\'e}raps}, {Erben}, {Eriksen}, {Heymans}, {Hoekstra}, {Joachimi},
  {Kitching}, {Klaes}, {Koopmans}, {K{\"o}hlinger}, {Roy}, {Sif{\'o}n},
  {Schneider}, {Sutherland}, {Viola}, \& {Vriend}}]{deJong+15_KiDS_paperI}
{de Jong} J.~T.~A. {et~al.}, 2015, \aap, 582, A62

\bibitem[{{Dekel} \& {Burkert}(2014)}]{Dekel_Burkert14}
{Dekel} A., {Burkert} A., 2014, \mnras, 438, 1870

\bibitem[{{Driver} {et~al}\mbox{.}(2011){Driver}, {Hill}, {Kelvin}, {Robotham},
  {Liske}, {Norberg}, {Baldry}, {Bamford}, {Hopkins}, {Loveday}, {Peacock},
  {Andrae}, {Bland-Hawthorn}, {Brough}, {Brown}, {Cameron}, {Ching}, {Colless},
  {Conselice}, {Croom}, {Cross}, {de Propris}, {Dye}, {Drinkwater}, {Ellis},
  {Graham}, {Grootes}, {Gunawardhana}, {Jones}, {van Kampen}, {Maraston},
  {Nichol}, {Parkinson}, {Phillipps}, {Pimbblet}, {Popescu}, {Prescott},
  {Roseboom}, {Sadler}, {Sansom}, {Sharp}, {Smith}, {Taylor}, {Thomas},
  {Tuffs}, {Wijesinghe}, {Dunne}, {Frenk}, {Jarvis}, {Madore}, {Meyer},
  {Seibert}, {Staveley-Smith}, {Sutherland}, \& {Warren}}]{Driver+11_GAMA}
{Driver} S.~P. {et~al.}, 2011, \mnras, 413, 971

\bibitem[{{Edge} {et~al}\mbox{.}(2014){Edge}, {Sutherland}, \& {The Viking
  Team}}]{Edge+14_VIKING-DR1}
{Edge} A., {Sutherland} W., {The Viking Team}, 2014, VizieR Online
Data
  Catalog, 2329, 0

\bibitem[{{Eisenstein} {et~al}\mbox{.}(2011){Eisenstein}, {Weinberg}, {Agol},
  {Aihara}, {Allende Prieto}, {Anderson}, {Arns}, {Aubourg}, {Bailey},
  {Balbinot}, \& et~al.}]{Eisenstein+11_SDSSIII}
{Eisenstein} D.~J. {et~al.}, 2011, \aj, 142, 72

\bibitem[{{Ferr{\'e}-Mateu} {et~al}\mbox{.}(2015){Ferr{\'e}-Mateu}, {Mezcua},
  {Trujillo}, {Balcells}, \& {van den Bosch}}]{Ferre-Mateu+15}
{Ferr{\'e}-Mateu} A., {Mezcua} M., {Trujillo} I., {Balcells} M.,
{van den
  Bosch} R.~C.~E., 2015, \apj, 808, 79

\bibitem[{{Ferr{\'e}-Mateu} {et~al}\mbox{.}(2017){Ferr{\'e}-Mateu}, {Trujillo},
  {Mart{\'{\i}}n-Navarro}, {Vazdekis}, {Mezcua}, {Balcells}, \&
  {Dom{\'{\i}}nguez}}]{Ferre-Mateu+17}
{Ferr{\'e}-Mateu} A., {Trujillo} I., {Mart{\'{\i}}n-Navarro} I.,
{Vazdekis} A.,
  {Mezcua} M., {Balcells} M., {Dom{\'{\i}}nguez} L., 2017, \mnras, 467, 1929

\bibitem[{{Ferr{\'e}-Mateu} {et~al}\mbox{.}(2012){Ferr{\'e}-Mateu}, {Vazdekis},
  {Trujillo}, {S{\'a}nchez-Bl{\'a}zquez}, {Ricciardelli}, \& {de la
  Rosa}}]{Ferre-Mateu+12}
{Ferr{\'e}-Mateu} A., {Vazdekis} A., {Trujillo} I.,
{S{\'a}nchez-Bl{\'a}zquez}
  P., {Ricciardelli} E., {de la Rosa} I.~G., 2012, \mnras, 423, 632

\bibitem[{{Ferreras} {et~al}\mbox{.}(2014){Ferreras}, {Trujillo},
  {M{\'a}rmol-Queralt{\'o}}, {P{\'e}rez-Gonz{\'a}lez}, {Cava}, {Barro},
  {Cenarro}, {Hern{\'a}n-Caballero}, {Cardiel},
  {Rodr{\'{\i}}guez-Zaur{\'{\i}}n}, \& {Cebri{\'a}n}}]{Ferreras+14}
{Ferreras} I. {et~al.}, 2014, \mnras, 444, 906

\bibitem[{{Gargiulo} {et~al}\mbox{.}(2016{\natexlab{a}}){Gargiulo},
  {Bolzonella}, {Scodeggio}, {Krywult}, {De Lucia}, {Guzzo}, {Garilli},
  {Grannet}, {de la Torre}, {Abbas}, {Adami}, {Arnouts}, {Bottini}, {Cappi},
  {Cucciati}, {Davidzon}, {Franzetti}, {Fritz}, {Haines}, {Hawken}, {Iovino},
  {Le Brun}, {Le F{\`e}vre}, {Maccagni}, {Ma{\l}ek}, {Marulli}, {Moutard},
  {Polletta}, {Pollo}, {Tasca}, {Tojeiro}, {Vergani}, {Zanichelli}, {Zamorani},
  {Bel}, {Branchini}, {Coupon}, {Ilbert}, \& {Moscardini}}]{Gargiulo+17_dense}
{Gargiulo} A. {et~al.}, 2016{\natexlab{a}}, ArXiv e-prints

\bibitem[{{Gargiulo} {et~al}\mbox{.}(2016{\natexlab{b}}){Gargiulo}, {Saracco},
  {Tamburri}, {Lonoce}, \& {Ciocca}}]{Gargiulo+16_dense}
{Gargiulo} A., {Saracco} P., {Tamburri} S., {Lonoce} I., {Ciocca}
F.,
  2016{\natexlab{b}}, \aap, 592, A132

\bibitem[{{Genzel} {et~al}\mbox{.}(2008){Genzel}, {Burkert}, {Bouch{\'e}},
  {Cresci}, {F{\"o}rster Schreiber}, {Shapley}, {Shapiro}, {Tacconi},
  {Buschkamp}, {Cimatti}, {Daddi}, {Davies}, {Eisenhauer}, {Erb}, {Genel},
  {Gerhard}, {Hicks}, {Lutz}, {Naab}, {Ott}, {Rabien}, {Renzini}, {Steidel},
  {Sternberg}, \& {Lilly}}]{Genzel+08}
{Genzel} R. {et~al.}, 2008, \apj, 687, 59

\bibitem[{{Guo} {et~al}\mbox{.}(2013){Guo}, {White}, {Angulo}, {Henriques},
  {Lemson}, {Boylan-Kolchin}, {Thomas}, \& {Short}}]{Guo+13_sims}
{Guo} Q., {White} S., {Angulo} R.~E., {Henriques} B., {Lemson} G.,
  {Boylan-Kolchin} M., {Thomas} P., {Short} C., 2013, \mnras, 428, 1351

\bibitem[{{Guo} {et~al}\mbox{.}(2011){Guo}, {White}, {Boylan-Kolchin}, {De
  Lucia}, {Kauffmann}, {Lemson}, {Li}, {Springel}, \& {Weinmann}}]{Guo+11_sims}
{Guo} Q. {et~al.}, 2011, \mnras, 413, 101

\bibitem[{{Hilz} {et~al}\mbox{.}(2013){Hilz}, {Naab}, \& {Ostriker}}]{Hilz+13}
{Hilz} M., {Naab} T., {Ostriker} J.~P., 2013, \mnras, 429, 2924

\bibitem[{{Hopkins} {et~al}\mbox{.}(2010){Hopkins}, {Croton}, {Bundy},
  {Khochfar}, {van den Bosch}, {Somerville}, {Wetzel}, {Keres}, {Hernquist},
  {Stewart}, {Younger}, {Genel}, \& {Ma}}]{Hopkins+10_Mergers_LCDM}
{Hopkins} P.~F. {et~al.}, 2010, \apj, 724, 915

\bibitem[{{Hopkins} {et~al}\mbox{.}(2009){Hopkins}, {Hernquist}, {Cox},
  {Keres}, \& {Wuyts}}]{Hopkins+09_DELGN_IV}
{Hopkins} P.~F., {Hernquist} L., {Cox} T.~J., {Keres} D., {Wuyts}
S., 2009,
  \apj, 691, 1424

\bibitem[{{Hsu} {et~al}\mbox{.}(2014){Hsu}, {Stockton}, \&
  {Shih}}]{Hsu+14_compacts}
{Hsu} L.-Y., {Stockton} A., {Shih} H.-Y., 2014, \apj, 796, 92

\bibitem[{{Ilbert} {et~al}\mbox{.}(2006){Ilbert}, {Arnouts}, {McCracken},
  {Bolzonella}, {Bertin}, {Le F{\`e}vre}, {Mellier}, {Zamorani}, {Pell{\`o}},
  {Iovino}, {Tresse}, {Le Brun}, {Bottini}, {Garilli}, {Maccagni}, {Picat},
  {Scaramella}, {Scodeggio}, {Vettolani}, {Zanichelli}, {Adami}, {Bardelli},
  {Cappi}, {Charlot}, {Ciliegi}, {Contini}, {Cucciati}, {Foucaud}, {Franzetti},
  {Gavignaud}, {Guzzo}, {Marano}, {Marinoni}, {Mazure}, {Meneux}, {Merighi},
  {Paltani}, {Pollo}, {Pozzetti}, {Radovich}, {Zucca}, {Bondi}, {Bongiorno},
  {Busarello}, {de La Torre}, {Gregorini}, {Lamareille}, {Mathez}, {Merluzzi},
  {Ripepi}, {Rizzo}, \& {Vergani}}]{Ilbert+06}
{Ilbert} O. {et~al.}, 2006, \aap, 457, 841

\bibitem[{{Kauffmann} {et~al}\mbox{.}(2003){Kauffmann}, {Heckman}, {White},
  {Charlot}, {Tremonti}, {Peng}, {Seibert}, {Brinkmann}, {Nichol}, {SubbaRao},
  \& {York}}]{Kauffmann+03}
{Kauffmann} G. {et~al.}, 2003, \mnras, 341, 54

\bibitem[{{Khochfar} \& {Burkert}(2003)}]{Khochfar_Burkert03}
{Khochfar} S., {Burkert} A., 2003, \apjl, 597, L117

\bibitem[{{Khochfar} \& {Silk}(2006)}]{Khochfar_Silk06}
{Khochfar} S., {Silk} J., 2006, \apjl, 648, L21

\bibitem[{{Kinney} {et~al}\mbox{.}(1996){Kinney}, {Calzetti}, {Bohlin},
  {McQuade}, {Storchi-Bergmann}, \& {Schmitt}}]{Kinney+96}
{Kinney} A.~L., {Calzetti} D., {Bohlin} R.~C., {McQuade} K.,
{Storchi-Bergmann}
  T., {Schmitt} H.~R., 1996, \apj, 467, 38

\bibitem[{{Komatsu} {et~al}\mbox{.}(2011){Komatsu}, {Smith}, {Dunkley},
  {Bennett}, {Gold}, {Hinshaw}, {Jarosik}, {Larson}, {Nolta}, {Page},
  {Spergel}, {Halpern}, {Hill}, {Kogut}, {Limon}, {Meyer}, {Odegard}, {Tucker},
  {Weiland}, {Wollack}, \& {Wright}}]{Komatsu+11_WMAP7}
{Komatsu} E. {et~al.}, 2011, \apjs, 192, 18

\bibitem[{{Kormendy}(1977)}]{Kormendy77_II}
{Kormendy} J., 1977, \apj, 218, 333

\bibitem[{{La Barbera} {et~al}\mbox{.}(2010){La Barbera}, {de Carvalho}, {de La
  Rosa}, {Lopes}, {Kohl-Moreira}, \& {Capelato}}]{SPIDER-I}
{La Barbera} F., {de Carvalho} R.~R., {de La Rosa} I.~G., {Lopes}
P.~A.~A.,
  {Kohl-Moreira} J.~L., {Capelato} H.~V., 2010, \mnras, 408, 1313

\bibitem[{{La Barbera} {et~al}\mbox{.}(2008){La Barbera}, {de Carvalho},
  {Kohl-Moreira}, {Gal}, {Soares-Santos}, {Capaccioli}, {Santos}, \&
  {Sant'anna}}]{LaBarbera_08_2DPHOT}
{La Barbera} F., {de Carvalho} R.~R., {Kohl-Moreira} J.~L., {Gal}
R.~R.,
  {Soares-Santos} M., {Capaccioli} M., {Santos} R., {Sant'anna} N., 2008,
  \pasp, 120, 681

\bibitem[{{La Barbera} {et~al}\mbox{.}(2013){La Barbera}, {Ferreras},
  {Vazdekis}, {de la Rosa}, {de Carvalho}, {Trevisan}, {Falc{\'o}n-Barroso}, \&
  {Ricciardelli}}]{LaBarbera+13_SPIDERVIII_IMF}
{La Barbera} F., {Ferreras} I., {Vazdekis} A., {de la Rosa} I.~G.,
{de
  Carvalho} R.~R., {Trevisan} M., {Falc{\'o}n-Barroso} J., {Ricciardelli} E.,
  2013, \mnras, 433, 3017

\bibitem[{{L{\"a}sker} {et~al}\mbox{.}(2013){L{\"a}sker}, {van den Bosch}, {van
  de Ven}, {Ferreras}, {La Barbera}, {Vazdekis}, \&
  {Falc{\'o}n-Barroso}}]{Lasker+13_IMF_compact}
{L{\"a}sker} R., {van den Bosch} R.~C.~E., {van de Ven} G.,
{Ferreras} I., {La
  Barbera} F., {Vazdekis} A., {Falc{\'o}n-Barroso} J., 2013, \mnras, 434, L31

\bibitem[{{Maddox} {et~al}\mbox{.}(2008){Maddox}, {Hewett}, {Warren}, \&
  {Croom}}]{Maddox+08}
{Maddox} N., {Hewett} P.~C., {Warren} S.~J., {Croom} S.~M., 2008,
\mnras, 386,
  1605

\bibitem[{{Mart\'in-Navarro} {et~al}\mbox{.}(2015){Mart\'in-Navarro}, {La
  Barbera}, {Vazdekis}, {Ferr{\'e}-Mateu}, {Trujillo}, \&
  {Beasley}}]{Martin-Navarro+15_IMF_relic}
{Mart\'in-Navarro} I., {La Barbera} F., {Vazdekis} A.,
{Ferr{\'e}-Mateu} A.,
  {Trujillo} I., {Beasley} M.~A., 2015, \mnras, 451, 1081

\bibitem[{{Muzzin} {et~al}\mbox{.}(2013){Muzzin}, {Marchesini}, {Stefanon},
  {Franx}, {Milvang-Jensen}, {Dunlop}, {Fynbo}, {Brammer}, {Labb{\'e}}, \& {van
  Dokkum}}]{Muzzin+13}
{Muzzin} A. {et~al.}, 2013, \apjs, 206, 8

\bibitem[{{Naab} {et~al}\mbox{.}(2009){Naab}, {Johansson}, \&
  {Ostriker}}]{Naab+09}
{Naab} T., {Johansson} P.~H., {Ostriker} J.~P., 2009, \apjl, 699,
L178

\bibitem[{{Peralta de Arriba} {et~al}\mbox{.}(2016){Peralta de Arriba},
  {Quilis}, {Trujillo}, {Cebri{\'a}n}, \& {Balcells}}]{PeraltadeArriba+16}
{Peralta de Arriba} L., {Quilis} V., {Trujillo} I., {Cebri{\'a}n}
M.,
  {Balcells} M., 2016, \mnras, 461, 156

\bibitem[{{Poggianti} {et~al}\mbox{.}(2013{\natexlab{a}}){Poggianti}, {Calvi},
  {Bindoni}, {D'Onofrio}, {Moretti}, {Valentinuzzi}, {Fasano}, {Fritz}, {De
  Lucia}, {Vulcani}, {Bettoni}, {Gullieuszik}, \&
  {Omizzolo}}]{Poggianti+13_low_z}
{Poggianti} B.~M. {et~al.}, 2013{\natexlab{a}}, \apj, 762, 77

\bibitem[{{Poggianti} {et~al}\mbox{.}(2013{\natexlab{b}}){Poggianti},
  {Moretti}, {Calvi}, {D'Onofrio}, {Valentinuzzi}, {Fritz}, \&
  {Renzini}}]{Poggianti+13_evol}
{Poggianti} B.~M., {Moretti} A., {Calvi} R., {D'Onofrio} M.,
{Valentinuzzi} T.,
  {Fritz} J., {Renzini} A., 2013{\natexlab{b}}, \apj, 777, 125

\bibitem[{{Quilis} \& {Trujillo}(2013)}]{Quilis_Trujillo13}
{Quilis} V., {Trujillo} I., 2013, \apjl, 773, L8

\bibitem[{{Roy} {et~al}\mbox{.}(2018){Roy}, {Napolitano}, {La Barbera},
  {Tortora}, {Getman}, {Radovich}, {Capaccioli}, {Brescia}, {Cavuoti}, {Longo},
  {Raj}, {Puddu}, {Covone}, {Amaro}, {Vellucci}, {Grado}, {Kuijken}, {Verdoes
  Kleijn}, \& {Valentijn}}]{Roy+18}
{Roy} N. {et~al.}, 2018, ArXiv e-prints

\bibitem[{{Saulder} {et~al}\mbox{.}(2015){Saulder}, {van den Bosch}, \&
  {Mieske}}]{Saulder+15_compacts}
{Saulder} C., {van den Bosch} R.~C.~E., {Mieske} S., 2015, \aap,
578, A134

\bibitem[{{Schlafly} \& {Finkbeiner}(2011)}]{Schlafly_Finkbeiner11}
{Schlafly} E.~F., {Finkbeiner} D.~P., 2011, \apj, 737, 103

\bibitem[{{Shih} \& {Stockton}(2011)}]{Shih_Stockton11}
{Shih} H.-Y., {Stockton} A., 2011, \apj, 733, 45

\bibitem[{{Stockton} {et~al}\mbox{.}(2014){Stockton}, {Shih}, {Larson}, \&
  {Mann}}]{Stockton+14_compacts}
{Stockton} A., {Shih} H.-Y., {Larson} K., {Mann} A.~W., 2014,
\apj, 780, 134

\bibitem[{{Stringer} {et~al}\mbox{.}(2015){Stringer}, {Trujillo}, {Dalla
  Vecchia}, \& {Martinez-Valpuesta}}]{Stringer+15_compacts}
{Stringer} M., {Trujillo} I., {Dalla Vecchia} C.,
{Martinez-Valpuesta} I.,
  2015, \mnras, 449, 2396

\bibitem[{{Taylor} {et~al}\mbox{.}(2010){Taylor}, {Franx}, {Glazebrook},
  {Brinchmann}, {van der Wel}, \& {van Dokkum}}]{Taylor+10_compacts}
{Taylor} E.~N., {Franx} M., {Glazebrook} K., {Brinchmann} J., {van
der Wel} A.,
  {van Dokkum} P.~G., 2010, \apj, 720, 723

\bibitem[{{Thomas} {et~al}\mbox{.}(2005){Thomas}, {Maraston}, {Bender}, \&
  {Mendes de Oliveira}}]{Thomas+05}
{Thomas} D., {Maraston} C., {Bender} R., {Mendes de Oliveira} C.,
2005, \apj,
  621, 673

\bibitem[{{Tortora} {et~al}\mbox{.}(2016){Tortora}, {La Barbera}, {Napolitano},
  {Roy}, {Radovich}, {Cavuoti}, {Brescia}, {Longo}, {Getman}, {Capaccioli},
  {Grado}, {Kuijken}, {de Jong}, {McFarland}, \&
  {Puddu}}]{Tortora+16_compacts_KiDS}
{Tortora} C. {et~al.}, 2016, \mnras, 457, 2845

\bibitem[{{Tortora} {et~al}\mbox{.}(2009){Tortora}, {Napolitano}, {Romanowsky},
  {Capaccioli}, \& {Covone}}]{Tortora+09}
{Tortora} C., {Napolitano} N.~R., {Romanowsky} A.~J., {Capaccioli}
M., {Covone}
  G., 2009, \mnras, 396, 1132

\bibitem[{{Tortora} {et~al}\mbox{.}(2018){Tortora}, {Napolitano}, {Roy},
  {Radovich}, {Getman}, {Koopmans}, {Verdoes Kleijn}, \&
  {Kuijken}}]{Tortora+18_KiDS_DMevol}
{Tortora} C., {Napolitano} N.~R., {Roy} N., {Radovich} M.,
{Getman} F.,
  {Koopmans} L.~V.~E., {Verdoes Kleijn} G.~A., {Kuijken} K.~H., 2018, \mnras,
  473, 969

\bibitem[{{Tortora} {et~al}\mbox{.}(2014){Tortora}, {Napolitano}, {Saglia},
  {Romanowsky}, {Covone}, \& {Capaccioli}}]{Tortora+14_DMevol}
{Tortora} C., {Napolitano} N.~R., {Saglia} R.~P., {Romanowsky}
A.~J., {Covone}
  G., {Capaccioli} M., 2014, \mnras, 445, 162

\bibitem[{{Tortora} {et~al}\mbox{.}(2013){Tortora}, {Romanowsky}, \&
  {Napolitano}}]{TRN13_SPIDER_IMF}
{Tortora} C., {Romanowsky} A.~J., {Napolitano} N.~R., 2013, \apj,
765, 8

\bibitem[{{Trenti} \& {Stiavelli}(2008)}]{Trenti_Stiavelli08}
{Trenti} M., {Stiavelli} M., 2008, \apj, 676, 767

\bibitem[{{Trujillo} {et~al}\mbox{.}(2001){Trujillo}, {Aguerri}, {Cepa}, \&
  {Guti{\'e}rrez}}]{Trujillo+01}
{Trujillo} I., {Aguerri} J.~A.~L., {Cepa} J., {Guti{\'e}rrez}
C.~M., 2001,
  \mnras, 321, 269

\bibitem[{{Trujillo} {et~al}\mbox{.}(2012){Trujillo}, {Carrasco}, \&
  {Ferr{\'e}-Mateu}}]{Trujillo+12_compacts}
{Trujillo} I., {Carrasco} E.~R., {Ferr{\'e}-Mateu} A., 2012, \apj,
751, 45

\bibitem[{{Trujillo} {et~al}\mbox{.}(2009){Trujillo}, {Cenarro}, {de
  Lorenzo-C{\'a}ceres}, {Vazdekis}, {de la Rosa}, \&
  {Cava}}]{Trujillo+09_superdense}
{Trujillo} I., {Cenarro} A.~J., {de Lorenzo-C{\'a}ceres} A.,
{Vazdekis} A., {de
  la Rosa} I.~G., {Cava} A., 2009, \apjl, 692, L118

\bibitem[{{Trujillo} {et~al}\mbox{.}(2007){Trujillo}, {Conselice}, {Bundy},
  {Cooper}, {Eisenhardt}, \& {Ellis}}]{Trujillo+07}
{Trujillo} I., {Conselice} C.~J., {Bundy} K., {Cooper} M.~C.,
{Eisenhardt} P.,
  {Ellis} R.~S., 2007, \mnras, 382, 109

\bibitem[{{Trujillo} {et~al}\mbox{.}(2014){Trujillo}, {Ferr{\'e}-Mateu},
  {Balcells}, {Vazdekis}, \& {S{\'a}nchez-Bl{\'a}zquez}}]{Trujillo+14}
{Trujillo} I., {Ferr{\'e}-Mateu} A., {Balcells} M., {Vazdekis} A.,
  {S{\'a}nchez-Bl{\'a}zquez} P., 2014, \apjl, 780, L20

\bibitem[{{Trujillo} {et~al}\mbox{.}(2011){Trujillo}, {Ferreras}, \& {de La
  Rosa}}]{Trujillo+11}
{Trujillo} I., {Ferreras} I., {de La Rosa} I.~G., 2011, \mnras,
415, 3903

\bibitem[{{Trujillo} {et~al}\mbox{.}(2006){Trujillo}, {F{\"o}rster Schreiber},
  {Rudnick}, {Barden}, {Franx}, {Rix}, {Caldwell}, {McIntosh}, {Toft},
  {H{\"a}ussler}, {Zirm}, {van Dokkum}, {Labb{\'e}}, \& {}}]{Trujillo+06}
{Trujillo} I. {et~al.}, 2006, \apj, 650, 18

\bibitem[{{Valentinuzzi} {et~al}\mbox{.}(2010){Valentinuzzi}, {Fritz},
  {Poggianti}, {Cava}, {Bettoni}, {Fasano}, {D'Onofrio}, {Couch}, {Dressler},
  {Moles}, {Moretti}, {Omizzolo}, {Kj{\ae}rgaard}, {Vanzella}, \&
  {Varela}}]{Valentinuzzi+10_WINGS}
{Valentinuzzi} T. {et~al.}, 2010, \apj, 712, 226

\bibitem[{{van der Wel} {et~al}\mbox{.}(2008){van der Wel}, {Holden}, {Zirm},
  {Franx}, {Rettura}, {Illingworth}, \& {Ford}}]{vanderWel+08}
{van der Wel} A., {Holden} B.~P., {Zirm} A.~W., {Franx} M.,
{Rettura} A.,
  {Illingworth} G.~D., {Ford} H.~C., 2008, \apj, 688, 48

\bibitem[{{van Dokkum} {et~al}\mbox{.}(2010){van Dokkum}, {Whitaker},
  {Brammer}, {Franx}, {Kriek}, {Labb{\'e}}, {Marchesini}, {Quadri}, {Bezanson},
  {Illingworth}, {Muzzin}, {Rudnick}, {Tal}, \& {Wake}}]{vanDokkum+10}
{van Dokkum} P.~G. {et~al.}, 2010, \apj, 709, 1018

\bibitem[{{Vazdekis} {et~al}\mbox{.}(2010){Vazdekis},
  {S{\'a}nchez-Bl{\'a}zquez}, {Falc{\'o}n-Barroso}, {Cenarro}, {Beasley},
  {Cardiel}, {Gorgas}, \& {Peletier}}]{Vazdekis10}
{Vazdekis} A., {S{\'a}nchez-Bl{\'a}zquez} P., {Falc{\'o}n-Barroso}
J.,
  {Cenarro} A.~J., {Beasley} M.~A., {Cardiel} N., {Gorgas} J., {Peletier}
  R.~F., 2010, \mnras, 404, 1639

\bibitem[{{Vulcani} {et~al}\mbox{.}(2014){Vulcani}, {Bamford},
  {H{\"a}u{\ss}ler}, {Vika}, {Rojas}, {Agius}, {Baldry}, {Bauer}, {Brown},
  {Driver}, {Graham}, {Kelvin}, {Liske}, {Loveday}, {Popescu}, {Robotham}, \&
  {Tuffs}}]{Vulcani+14}
{Vulcani} B. {et~al.}, 2014, \mnras, 441, 1340

\bibitem[{{Vulcani} {et~al}\mbox{.}(2011){Vulcani}, {Poggianti},
  {Arag{\'o}n-Salamanca}, {Fasano}, {Rudnick}, {Valentinuzzi}, {Dressler},
  {Bettoni}, {Cava}, {D'Onofrio}, {Fritz}, {Moretti}, {Omizzolo}, \&
  {Varela}}]{Vulcani+11}
{Vulcani} B. {et~al.}, 2011, \mnras, 412, 246

\bibitem[{{Wang} {et~al}\mbox{.}(2017){Wang}, {Luo}, {Shen}, {Hou}, {Kong},
  {Song}, {Zhang}, {Hong}, {Cao}, {Hou}, {Wang}, {Zhang}, \& {Zhao}}]{Wang+17}
{Wang} L.-L. {et~al.}, 2017, ArXiv e-prints

\bibitem[{{Wellons} {et~al}\mbox{.}(2016){Wellons}, {Torrey}, {Ma},
  {Rodriguez-Gomez}, {Pillepich}, {Nelson}, {Genel}, {Vogelsberger}, \&
  {Hernquist}}]{Wellons+15_lower_z}
{Wellons} S. {et~al.}, 2016, \mnras, 456, 1030

\bibitem[{{Y{\i}ld{\i}r{\i}m} {et~al}\mbox{.}(2015){Y{\i}ld{\i}r{\i}m}, {van
  den Bosch}, {van de Ven}, {Husemann}, {Lyubenova}, {Walsh}, {Gebhardt}, \&
  {G{\"u}ltekin}}]{Yildirim+15}
{Y{\i}ld{\i}r{\i}m} A., {van den Bosch} R.~C.~E., {van de Ven} G.,
{Husemann}
  B., {Lyubenova} M., {Walsh} J.~L., {Gebhardt} K., {G{\"u}ltekin} K., 2015,
  \mnras, 452, 1792

\end{thebibliography}

\appendix

\section{Completeness}\label{app:completeness}

\begin{figure*}
\includegraphics[trim= 0mm -1.5mm 5mm 0mm, width=0.32\textwidth]{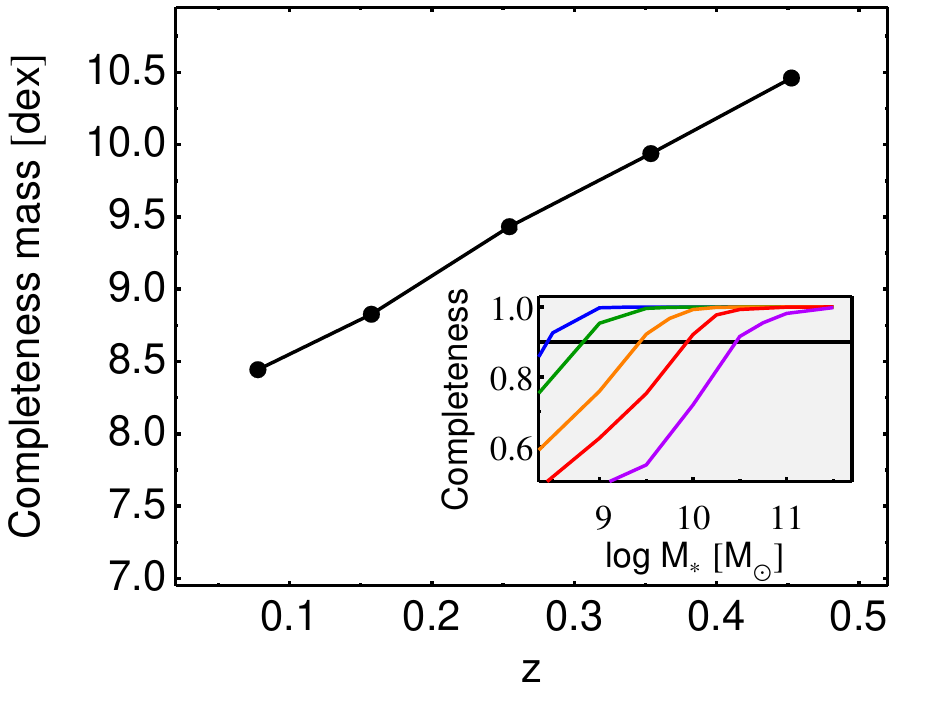}
\includegraphics[width=0.6\textwidth]{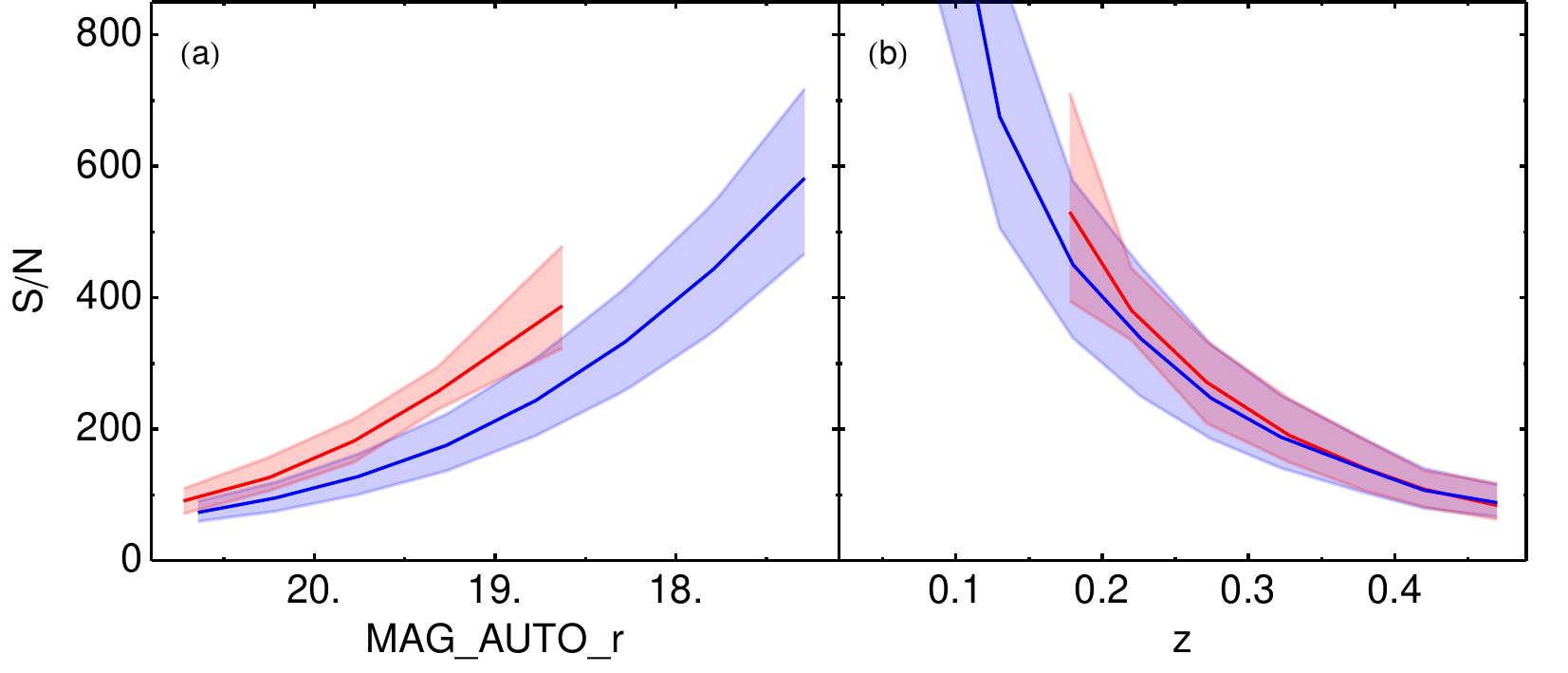}
\caption{Mass completeness of the high--\SN\ sample and typical
\SN\ values. In the left panel we show the mass completeness. In
the main panel the 90\% completeness mass is plotted in terms of
the redshift. In the inset we present the completeness as a
function of stellar mass for redshift bins, normalized to the
range $(0, \, 1)$, with the value $1$ corresponding to the 100\%
completeness. Five redshift bins are adopted: $z < 0.1$ (blue),
$0.1 < z \leq 0.2$ (green), $0.2 < z \leq 0.3$ (orange), $0.3 < z
\leq 0.4$ (red) and $0.4 < z \leq 0.5$ (purple). The solid
horizontal line correspond to the 90\% completeness. Instead, in
the right panels, we present the \SN\ as a function of \Mautor\
(panel a) and redshift (panel b). Median values are plotted as
solid lines, and shaded regions show 16--84th quantiles of the
distribution in each bin. Blue (red) lines with shaded regions are
for the high--\SN\ sample (\UP\
sample).}\label{fig:Mstar_completeness}
\end{figure*}

We evaluate the completeness of the high--\SN\ sample following
the approach discussed in \cite{Tortora+16_compacts_KiDS} and
\cite{Roy+18}. For the magnitude completeness we compute the
fraction of detected galaxies of the high--\SN\ sample with
respect to the number of galaxies of the deeper and complete
sample of $\sim 5$ million galaxies (see \Sec\ref{sec:sample}), at
magnitudes brighter than $mag_{\rm r,0}$. The $mag_{\rm r,0}$
value corresponding to a fraction of 90\% is by definition our
completeness magnitude. For the mass completeness we follow a
similar procedure. For the five redshift bins $0 \leq z < 0.1$,
$0.1 < z \leq 0.2$, $0.2 < z \leq 0.3$, $0.3 < z \leq 0.4$ and
$0.4 < z \leq 0.5$, we compute the fraction of detected galaxies
of the high--\SN\ sample with respect to the total number of
galaxies, with stellar masses larger than a value $\mst_{0}$. And
then we calculate the $\mst_{0}$ value which corresponds to the
$90\%$ completeness. For brevity, we only present the results for
the mass completeness in the left panel of
\Fig\ref{fig:Mstar_completeness}, where we show both the 90\%
completeness mass as a function of the redshift (main panel) and
the completeness mass in terms of stellar mass (inset panel). The
sample of high-\SN\ galaxies is complete down to a magnitude of
$\Mautor \sim 21$ and a stellar mass of $\sim 3 \times 10^{8} \,
\Msun$ up to redshift $z=0.1$ and $\sim 3 \times 10^{10} \, \Msun$
up to $z \sim 0.5$.

Due to their rare nature, some \UCMGs\ should potentially escape
our selection, since for example have a magnitude and mass
completeness which are different from those of normal-sized
galaxies, which are predominantly populating the samples just used
for the mass completeness calculation. For this reason, in the
right panels of \Fig\ref{fig:Mstar_completeness} we compare the
average \SN\ for a) the \UCMG\ candidates in the \UP\ sample and
b) the whole galaxy population (the control sample) within the
same mass and redshift ranges, i.e. $\mst
> 8 \times 10^{10}\, \rm \Msun$ and $z < 0.5$. The median
\SN\ is plotted in terms of \Mautor\ and redshift. The \SN\ is the
parameter which establishes if a galaxy is detected and if has
reliable structural parameters. At fixed magnitude, the median
\SN\ of the \UCMG\ candidates is larger than the \SN\ value of the
control sample, while the \SN s are closer if plotted in terms of
redshift. This demonstrates that, at fixed magnitude or redshift,
\UCMGs\ have similar chances to be detected of normal-sized
galaxies at similar magnitude or redshift. This is not surprising
if we consider that our objects are luminous and have a light
profile which is very concentrated, generating a larger \SN\ per
unit area. Thus, this confirms that our sample of \UCMGs\ is
complete down to a magnitude of $\Mautor \sim 21$ and a stellar
mass of $\sim 3 \times 10^{10} \, \Msun$ up to $z \sim 0.5$.

\section{Systematics and statistical uncertainties in the size
measurement}\label{app:syst_uncertainties}

\subsection{Simulated galaxies and systematics}

\begin{figure*}
\centering
\includegraphics[width=0.9\textwidth]{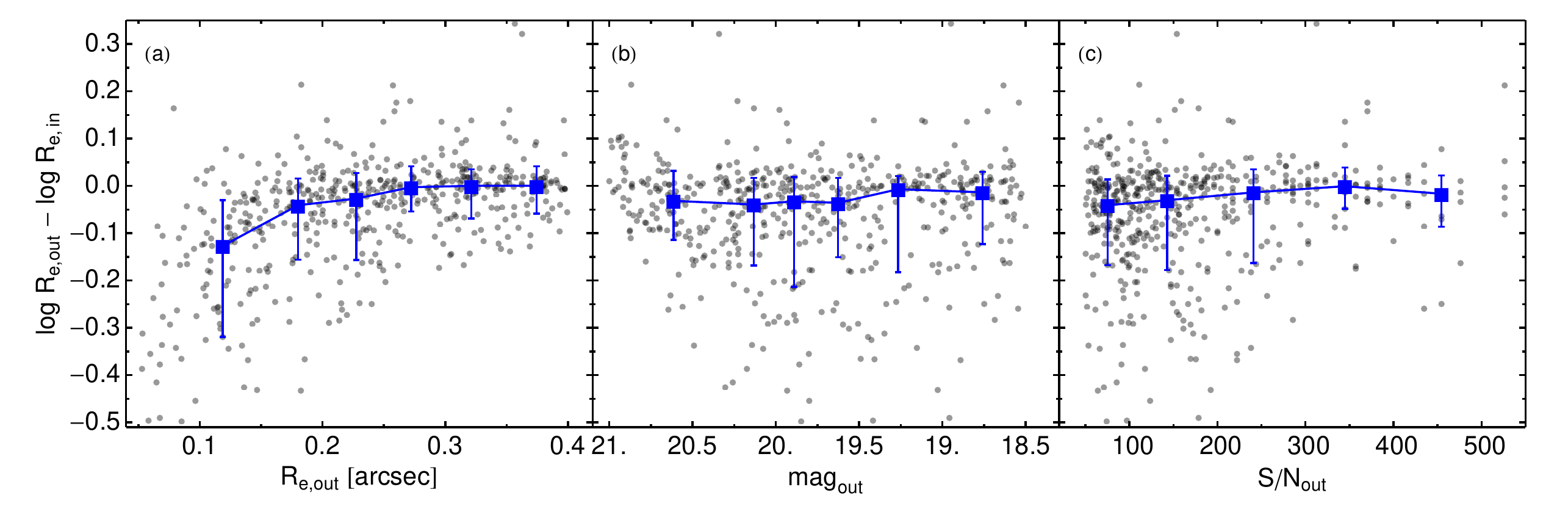}
\caption{Differences between input and output \Re\ (in logarithmic
scale) as a function of different output quantities: \Re\ (panel
a), S\'ersic magnitude (panel b) and \SN\ (panel c). Datapoints
for single mock galaxies are plotted as gray circles. Median
values are plotted as filled blue squares, and error bars show
16--84th quantiles of the distribution in each bin of the quantity
plotted on the x-axis.} \label{fig:simulations}
\end{figure*}

To assess the reliability of the effective radii adopted for the
\UCMG\ selection, we have generated simulated galaxies and we have
run \twodphot\ on them.

In particular, we have generated mock galaxy images with a
Gaussian background noise, given by the background rms measured
for the KiDS images. Then we added artificial galaxies whose
physical parameters, i.e., magnitude $mag_{\rm S}$, S\'ersic index
$n$, effective radius $\Re$ and axis ratio $q$ were assigned based
on a grid of values, chosen according to the range of values of
the \UCMGs\ found in this paper. We have uniformly sampled the
parameters in the following intervals: $0.05 \leq \Re \leq 0.4$
arcseconds, $1\leq n \leq 9$, $0.2\leq q\leq 1$, and $18.5 \leq
m_{\rm S}\leq 21$ mag.  Such mock observations are generated in
different seeing conditions. We have simulated about 400 galaxies.

We have then run \twodphot\ on the mock images with the same setup
used for the real images (see \Sec\ref{sec:sample};
\citealt{Roy+18}). The relative differences between the measured
value of \Re\ and the input value adopted to generate the
simulated galaxies are shown in \Fig\ref{fig:simulations} as a
function of the the output \Re, magnitude and \SN. The figure
shows that the input and output values are well in agreement, with
an average difference of $\sim -0.025$ dex, corresponding to an
average underestimate of about $-6\%$. However, the agreement is
quite better at $\Re \gsim 0.2''$, instead it can reach $\delta
\log \Re \sim -0.15$ dex ($\sim -30\%$) in the smallest galaxies.
We do not observe any trend of this discrepancy with magnitude,
which also suggest that we should expect a negligible impact in
terms of redshift, since the magnitude is strictly correlated with
redshifts in real galaxies.

If we correct our measured sizes for this systematics and apply
the \UCMG\ selection, then $\sim 9\%$ of the photometrically
selected candidates at $z<0.5$ misses the compactness criterion,
and this fraction is reduced to $\sim 5\%$ at $z<0.4$. These
changes are smaller that the typical uncertainties on \Re\ (see
next section) and on number densities (due to Poisson noise,
cosmic variance and errors on \Re\ and \mst). The number densities
in \Fig\ref{fig: abundances} are negligibly affected.

\subsection{Statistical uncertainties}

\begin{figure}
\centering
\includegraphics[width=0.45\textwidth]{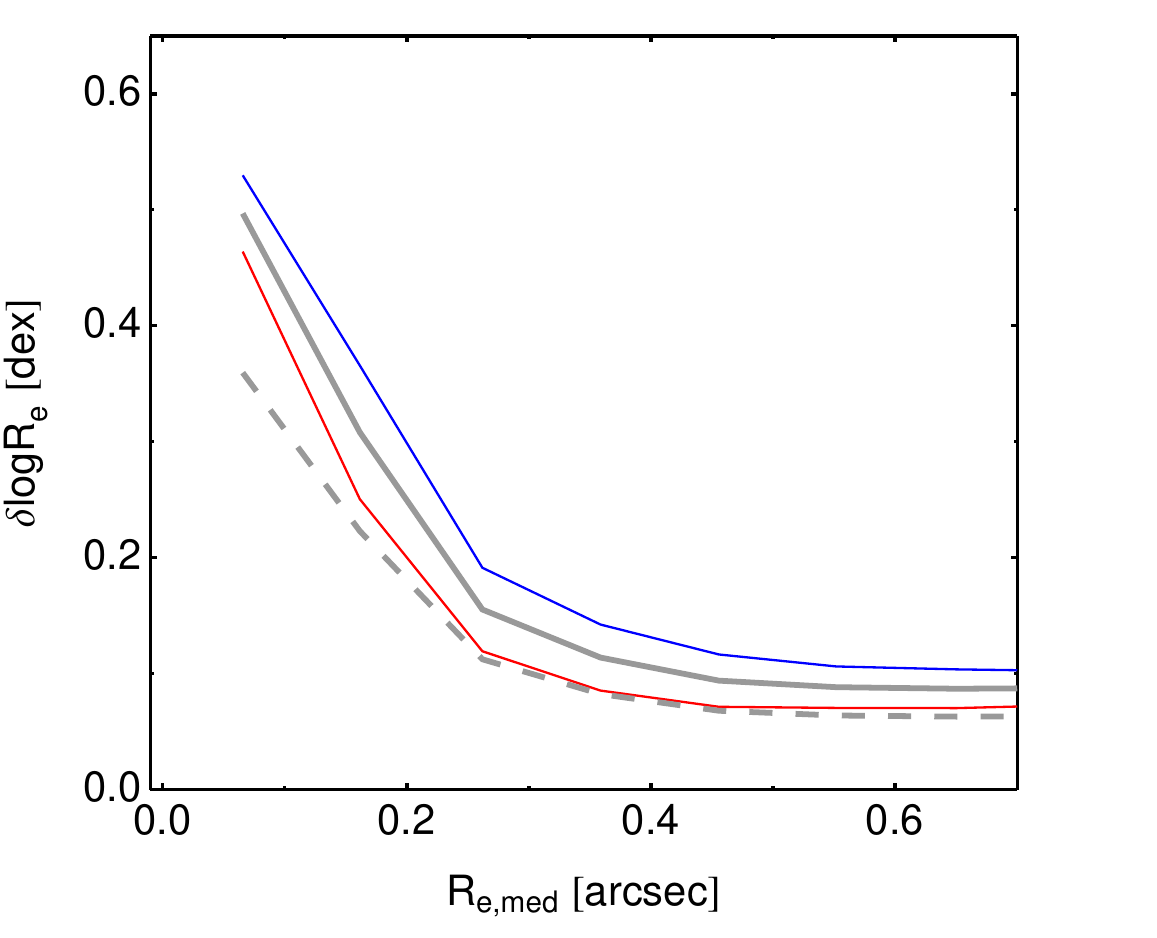}
\caption{Average uncertainties on structural parameters calculated
in bins of the median of the g-, r- and i-band effective radii.
Solid blue (red) lines are determined from the differences between
$g$ and $r$ ($r$ and $i$), while their average value is provided
as solid gray line. The gray dashed line is for the error on the
median \Re\ value.} \label{fig:Re_errors}
\end{figure}

In this section we calculate the uncertainties for the measured
effective radii, following the method explained in \cite{SPIDER-I}
and \cite{Roy+18}. We consider the full sample of galaxies with
masses $\mst
> 8 \times 10^{10}\, \rm \Msun$ and no cut on \Re. We bin the
differences in the $\log \Re$ between contiguous KiDS wavebands
($g$ and $r$, $r$ and $i$ bands) with respect to the median \Re\
from the effective radii in the three bands, which is the quantity
we use in this paper to select the most compact galaxies. From the
distribution of the differences in each \Re\ bin, we calculate the
median absolute deviation (MAD). Measurement errors on $\log \Re$
are computed as $\sigma = MAD/0.6745$ of the corresponding
differences in that bin. The results are shown in
\Fig\ref{fig:Re_errors}.

The error estimate for the median \Re\ (calculated from \Re\
values in the 3 bands) is finally calculated as $1.25 \sigma /
\sqrt{3}$, and plotted as dashed gray line. The uncertainty on the
size stays constant and $\sim 0.1$ dex (i.e. $\sim 20\%$) at $\Re
\gsim 0.3''$, while it reaches a value $\sim 0.3$ dex (i.e., $\sim
80\%$) in the smallest galaxies with median $\Re\sim 0.05''$. We
have used this source of error, together with the uncertainty for
stellar mass, to quantify the impact on the errors of number
density discussed in \Sec\ref{sec:number_counts}.

\section{Spectroscopic sample with redshifts from the literature}\label{app:zspec_sample}

We have collected and discussed in the main text a sample of
\UCMGs\ with spectroscopic redshifts from the literature, which we
named \USS. We have gathered these spectroscopic redshifts from
SDSS (\citealt{Ahn+12_SDSS_DR9, Ahn+14_SDSS_DR10}), GAMA
(\citealt{Driver+11_GAMA}), which overlap the KiDS fields in the
Northern cap, and 2dFLenS (\citealt{Blake+16_2dflens}), which
observed in the Southern hemisphere, with few tiles overlapping
with our northern fields. We have found 46 confirmed \UCMGs\ at
$\zs < 1$, using \modMnozpt\ masses, and 27 using \modMzpt\
values. We show the basic photometric and structural parameters
for such 46 candidates in the spectroscopically selected sample
\USS\ in \Tabs\ref{tab:phot_parameters_sample_lit} and
\ref{tab:struc_parameters_sample_lit}. In
\Tab\ref{tab:phot_parameters_sample_lit} we show r-band Kron
magnitude, aperture magnitudes used in the SED fitting,
spectroscopic redshifts and stellar masses. S\'ersic structural
parameters from the \twodphot\ fit of g-, r- and i-band KiDS
surface photometry, as such as $\chi^{2}$s and \SN s, are
presented in \Tab\ref{tab:struc_parameters_sample_lit}.

\begin{table*}
\centering \caption{Integrated photometry for the sample of
\UCMGs\ with redshifts from the literature. Columns are as in
Table \ref{tab:phot_parameters}. \UCMGs\ are ordered by Right
Ascension. The source of spectroscopic redshifts is reported in
the notes.}\label{tab:phot_parameters_sample_lit}
\resizebox{\textwidth}{!}{\begin{tabular}{ccccccccc} \hline
%
\rm ID & name &  \Mautor & $u_{\rm 6''}$ & $g_{\rm 6''}$ & $r_{\rm 6''}$ & $i_{\rm 6''}$ & \zs & $\log \mst/\Msun$ \\
\hline
1   &   KIDS J084320.59-000543.77   &   18.52   &   21.55   $\pm$   0.06    &   19.71   $\pm$   0.001   &   18.53   $\pm$   0.002   &   18.12   $\pm$   0.005   &   0.24$^{{\it 2   }}$ &   10.93   \\
2   &   KIDS J085344.88+024948.47   &   18.49   &   21.63   $\pm$   0.07    &   19.7    $\pm$   0.001   &   18.5    $\pm$   0.002   &   18.08   $\pm$   0.005   &   0.23$^{{\it 2   }}$ &   10.93   \\
3   &   KIDS J085846.16+020942.62   &   21.27   &   23.08   $\pm$   0.27    &   22.72   $\pm$   0.08    &   21.24   $\pm$   0.021   &   20.     $\pm$   0.023   &   0.74$^{{\it 1   }}$ &   11.49   \\
4   &   KIDS J090324.20+022645.50   &   17.25   &   20.24   $\pm$   0.02    &   18.34   $\pm$   0.001   &   17.34   $\pm$   0.001   &   16.98   $\pm$   0.001   &   0.19$^{{\it 2   }}$ &   11. \\
5   &   KIDS J090935.74+014716.81   &   18.68   &   22.52   $\pm$   0.17    &   20.15   $\pm$   0.001   &   18.75   $\pm$   0.002   &   18.23   $\pm$   0.006   &   0.22$^{{\it 2   }}$ &   11.02   \\
6   &   KIDS J102653.56+003329.15   &   17.39   &   20.49   $\pm$   0.02    &   18.52   $\pm$   0.001   &   17.45   $\pm$   0.001   &   17.04   $\pm$   0.002   &   0.17$^{{\it 1   }}$ &   11.17   \\
7   &   KIDS J103157.23+001041.21   &   20.73   &   23.31   $\pm$   0.41    &   22.34   $\pm$   0.06    &   20.68   $\pm$   0.014   &   19.77   $\pm$   0.017   &   0.53$^{{\it 1   }}$ &   11.3    \\
8   &   KIDS J112825.16-015303.29   &   20.94   &   23.9    $\pm$   0.57    &   22.56   $\pm$   0.06    &   20.91   $\pm$   0.015   &   20.19   $\pm$   0.035   &   0.46$^{{\it 1   }}$ &   10.94   \\
9   &   KIDS J113612.68+010316.86   &   19.01   &   22.07   $\pm$   0.08    &   20.26   $\pm$   0.001   &   19.02   $\pm$   0.003   &   18.59   $\pm$   0.005   &   0.22$^{{\it 2   }}$ &   10.97   \\
10  &   KIDS J114650.20+003710.25   &   20.27   &   23.23   $\pm$   0.3     &   21.59   $\pm$   0.03    &   20.28   $\pm$   0.01    &   19.66   $\pm$   0.019   &   0.68$^{{\it 1   }}$ &   11.31   \\
11  &   KIDS J115652.47-002340.77   &   18.83   &   21.98   $\pm$   0.09    &   20.06   $\pm$   0.001   &   18.83   $\pm$   0.003   &   18.08   $\pm$   0.006   &   0.26$^{{\it 2   }}$ &   11.14   \\
12  &   KIDS J120818.93+004600.16   &   17.74   &   20.65   $\pm$   0.03    &   18.88   $\pm$   0.001   &   17.93   $\pm$   0.001   &   17.56   $\pm$   0.002   &   0.18$^{{\it 2   }}$ &   10.92   \\
13  &   KIDS J120902.53-010503.08   &   18.83   &   22.68   $\pm$   0.21    &   20.16   $\pm$   0.001   &   18.82   $\pm$   0.003   &   18.36   $\pm$   0.008   &   0.27$^{{\it 2   }}$ &   11.04   \\
14  &   KIDS J121152.97-014439.23   &   18.6    &   21.64   $\pm$   0.08    &   19.79   $\pm$   0.001   &   18.65   $\pm$   0.003   &   18.23   $\pm$   0.005   &   0.23$^{{\it 2   }}$ &   10.96   \\
15  &   KIDS J121424.90-020053.72   &   20.57   &   22.72   $\pm$   0.17    &   21.87   $\pm$   0.03    &   20.59   $\pm$   0.012   &   19.51   $\pm$   0.019   &   0.7$^{{\it  1   }}$ &   10.92   \\
16  &   KIDS J121555.27+022828.13   &   20.56   &   23.36   $\pm$   0.32    &   22.21   $\pm$   0.04    &   20.53   $\pm$   0.012   &   19.81   $\pm$   0.017   &   0.47$^{{\it 1   }}$ &   10.97   \\
17  &   KIDS J123254.29+002243.41   &   21.13   &   22.38   $\pm$   0.12    &   22.19   $\pm$   0.04    &   21.08   $\pm$   0.019   &   19.89   $\pm$   0.019   &   0.85$^{{\it 1   }}$ &   10.98   \\
18  &   KIDS J140620.09+010643.00   &   19.16   &   22.55   $\pm$   0.13    &   20.68   $\pm$   0.01    &   19.19   $\pm$   0.004   &   18.7    $\pm$   0.009   &   0.37$^{{\it 2   }}$ &   11.28   \\
19  &   KIDS J140820.77+023348.62   &   20.12   &   23.07   $\pm$   0.27    &   21.76   $\pm$   0.04    &   20.14   $\pm$   0.008   &   19.35   $\pm$   0.015   &   0.6$^{{\it  1   }}$ &   11.07   \\
20  &   KIDS J141039.93+000415.09   &   20.54   &   23.6    $\pm$   0.39    &   22.08   $\pm$   0.04    &   20.5    $\pm$   0.012   &   19.74   $\pm$   0.024   &   0.54$^{{\it 1   }}$ &   10.96   \\
21  &   KIDS J141108.94-003647.51   &   19.22   &   22.27   $\pm$   0.14    &   20.57   $\pm$   0.01    &   19.2    $\pm$   0.004   &   18.74   $\pm$   0.015   &   0.29$^{{\it 2   }}$ &   10.93   \\
22  &   KIDS J141200.92-002038.65   &   19.19   &   22.94   $\pm$   0.27    &   20.76   $\pm$   0.02    &   19.21   $\pm$   0.005   &   18.69   $\pm$   0.015   &   0.28$^{{\it 2   }}$ &   11.08   \\
23  &   KIDS J141415.53+000451.51   &   18.99   &   22.86   $\pm$   0.17    &   20.41   $\pm$   0.001   &   19.0    $\pm$   0.003   &   18.5    $\pm$   0.006   &   0.3$^{{\it  2   }}$ &   11.07   \\
24  &   KIDS J141417.33+002910.20   &   18.77   &   21.73   $\pm$   0.07    &   20.04   $\pm$   0.001   &   18.77   $\pm$   0.003   &   18.34   $\pm$   0.006   &   0.3$^{{\it  2   }}$ &   11.03   \\
25  &   KIDS J141728.44+010626.61   &   17.9    &   20.94   $\pm$   0.04    &   19.06   $\pm$   0.001   &   17.98   $\pm$   0.002   &   17.59   $\pm$   0.003   &   0.18$^{{\it 2   }}$ &   10.96   \\
26  &   KIDS J141828.24-013436.27   &   18.82   &   21.13   $\pm$   0.07    &   19.9    $\pm$   0.001   &   18.8    $\pm$   0.003   &   18.39   $\pm$   0.005   &   0.43$^{{\it 2   }}$ &   11.28   \\
27  &   KIDS J142033.15+012650.38   &   19.38   &   23.58   $\pm$   0.38    &   20.79   $\pm$   0.02    &   19.37   $\pm$   0.005   &   18.89   $\pm$   0.011   &   0.32$^{{\it 2   }}$ &   10.92   \\
28  &   KIDS J142041.17-003511.27   &   18.95   &   22.4    $\pm$   0.14    &   20.37   $\pm$   0.001   &   19.01   $\pm$   0.003   &   18.51   $\pm$   0.005   &   0.25$^{{\it 2   }}$ &   10.96   \\
29  &   KIDS J142606.67+015719.28   &   19.33   &   22.97   $\pm$   0.22    &   20.69   $\pm$   0.01    &   19.3    $\pm$   0.005   &   18.86   $\pm$   0.01    &   0.35$^{{\it 2   }}$ &   11.14   \\
30  &   KIDS J143155.56-000358.65   &   19.34   &   22.74   $\pm$   0.18    &   20.73   $\pm$   0.02    &   19.32   $\pm$   0.004   &   18.82   $\pm$   0.007   &   0.34$^{{\it 2   }}$ &   11.05   \\
31  &   KIDS J143419.53-005231.62   &   19.14   &   22.64   $\pm$   0.17    &   20.79   $\pm$   0.01    &   19.13   $\pm$   0.004   &   18.57   $\pm$   0.005   &   0.46$^{{\it 2   }}$ &   10.96   \\
32  &   KIDS J143459.11-010154.63   &   19.37   &   22.95   $\pm$   0.25    &   20.7    $\pm$   0.01    &   19.36   $\pm$   0.004   &   18.88   $\pm$   0.015   &   0.28$^{{\it 2   }}$ &   10.92   \\
33  &   KIDS J143616.24+004801.40   &   19.24   &   22.78   $\pm$   0.25    &   20.62   $\pm$   0.01    &   19.24   $\pm$   0.004   &   18.76   $\pm$   0.009   &   0.29$^{{\it 2   }}$ &   11.08   \\
34  &   KIDS J143805.25-012729.78   &   19.29   &   22.74   $\pm$   0.19    &   20.64   $\pm$   0.01    &   19.29   $\pm$   0.004   &   18.73   $\pm$   0.007   &   0.29$^{{\it 2   }}$ &   10.94   \\
35  &   KIDS J144138.27-011840.93   &   19.35   &   23.62   $\pm$   0.48    &   20.78   $\pm$   0.01    &   19.35   $\pm$   0.004   &   18.83   $\pm$   0.008   &   0.29$^{{\it 2   }}$ &   11. \\
36  &   KIDS J144924.11-013845.59   &   19.4    &   22.79   $\pm$   0.24    &   20.82   $\pm$   0.02    &   19.39   $\pm$   0.005   &   18.89   $\pm$   0.009   &   0.27$^{{\it 2   }}$ &   10.98   \\
37  &   KIDS J145356.13+001849.32   &   20.32   &   23.24   $\pm$   0.3     &   22.06   $\pm$   0.04    &   20.32   $\pm$   0.009   &   19.68   $\pm$   0.026   &   0.42$^{{\it 1   }}$ &   11.16   \\
38  &   KIDS J145507.26+013458.22   &   21.     &   23.45   $\pm$   0.35    &   22.56   $\pm$   0.06    &   20.92   $\pm$   0.018   &   19.89   $\pm$   0.022   &   0.65$^{{\it 1   }}$ &   11.56   \\
39  &   KIDS J145638.63+010933.24   &   19.66   &   23.21   $\pm$   0.26    &   21.31   $\pm$   0.02    &   19.63   $\pm$   0.006   &   19.09   $\pm$   0.01    &   0.42$^{{\it 1   }}$ &   11.02   \\
40  &   KIDS J155133.16+005709.77   &   19.37   &   24.82   $\pm$   1.76    &   20.95   $\pm$   0.02    &   19.34   $\pm$   0.005   &   18.86   $\pm$   0.012   &   0.42$^{{\it 1   }}$ &   11.05   \\
\hline
41  &   KIDS J021342.59-325755.18   &   21.33   &   23.58   $\pm$   0.43    &   22.73   $\pm$   0.08    &   21.3    $\pm$   0.022   &   20.32   $\pm$   0.034   &   0.75$^{{\it 3   }}$ &   10.97   \\
42  &   KIDS J031536.71-301046.04   &   21.85   &   23.36   $\pm$   0.46    &   23.29   $\pm$   0.1     &   21.77   $\pm$   0.029   &   20.57   $\pm$   0.032   &   0.71$^{{\it 3   }}$ &   11.27   \\
43  &   KIDS J220453.48-311200.94   &   19.32   &   22.9    $\pm$   0.23    &   20.84   $\pm$   0.01    &   19.34   $\pm$   0.004   &   18.87   $\pm$   0.005   &   0.26$^{{\it 3   }}$ &   10.96   \\
44  &   KIDS J222201.71-320447.81   &   17.71   &   20.04   $\pm$   0.01    &   18.6    $\pm$   0.001   &   17.82   $\pm$   0.001   &   17.48   $\pm$   0.002   &   0.19$^{{\it 3   }}$ &   10.92   \\
45  &   KIDS J231410.93-324101.31   &   19.26   &   22.59   $\pm$   0.16    &   20.56   $\pm$   0.001   &   19.26   $\pm$   0.004   &   18.75   $\pm$   0.006   &   0.29$^{{\it 3   }}$ &   10.97   \\
46  &   KIDS J235130.04-311228.42   &   20.12   &   22.79   $\pm$   0.14    &   21.56   $\pm$   0.03    &   20.09   $\pm$   0.007   &   19.32   $\pm$   0.01    &   0.59$^{{\it 3   }}$ &   11. \\
\hline
\end{tabular}}
        \begin{tablenotes}
      \small
      \item $^{\it 1}$ \citealt{Eisenstein+11_SDSSIII}; $^\textit{2}$ \citealt{Dawson+13_GAMA}; $^\textit{3}$
      \citealt{Blake+16_2dflens}.
        \end{tablenotes}
\end{table*}

\begin{table*}
\centering \caption{Structural parameters derived from running
\twodphot\ on g-, r- and i-bands for the sample of galaxies with
spectroscopic redshifts from the literature. Columns are as in
Table
\ref{tab:struc_parameters}.}\label{tab:struc_parameters_sample_lit}
\resizebox{\textwidth}{!}{
\begin{tabular}{cccccccccccccccccccccc}
\hline
\rm & \multicolumn{7}{c}{g-band} & \multicolumn{7}{c}{r-band} & \multicolumn{7}{c}{i-band} \\
 \cmidrule(lr){2-8} \cmidrule(lr){9-15} \cmidrule(lr){16-22}
\rm ID & \Te\ & \Re\ & n & q & $\chi^{2}$ & $\chi^{\prime 2}$ & \SN\ & \Te\ & \Re\ & n & q & $\chi^{2}$ & $\chi^{\prime 2}$ & \SN\ & \Te\ & \Re\ & n & q & $\chi^{2}$ & $\chi^{\prime 2}$ & \SN\ \\
\hline
1   &   0.29    &   1.12    &   4.4     &   0.58    &   1.  &   1.1 &   190.    &   0.26    &   1.01    &   5.59    &   0.61    &   1.2 &   1.7 &   506.    &   0.33    &   1.25    &   8.48    &   0.68    &   1.  &   1.  &   203.        \\
2   &   0.39    &   1.44    &   3.83    &   0.46    &   1.  &   1.  &   185.    &   0.34    &   1.25    &   4.13    &   0.44    &   1.1 &   1.5 &   443.    &   0.34    &   1.26    &   4.  &   0.42    &   1.1 &   1.1 &   190.    \\
3   &   0.09    &   0.64    &   6.13    &   0.32    &   1.  &   0.9 &   14. &   0.18    &   1.3 &   6.64    &   0.66    &   1.  &   1.  &   58. &   0.26    &   1.89    &   6.67    &   0.54    &   1.  &   0.9 &   51. \\
4   &   0.46    &   1.45    &   4.34    &   0.24    &   1.  &   1.4 &   492.    &   0.23    &   0.73    &   7.04    &   0.29    &   1.3 &   2.9 &   1003.   &   0.54    &   1.7 &   4.82    &   0.26    &   1.1 &   1.3 &   641.    \\
5   &   0.56    &   1.96    &   9.95    &   0.81    &   0.8 &   0.9 &   110.    &   0.14    &   0.48    &   10.07   &   0.76    &   1.1 &   1.8 &   357.    &   0.3 &   1.05    &   9.97    &   0.77    &   1.  &   1.  &   152.    \\
6   &   0.43    &   1.26    &   2.7     &   0.29    &   1.1 &   11.5    &   360.    &   0.32    &   0.95    &   3.64    &   0.29    &   1.1 &   25.8    &   1092.   &   0.34    &   1.01    &   3.18    &   0.29    &   1.  &   9.6 &   464.    \\
7   &   0.22    &   1.38    &   6.93    &   0.65    &   1.  &   1.1 &   18. &   0.22    &   1.42    &   6.05    &   0.86    &   1.  &   1.  &   84. &   0.5 &   3.19    &   6.81    &   0.96    &   1.  &   1.  &   69. \\
8   &   0.31    &   1.78    &   8.8     &   0.21    &   1.  &   1.1 &   16. &   0.25    &   1.46    &   8.54    &   0.44    &   1.  &   1.  &   74. &   0.21    &   1.22    &   3.66    &   0.59    &   1.  &   1.3 &   32. \\
9   &   0.29    &   1.02    &   4.03    &   0.26    &   1.1 &   1.  &   130.    &   0.14    &   0.48    &   7.96    &   0.27    &   1.1 &   1.2 &   327.    &   0.11    &   0.4 &   8.07    &   0.25    &   1.  &   1.  &   188.    \\
10  &   0.11    &   0.78    &   8.54    &   0.81    &   1.  &   1.  &   36. &   0.2 &   1.41    &   9.26    &   0.99    &   1.1 &   1.5 &   101.    &   0.85    &   5.98    &   0.97    &   1.  &   1.  &   1.  &   52. \\
11  &   0.37    &   1.47    &   4.79    &   0.38    &   1.  &   1.  &   140.    &   0.2 &   0.79    &   6.53    &   0.4 &   1.  &   1.2 &   381.    &   0.26    &   1.03    &   8.63    &   0.38    &   1.  &   0.9 &   163.    \\
12  &   0.5     &   1.49    &   7.65    &   0.38    &   1.  &   8.  &   210.    &   0.45    &   1.34    &   7.52    &   0.41    &   1.1 &   23.2    &   673.    &   0.72    &   2.14    &   7.51    &   0.45    &   1.  &   11.1    &   357.    \\
13  &   0.36    &   1.49    &   2.64    &   0.3     &   1.  &   0.9 &   127.    &   0.35    &   1.47    &   2.88    &   0.28    &   1.1 &   1.5 &   410.    &   0.35    &   1.46    &   2.42    &   0.27    &   1.  &   0.9 &   128.    \\
14  &   0.52    &   1.94    &   8.65    &   0.52    &   1.  &   1.1 &   154.    &   0.38    &   1.42    &   7.59    &   0.61    &   1.  &   1.3 &   363.    &   0.25    &   0.93    &   8.95    &   0.59    &   1.  &   1.  &   193.    \\
15  &   0.07    &   0.53    &   7.23    &   0.18    &   1.  &   0.9 &   29. &   0.35    &   2.53    &   9.09    &   0.61    &   1.  &   1.  &   80. &   0.2 &   1.42    &   9.33    &   0.55    &   1.  &   1.  &   51. \\
16  &   0.17    &   1.01    &   0.69    &   0.14    &   1.  &   0.9 &   29. &   0.2 &   1.19    &   3.6 &   0.51    &   1.  &   1.  &   97. &   0.18    &   1.04    &   4.96    &   0.49    &   1.  &   1.  &   69. \\
17  &   0.13    &   1.      &   7.39    &   0.62    &   1.  &   1.  &   30. &   0.1 &   0.77    &   6.01    &   0.62    &   1.  &   1.  &   66. &   0.17    &   1.32    &   3.77    &   0.73    &   1.  &   1.  &   67. \\
18  &   0.32    &   1.64    &   6.76    &   0.29    &   1.  &   1.2 &   85. &   0.27    &   1.36    &   7.52    &   0.33    &   1.1 &   1.6 &   276.    &   0.25    &   1.27    &   9.23    &   0.35    &   1.  &   1.2 &   115.    \\
19  &   0.17    &   1.17    &   4.88    &   0.35    &   1.  &   0.9 &   25. &   0.11    &   0.76    &   9.27    &   0.66    &   1.  &   1.3 &   121.    &   0.57    &   3.79    &   6.84    &   0.48    &   1.  &   1.1 &   70. \\
20  &   0.18    &   1.12    &   5.27    &   0.28    &   1.  &   1.  &   29. &   0.18    &   1.17    &   3.97    &   0.47    &   1.  &   1.9 &   95. &   0.36    &   2.26    &   7.23    &   0.47    &   1.  &   1.1 &   49. \\
21  &   0.4     &   1.76    &   2.8     &   0.56    &   1.  &   1.1 &   76. &   0.3 &   1.32    &   3.13    &   0.45    &   1.  &   1.1 &   261.    &   0.25    &   1.1 &   4.71    &   0.4 &   1.  &   0.9 &   75. \\
22  &   0.34    &   1.44    &   5.      &   0.33    &   1.  &   0.9 &   52. &   0.32    &   1.35    &   6.3 &   0.39    &   1.  &   1.  &   217.    &   0.33    &   1.41    &   6.13    &   0.42    &   1.  &   1.  &   66. \\
23  &   0.38    &   1.69    &   3.99    &   0.46    &   1.  &   1.  &   108.    &   0.31    &   1.4 &   4.26    &   0.42    &   1.  &   1.2 &   316.    &   0.3 &   1.33    &   5.03    &   0.42    &   1.  &   0.9 &   169.    \\
24  &   0.31    &   1.36    &   5.12    &   0.81    &   1.  &   1.  &   142.    &   0.32    &   1.41    &   4.72    &   0.85    &   1.  &   1.2 &   383.    &   0.27    &   1.18    &   7.84    &   0.88    &   1.  &   1.  &   173.    \\
25  &   0.54    &   1.63    &   3.35    &   0.35    &   1.  &   1.1 &   244.    &   0.49    &   1.47    &   3.92    &   0.31    &   1.1 &   1.5 &   555.    &   0.45    &   1.36    &   4.74    &   0.33    &   1.  &   1.1 &   294.    \\
26  &   0.22    &   1.22    &   3.66    &   0.52    &   1.  &   1.8 &   168.    &   0.23    &   1.3 &   3.95    &   0.58    &   1.  &   6.9 &   399.    &   0.24    &   1.36    &   3.15    &   0.56    &   1.  &   2.8 &   232.    \\
27  &   0.19    &   0.9     &   3.87    &   0.15    &   1.  &   0.9 &   72. &   0.22    &   1.02    &   4.04    &   0.17    &   1.  &   1.1 &   237.    &   0.23    &   1.07    &   3.67    &   0.21    &   1.  &   1.  &   100.    \\
28  &   0.37    &   1.42    &   6.64    &   0.64    &   1.1 &   1.  &   94. &   0.31    &   1.23    &   4.76    &   0.62    &   1.  &   1.3 &   299.    &   0.34    &   1.34    &   5.67    &   0.61    &   1.  &   0.9 &   156.    \\
29  &   0.28    &   1.39    &   7.43    &   0.35    &   1.  &   1.  &   77. &   0.18    &   0.89    &   8.44    &   0.3 &   1.5 &   1.2 &   244.    &   0.28    &   1.37    &   6.47    &   0.25    &   1.  &   0.9 &   115.    \\
30  &   0.26    &   1.26    &   4.24    &   0.7     &   0.9 &   0.9 &   69. &   0.28    &   1.36    &   3.31    &   0.78    &   1.  &   1.1 &   272.    &   0.3 &   1.47    &   2.89    &   0.7 &   1.  &   0.9 &   174.    \\
31  &   0.27    &   1.56    &   2.84    &   0.29    &   1.  &   1.  &   83. &   0.23    &   1.37    &   3.21    &   0.26    &   1.2 &   1.2 &   297.    &   0.2 &   1.2 &   3.29    &   0.3 &   1.  &   1.  &   199.    \\
32  &   0.17    &   0.71    &   6.34    &   0.53    &   1.  &   1.  &   82. &   0.19    &   0.84    &   5.21    &   0.5 &   1.  &   1.1 &   249.    &   0.19    &   0.8 &   7.52    &   0.34    &   1.  &   1.  &   72. \\
33  &   0.51    &   2.26    &   5.63    &   0.53    &   1.  &   1.  &   81. &   0.33    &   1.47    &   7.59    &   0.56    &   1.  &   1.3 &   255.    &   0.3 &   1.33    &   8.73    &   0.5 &   1.  &   0.9 &   108.    \\
34  &   0.37    &   1.6     &   4.8     &   0.37    &   1.  &   1.1 &   95. &   0.28    &   1.19    &   4.07    &   0.38    &   1.  &   1.4 &   259.    &   0.26    &   1.11    &   4.11    &   0.38    &   1.  &   1.5 &   149.    \\
35  &   0.37    &   1.61    &   6.28    &   0.28    &   1.  &   0.9 &   89. &   0.32    &   1.4 &   4.73    &   0.29    &   1.  &   1.2 &   246.    &   0.32    &   1.42    &   6.48    &   0.29    &   1.  &   0.9 &   137.    \\
36  &   0.35    &   1.43    &   5.48    &   0.23    &   1.  &   1.1 &   74. &   0.27    &   1.12    &   6.38    &   0.39    &   1.1 &   1.7 &   216.    &   0.37    &   1.51    &   5.81    &   0.33    &   1.  &   1.2 &   128.    \\
37  &   0.22    &   1.2     &   6.55    &   0.33    &   1.  &   0.9 &   23. &   0.36    &   1.99    &   7.11    &   0.47    &   1.  &   1.  &   109.    &   0.23    &   1.3 &   6.66    &   0.44    &   1.  &   1.  &   39. \\
38  &   0.17    &   1.16    &   3.9     &   0.27    &   1.  &   1.1 &   20. &   0.14    &   0.98    &   5.13    &   0.4 &   1.  &   1.  &   78. &   0.16    &   1.08    &   4.23    &   0.4 &   1.  &   1.  &   64. \\
39  &   0.29    &   1.6     &   5.37    &   0.54    &   1.  &   1.  &   56. &   0.14    &   0.78    &   6.9 &   0.41    &   1.  &   1.3 &   198.    &   0.22    &   1.23    &   3.24    &   0.51    &   1.  &   0.9 &   107.    \\
40  &   0.14    &   0.76    &   6.14    &   0.28    &   1.1 &   1.  &   54. &   0.09    &   0.51    &   4.83    &   0.32    &   1.  &   1.3 &   239.    &   0.13    &   0.74    &   4.45    &   0.28    &   1.  &   1.  &   105.    \\
\hline
41  &   0.1     &   0.75    &   6.34    &   0.44    &   1.  &   1.1 &   14. &   0.22    &   1.63    &   3.12    &   0.63    &   1.  &   0.9 &   56. &   0.13    &   0.94    &   3.39    &   0.33    &   1.  &   0.9 &   36. \\
42  &   0.17    &   1.21    &   3.97    &   0.76    &   1.  &   1.  &   15. &   0.18    &   1.28    &   3.18    &   0.35    &   1.  &   1.  &   55. &   0.22    &   1.59    &   1.98    &   0.46    &   1.  &   1.  &   46. \\
43  &   0.34    &   1.35    &   6.48    &   0.34    &   1.  &   1.  &   74. &   0.35    &   1.38    &   6.36    &   0.31    &   1.1 &   1.3 &   282.    &   0.44    &   1.76    &   3.91    &   0.29    &   1.  &   1.  &   207.    \\
44  &   0.49    &   1.58    &   7.18    &   0.53    &   1.2 &   2.5 &   349.    &   0.4 &   1.27    &   2.09    &   0.64    &   1.3 &   2.1 &   694.    &   0.4 &   1.29    &   1.73    &   0.65    &   1.1 &   4.  &   425.    \\
45  &   0.36    &   1.59    &   4.71    &   0.46    &   1.  &   0.9 &   106.    &   0.29    &   1.29    &   5.14    &   0.43    &   1.  &   1.2 &   286.    &   0.31    &   1.34    &   3.52    &   0.43    &   1.  &   1.  &   159.    \\
46  &   0.18    &   1.23    &   6.79    &   0.66    &   1.  &
0.9 &   46. &   0.11    &   0.75    &   8.18    &   0.71    &   1.
&   1.3 &   160.    &   0.13    &   0.85    &   8.26    &   0.74
&   1.1 &   0.9 &   108.    \\
\hline
\end{tabular}
}
\end{table*}

\section{Number densities across the KiDS area}\label{app:syst}

In order to investigate the homogeneity of our density estimates
across the KiDS field, quantifying the impact of Poisson noise and
Cosmic Variance, we divide the sample of \UCMG\ candidates from
\UP\ in different subsamples, calculate the densities as discussed
in the paper and show the results in \Fig\ref{fig:
densities_systematics}.

We start showing the results for uncorrected and corrected
densities calculated using in turn DR1/2 and DR3 tiles (panels a
and e). Densities calculated with \UCMG\ candidates in DR3 tiles
are, on average, $\sim 0.2$ dex larger than those found in DR1/2
tiles, reaching a maximum difference of $\sim 0.4$ dex. More
moderate changes are found among North and South fields (penels b
and f), with the former producing larger densities, this
discrepancy is larger at lower redshift, but stays below $0.2$
dex. If the KiDS patch is divided in East and West fields (the
separation is set at $\rm RA = 180.5 \, \rm deg$, panels c and g),
we observe differences in the lowest redshift bin, as well as in
the case of 3 random areas selected in the Northern cap (panels d
and h). Most of the differences observed are easily accounted by
Poisson noise and Cosmic Variance. This is the case of the lowest
redshift bin, and holds mainly for the results discussed in panels
d and h, where the strong differences observed are clearly due to
very poor statistics (one or no \UCMGs\ at all are found in this
redshift bin in the three random selected areas in panels d and
h). However, we cannot exclude some of the discrepancies among
DR1/2 and DR3 or among North and South fields to be caused by data
inhomogeneities. One possible source of such discrepancies should
be related to structural parameter determination. For \UCMG\
candidates in DR1/2 and DR3, structural parameters were determined
at different epochs, using inhomogeneous KiDS tiles. Although
these differences do not produce a significant change in the
overall number densities, we will further investigate these issues
in future analysis of next KiDS data releases.

\begin{figure*}
\includegraphics[width=0.9\textwidth]{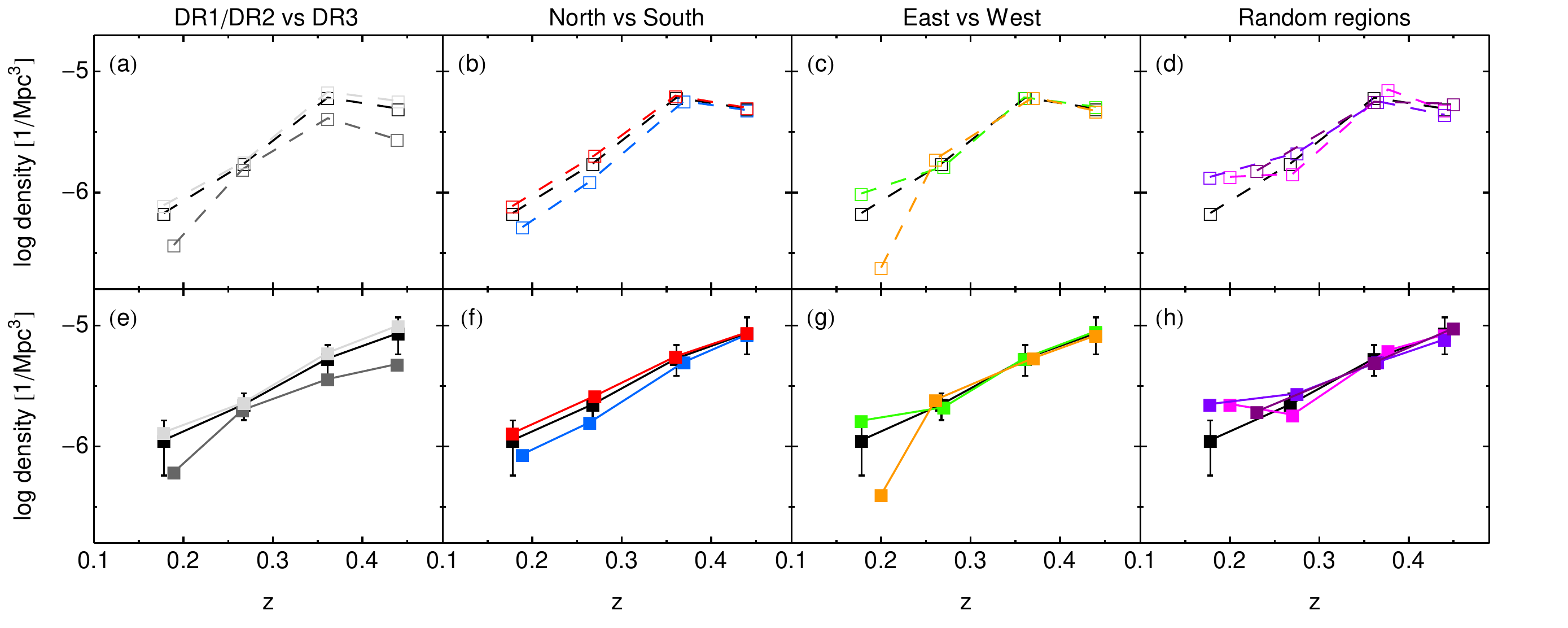}
\caption{Internal consistency of Number densities. Following
\Fig\ref{fig: abundances}, open (filled) black squares, with
dashed (solid) lines, plot the number density before (after)
correction for systematics, for the sample assuming \modMnozpt\
masses. These are for the sample selected across the whole
KiDS--DR1/2/3 area. We plot number densities for the following
subsamples of \UCMGs: a,e) DR1/2 (dark gray) vs. DR3 (light gray);
b,f) North (red) vs. South (blue) fields; c,g) East (green) vs.
West (orange) fields; d,h) three random regions in the North
hemisphere containing $\sim 30$ tiles, corresponding to an
effective area of $\sim 23\sqd$ each (pink, purple and violet).
Northern (Southern) fields have DEC$>-5$ ($< -5$) deg. East (West)
stays for regions with RA $> 180.5$ ($< 180.5$) deg.}\label{fig:
densities_systematics}
\end{figure*}

\end{document}